%% file: paper.tex
\pdfoutput=1
\documentclass[preprint,12pt]{elsarticle}

\usepackage{mathtools}
\usepackage{amssymb}
\usepackage{amsmath}
\usepackage[T1]{fontenc}
\usepackage{ltablex} 
\usepackage{float}
\usepackage{xcolor}

\newcommand{\hatpc}{HA-TPCs~}

\usepackage{siunitx}
\DeclareSIUnit{\clight}{\ensuremath{c}}
\usepackage{tikz}
\usepackage{upgreek}
\usepackage{gensymb}
\usepackage{comment}

\usepackage{lineno}

\usepackage{subcaption}
\usepackage{mwe}
\usepackage{pdfpages}

\usepackage{hyperref}
\hypersetup{colorlinks,allcolors=black}
\usepackage{comment}
\usepackage{physics}
\usepackage{accents}

\usepackage[section]{placeins}

\graphicspath{{figures/}{figures/dEdx/}}

\journal{Nuclear Instruments and Methods}

\begin{document}

\begin{frontmatter}





\title{Performance of the High-Angle Time Projection Chambers in the Upgraded T2K Off-Axis Near Detector}
\input{author.tex}

\begin{abstract}
The off-axis magnetic near detector of the T2K experiment has undergone a significant upgrade, including the construction and installation of two new Time Projection Chambers featuring innovative resistive Micromegas technology and a field cage composed of thin composite walls. This paper provides a detailed description of the new components of the chambers, including the gas system, gas monitoring chambers, and data acquisition system. Additionally, it reports the results of extensive testing using both neutrino beams and cosmic rays, with comparisons between data and Monte Carlo simulations. The new detectors achieve improved spatial resolution and enhanced particle identification capabilities which are crucial for the precision goals of the T2K experiment.
\end{abstract}
\begin{keyword}
\texorpdfstring{Time Projection Chambers \sep Resistive Micromegas \sep Field Cage \sep spatial and momentum resolution \sep dE/dx resolution}{Time Projection Chambers, Resistive Micromegas, Field Cage, spatial and momentum resolution, dE/dx resolution}  



\end{keyword}

\end{frontmatter}
\newpage
\tableofcontents
\newpage
\section{Introduction}
\input{introduction.tex}

\label{sec:introduction}
\section{Field Cage Description and Performance}
\label{sec:FieldCage}
\input{FieldCage.tex}
\section{Readout Description and Performance}
\label{sec:ERAMs}
\input{ERAM.tex}

\section{Electronics }
\label{sec:Electronics Noise}
\input{ElectronicsNoise.tex}

\section{Slow Control and DAQ}
\label{sec:DAQ}
\input{DAQ.tex}
\section{Gas System}
\label{sec:GasSystem}
\input{GasSystem.tex}
\section{Gas Monitoring Chamber}
\label{sec:GMC}
\input{GMC.tex}

\section{HA-TPC Reconstruction and Simulation}
\label{sec:HATRecoSim}
\input{HATRecoSim.tex}
\section{HA-TPC Performance}
\label{sec:HATPerformance}
\input{HATPerformance.tex}

\section{Conclusions}
\label{sec:conclusion}
\input{conclusions.tex}
\section*{Acknowledgments}
This paper is dedicated to the memory of our colleague Denis Calvet, whose contribution to this work and to the collaboration was invaluable.\\ 
We acknowledge the support of CEA and CNRS/IN2P3, France; CERN; DFG, Germany; INFN, Italy.\\ 
We thank the ND280/T2K collaboration for the continuous support and  for active software development without which the studies of the HA-TPC performances wouldn’t have been possible.\\
We would like to thank Marc Lefevre (IRFU/DEDIP) for his support in the fabrication of mechanical parts. We are also grateful to Sandrine Javello (IRFU/DEDIP), whose contribution went well beyond her essential role in administrative support for the IRFU antenna at CERN, and who was deeply involved in handling the logistics. Finally, we wish to acknowledge the support of CERN services, in particular the logistics team, with special thanks to Stephanie Krattinger.\\ 
In addition, the participation of individual researchers has been supported by the European Union’s Horizon 2020 Research and Innovation Programme under the grant number RISE-GA822070-JENNIFER2 and the Horizon Europe Marie Sklodowska-Curie Staff Exchange project JENNIFER3 Grant Agreement no.101183137 as well as by the French "Agence nationale pour la recherche" under grant numbers ANR-19-CE31-0001 and ANR-21-CE31-0008.\\
The research leading to these results has received funding from the Spanish Ministry of Science and Innovation \textnormal{PID2022-136297NB\_I00/AEI\allowbreak/10.13039\allowbreak/501100011033\allowbreak/FEDER,\allowbreak\ UE}.  With the support from the Secretariat for Universities and Research of the Ministry of Business and Knowledge of the Government of Catalonia and the European Social Fund (2022FI$_{B}$ 00336) and AGAUR 2021-SGR-01506. This study was supported by MICIIN, with funding from the European Union NextGenerationEU(PRTR-C17.I1), and by Generalitat de Catalunya. IFAE is partially funded by the CERCA program of the Generalitat de Catalunya.






\bibliographystyle{elsarticle-num}
\bibliography{bibliography}

\end{document}

%% file: author.tex
\author[saclay]{K.~Aivazelis}
\author[saclay]{D.~Attié}
\author[lpnhe]{P.~Billoir}
\author[lpnhe]{A.~Blanchet}
\author[padova]{G.~Bortolato}
\author[saclay]{S.~Bolognesi}
\author[saclay]{R.~Boullon}
\author[Bari]{N.~F.~Calabria}
\author[saclay]{D.~Calvet\fnref{fnref4}\textsuperscript{\dag}}
\author[IFAE]{M.P.~Casado\fnref{fnref3}}
\author[Bari]{M.~G.~Catanesi}
\author[padova]{M.~Cicerchia}
\author[padova]{G.~Cogo}
\author[padova]{G.~Collazuol\fnref{fnref2}}
\author[saclay]{P.~Colas}
\author[saclay]{D.~Cotte}
\author[padova]{D.~D'Ago}
\author[lpnhe]{C.~Dalmazzone}
\author[saclay]{T.~Daret}
\author[cern]{R.~de Oliveira}
\author[saclay]{A.~Delbart}
\author[lpnhe]{J.~Dumarchez}
\author[Warsaw]{K.~Dygnarowicz}
\author[saclay]{S.~Emery-Schrenk}
\author[saclay]{A.~Ershova}
\author[saclay]{G.~Eurin}
\author[padova]{M.~Feltre}
\author[padova]{C.~Forza}
\author[lpnhe]{A.~N.~Gacino Olmedo}
\author[legnaro]{A.~Gambalonga}
\author[lpnhe]{C.~Giganti}
\author[legnaro]{F.~Gramegna}
\author[cern]{P.~Granger}
\author[cern]{R.~Guida}
\author[lpnhe]{M.~Guigue}
\author[saclay]{S.~Hassani\fnref{fnref1}}
\author[saclay]{D.~Henaff}
\author[padova]{F.~Iacob}
\author[IFAE]{C.~Jesús-Valls}
\author[saclay]{S.~Joshi}
\author[Warsaw]{R.~Kurjata}
\author[padova]{M.~Lamoureux}
\author[saclay]{J.~F.~Laporte}
\author[saclay]{M.~Lehuraux}
\author[padova]{S.~Levorato}
\author[padova]{A.~Longhin}
\author[IFAE]{T.~Lux}
\author[Bari]{L.~Magaletti}
\author[legnaro]{T.~Marchi}
\author[padova]{D.~Marchesini}
\author[padova]{L.~Mareso}
\author[padova]{M.~Mattiazzi}
\author[padova]{M.~Mezzetto}
\author[cern]{B.~Mehl}
\author[IFAE]{E.~Miller}
\author[lpnhe]{L.~Mellet}
\author[cern]{L.~Munteanu}
\author[lpnhe]{Q.~V.~Nguyen}
\author[Bari]{N.~Ospina}
\author[lpnhe]{Y.~Orain}
\author[padova]{R.~Palumbo}
\author[Bari]{C.~Pastore}
\author[lpnhe]{J.-M.~Parraud}
\author[lpnhe]{E.~Pierre}
\author[IFAE]{C.~Pio}
\author[cern]{O.~Pizzirusso}
\author[lpnhe]{B.~Popov}
\author[padova]{F.~Pupilli}
\author[Bari]{E.~Radicioni}
\author[saclay]{Ch.~Riccio}
\author[padova]{L.~Rinaldi}
\author[saclay]{F.~Rossi}
\author[aachen]{S.~Roth}
\author[lpnhe]{L.~Russo}
\author[lpnhe]{S.~Russo}
\author[Warsaw]{A.~Rychter}
\author[lpnhe]{W.~Saenz Arevalo}
\author[padova]{L.~Scomparin}
\author[saclay]{Ph.~Schune}
\author[aachen]{D.~Smyczek}
\author[Bari]{R.~Spina}
\author[aachen]{J.~Steinmann}
\author[lpnhe]{S.~Suvorov}
\author[aachen]{N.~Thamm}
\author[lpnhe]{D.~Terront}
\author[cern]{A.~Teixeira}
\author[lpnhe]{F.~Toussenel}
\author[Bari]{V.~Valentino}
\author[IFAE]{D.~Vargas}
\author[IFAE]{M.~Varghese}
\author[saclay]{G.~Vasseur}
\author[saclay]{C.~Vuillemin}
\author[lpnhe]{U.~Virginet}
\author[saclay]{Ch.~Winterstein}
\author[lpnhe]{U.~Yevarouskaya}
\author[Warsaw]{M.~Ziembicki}
\author[lpnhe]{M.~Zito}

\address[saclay]{IRFU, CEA, Universit\'e Paris-Saclay, Gif-sur-Yvette, France}
\address[lpnhe]{LPNHE Paris, Sorbonne Universit\'e, CNRS/IN2P3, Paris 75252, France}
\address[padova]{INFN Sezione di Padova and Universit\`a di Padova, Dipartimento di Fisica e Astronomia, Padova, Italy}
\address[cern]{CERN, European Organization for Nuclear Research, Geneva, Switzerland}
\address[Warsaw]{Warsaw University of Technology, Poland}
\address[Bari]{INFN , Universita’ e Politecnico di Bari}
\address[IFAE]{Institut de Física d’Altes Energies (IFAE) - The Barcelona Institute of Science and Technology (BIST), Campus UAB, 08193 Bellaterra (Barcelona), Spain}
\address[legnaro]{INFN: Laboratori Nazionali di Legnaro (LNL), Padova , Italy}
\address[aachen]{III. Physikalisches Institut, RWTH Aachen University, Aachen, Germany}


\cortext[cor1]{Corresponding author}
\fntext[fnref1]{samira.hassani@cea.fr}
\fntext[fnref2]{gianmaria.collazuol@unipd.it}

\fntext[fnref4]{\it Deceased. In memory of Denis Calvet, whose work and spirit continue to inspire us.} 

\fntext[fnref3]{Also at Universidad Autónoma de Barcelona}

%% file: introduction.tex
The T2K (Tokai to Kamioka) experiment~\cite{T2K:2011qtm} is a long-baseline neutrino oscillation experiment being conducted in Japan. It investigates neutrino oscillation parameters using a high-intensity muon (anti)neutrino beam, with an energy peak of approximately 600 MeV, produced at the \mbox{J-PARC} facility. The beam is first measured 280 m from the production target by a set of near detectors, including the off-axis magnetic detector ND280, which monitor and constrain systematic uncertainties related to the neutrino flux and neutrino interaction models before oscillations occur. The beam then travels 295 km to the far detector, Super-Kamiokande, which observes the disappearance of muon (anti)neutrinos and the appearance of electron (anti)neutrinos. T2K has provided the first indications of CP violation in the leptonic sector~\cite{Abe:2019vii}. To overcome statistical limitations and further investigate this phenomenon, a new data-taking phase began in 2024, featuring higher beam power and upgraded detectors.

\begin{figure}[htb]
  \centering
  \includegraphics[width=\linewidth]{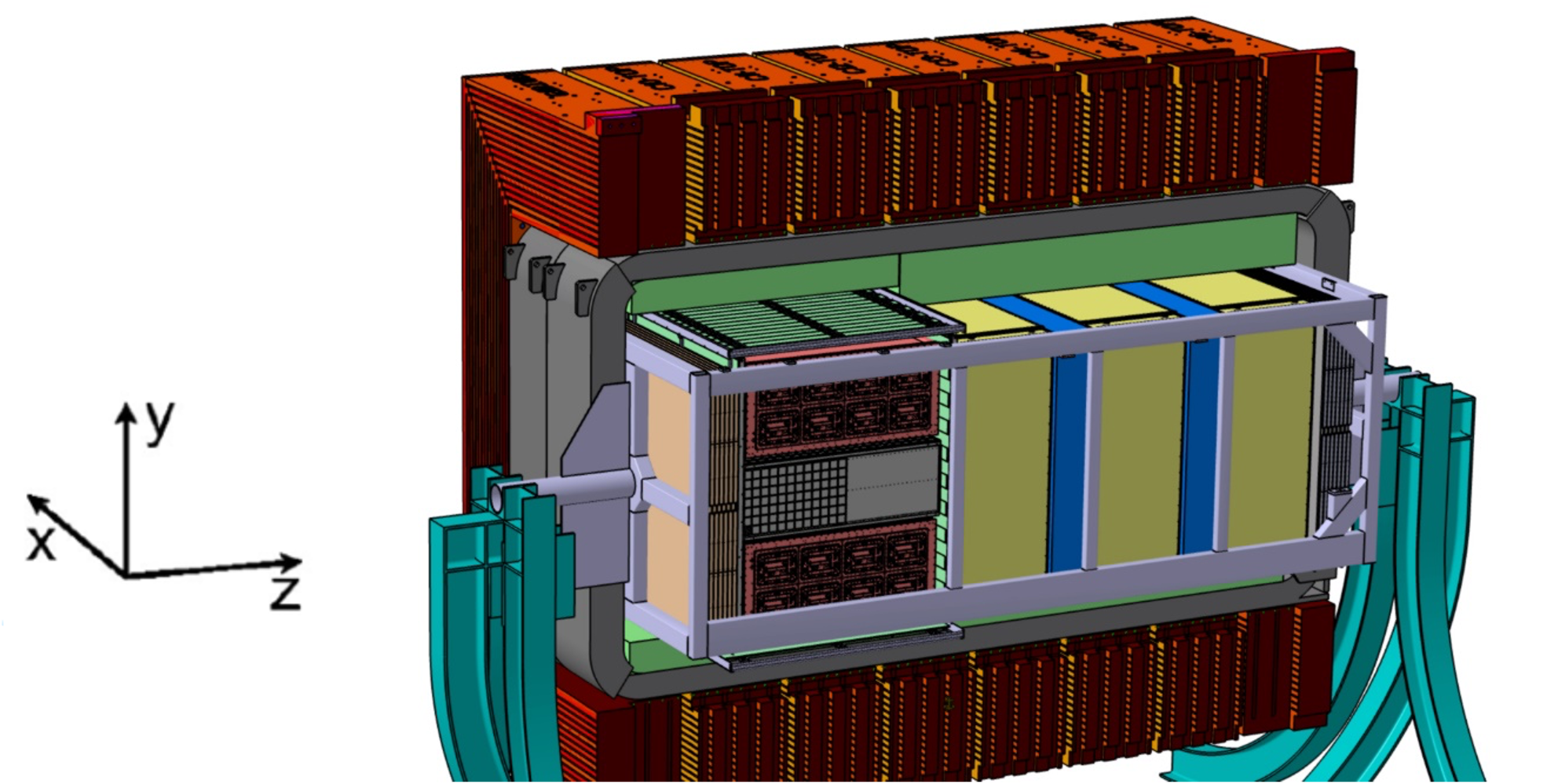}
  \caption{Schematic view of the upgraded ND280 installed in the opened magnet. The newly installed subdetectors are located in the left part of the drawing: two horizontal top and bottom HA-TPCs (black boxes) surround the new highly segmented Super-FGD active target (grey box). The top TOF panel is also visible. The original tracking system of the ND280 -- three vertical TPCs (yellow boxes) and two FGDs (blue boxes) -- is also shown. The \mbox{J-PARC} neutrino beam is coming from the left. Cosmic rays traverse the detector vertically. The coordinate reference system used for ND280 is also shown. The $x$-axis is parallel to the magnetic field, the $y$-axis is opposite to gravity, and the $z$-axis is along the incoming neutrino beam.} 
  \label{fig:sketch_ND280upgrade}
\end{figure}

The ND280 detector has recently been upgraded with several new subdetectors installed inside a large magnet upstream of the existing setup, as illustrated in Fig.~\ref{fig:sketch_ND280upgrade}. This upgraded system includes the Super-FGD (Fine Grain Detector), a highly segmented active target, positioned between two high-angle time projection chambers (HA-TPCs), all surrounded by six time-of-flight detector panels (ToF). The installation of these components at \mbox{J-PARC} was finalized in May 2024, followed by the collection of the first neutrino data in June 2024. The coordinate reference system used for ND280 is shown in Fig.~\ref{fig:sketch_ND280upgrade}. The $x$-axis is parallel to the main component of the magnetic field, the $y$-axis is opposite to the direction of gravity, and the orthogonal $z$-axis is aligned with the incoming neutrino beam direction.

The HA-TPC incorporates new field cage design aimed at reducing dead space and maximizing the available tracking volume. Its readout system utilizes the advanced resistive Micromegas technology~\cite{Attie:2022smn}, offering enhanced performance and reliability. 
The HA-TPC comprises a top and a bottom TPC, each featuring two endplates. Every endplate hosts eight resistive Micromegas detectors that share the same drift volume, resulting in a total of 32 detectors. 
The detector characterization and performance are detailed in Refs.~\cite{Attie:2022smn, Ambrosi:2023smx, Attie:2021yeh, Attie:2019hua}. The detector design of the HA-TPCs was successfully validated through a series of cosmic tests and test-beam campaigns conducted at CERN and DESY. These tests not only confirmed the reliability of the detector technologies but also provided valuable information on their performance. A spatial resolution better than 800~\textmu m and a $\mathrm{d}E/\mathrm{d}x$ resolution better than 10\% were achieved across all incident angles and drift distances~\cite{Attie:2022smn}.

This paper presents a detailed description of the new chamber components — including the field cage, the new readout system based on the resistive Micromegas technology, gas system, gas monitoring chambers, and data acquisition system — and evaluates the performance of the two new HA-TPCs using cosmic ray and neutrino beam data collected at J-PARC. It also reports the results of extensive tests, comparing experimental data with Monte Carlo simulations.

The paper is structured as follows. Section~\ref{sec:FieldCage} describes the design, production, and electric field uniformity of the field cage. The production and performance of the new readout modules are detailed in Section~\ref{sec:ERAMs}. Sections~\ref{sec:Electronics} and~\ref{sec:DAQ} present the architecture of the readout electronics and the data acquisition system, respectively. The gas system is introduced in Section~\ref{sec:GasSystem}, followed by the gas monitoring chamber in Section~\ref{sec:GMC}. Section~\ref{sec:HATRecoSim} outlines the reconstruction and simulation framework developed for the HA-TPCs. Building on this, Section~\ref{sec:HATPerformance} provides a detailed assessment of the detector performance using cosmic rays and neutrino data, compared with simulation. Finally, conclusions are summarized in Section~\ref{sec:conclusion}.

%% file: FieldCage.tex
The field cages (FC) of the HA-TPCs are lightweight composite structures, designed to satisfy demanding mechanical and electrical requirements while minimizing inactive volume and material budget compared to the vertical TPCs of ND280~\cite{Abgrall:2010hi}. Each cage has a box-shaped geometry with inner dimensions of about $x \times y \times z = 1012 \times 705 \times 1750$~mm$^3$, using the coordinate system reported in Fig.~\ref{fig:sketch_ND280upgrade}. The cage walls have a thickness of approximately 39~mm and they consist of several layers, with the innermost layer designed to establish a uniform electric field across the drift volume. This is achieved using two sets of copper strips: 197 ``field strips'' facing the active volume and 196 ``mirror strips'' on the opposite side, connected via two voltage dividers.

\begin{figure*}
  \centering
  \includegraphics[width=0.8\linewidth]{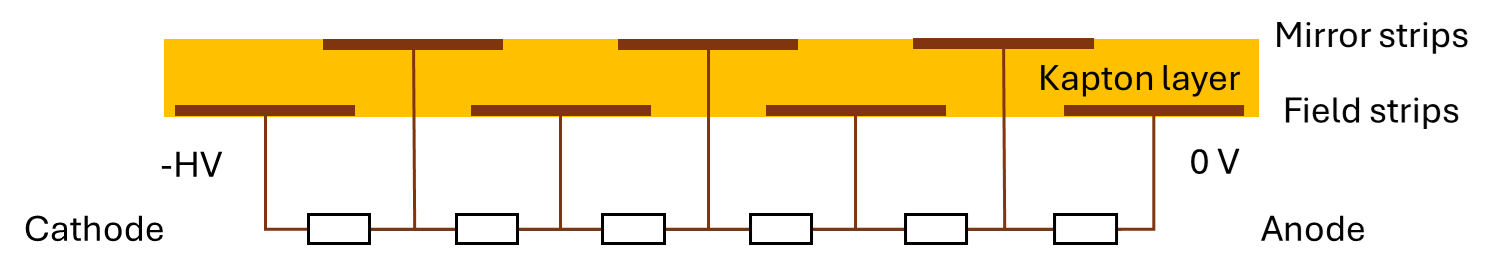}
  \caption{Cross section of the field cage, showing a schematic view of the copper strips connected by voltage dividers from cathode to anode. (not to scale)}
  \label{fig:sketch_strips}
\end{figure*}

Each copper strip is 3\,mm wide, with a 5\,mm pitch, and strips in the two layers are staggered to overlap by 0.5\,mm, ensuring field uniformity. The voltage dividers employ redundant 5\,M$\Omega$ SMD resistors, connecting alternating field and mirror strips with a total resistance of 1\,G$\Omega$ per TPC half. A schematic of this configuration is shown in Fig.~\ref{fig:sketch_strips}.

The field cages were required to maintain an electric field uniformity better than $10^{-3}$ relative inhomogeneity at 15\,mm from the walls, while withstanding overpressure, gravitational, and thermal loads. The selected composite materials offer both structural integrity and low permeability to atmospheric gases, with a strict limit of 10\,ppm oxygen inside the drift volume.

Five full-scale field cages were produced. The first prototype (FC0) exhibited a local insulation defect, resolved by refining the assembly process near the strip foils. The subsequent four cages fully met all specifications and were installed in the two HA-TPC modules. The final design ensures high-voltage stability, mechanical robustness, and operational reliability within the T2K near detector environment.
A brief overview of the field cage characteristics and basic tests is presented here, while a more detailed description is provided in~\cite{MatteoPhD}.

\subsection{Metrology and mechanical performance}
A comprehensive metrology and mechanical evaluation of the HA-TPC field cages was performed both at the Nexus Projectes S.L. production facility and CERN to validate their dimensional accuracy and mechanical integrity. Key parameters measured include surface planarity, wall parallelism and orthogonality, cathode planarity, flange quality, and structural response to mechanical stresses and internal overpressure.

\subsubsection{Metrology Measurements}

A high-precision 3D laser scan, accurate to $80~\mu$m, confirmed that deviations from the nominal CAD model remained within $\pm500~\mu$m in critical areas, with isolated deformations up to $800~\mu$m attributable to curing-induced stresses and orientation changes during operation. Wall parallelism was verified by measuring inter-wall distances at 15 reference points, showing deviations no greater than $0.93$~mm on the largest sides — a remarkable result given the structural complexity.

Cathode planarity, assessed independently at CERN, revealed systematic deformations up to $0.8$~mm, likely due to the copper layering process. Flange planarity measurements indicated a global tilt of about $0.3$~mm without significant local irregularities.

\subsubsection{Mechanical Performance Tests}

Two load tests were conducted on HA-TPC assemblies (\textit{FC0+FC1} and \textit{FC1+FC2}) with weights up to $200$~kg applied to the upper flanges. Both setups showed elastic behavior, with no permanent deformation. Deflection rates were consistent: $0.527\pm0.007~\mu$m/kg for \textit{FC0+FC1} and $0.57\pm0.03~\mu$m/kg for \textit{FC1+FC2}, confirming reproducibility and compliance with safety requirements.

Overpressure tests (up to $20$--$30$~mbar) revealed larger deflections on wide surfaces (approx. $50~\mu$m/mbar) than on narrow sides (approx. $2.3~\mu$m/mbar). Results were broadly compatible with finite element simulations, though measured deformations exceeded predictions, likely due to partial constraint effects at the flanges.

\subsubsection{Electrical Performance}

Electrical integrity was validated by measuring individual resistor values at $70$~V and $150$~V, confirming uniformity with no anomalies (e.g., $R_{FC4}=5.08818\pm0.00007$~M$\Omega$). Full voltage divider tests at $10$~kV demonstrated identical behavior between field cages within the sensitivity of the insulation tester. Additional tests quantified leakage currents to the external shielding, ensuring operational safety and verifying insulation effectiveness.

In summary, the HA-TPC field cages met all mechanical and electrical requirements for precision operation, with reliable structural and metrological properties achieved through carefully controlled production and validation protocols.

\subsection{Electric field in HA-TPCs}
\label{FieldCageElectricField}
The field cages of the HA-TPCs are designed to maintain a uniform electric field in the active drift volume, targeting a field homogeneity better than \(10^{-3}\) at 15~mm from the walls, corresponding to \( E_\perp/E_\parallel < 10^{-3} \). The initial design met these specifications~\cite{abe2019t2k}, but modifications were introduced to mitigate electrical breakdown risks and improve long-term reliability.

Two changes were made near the cathode:  
(i) increasing the cathode-to-wall separation to over 12~mm, and  
(ii) shifting the first field strip 8~mm toward the anode (along the ND280 $x$-axis as depicted in Fig.~\ref{fig:sketch_ND280upgrade}), decoupling it from the flange plane.  
As a result, a 12~mm gap was introduced in the \(y\)-\(z\) plane, and the first strip was offset from the cathode-wall intersection.

These modifications enhanced insulation by increasing the path length of potential leakage currents from high-voltage areas to external shielding. Given the HA-TPCs' multi-year operation and the potential for insulation degradation from charge buildup or aging, all high-voltage leakage paths were extended to maintain safe voltage-drop-to-distance ratios.

Further insulation was added by machining a 30~mm groove into each cathode flange, supplementing the two 8~mm O-ring grooves. This groove acts as a gas and surface charge labyrinth, lengthening potential leakage paths.

\begin{figure*}
  \centering
  \includegraphics[width=0.5\linewidth]{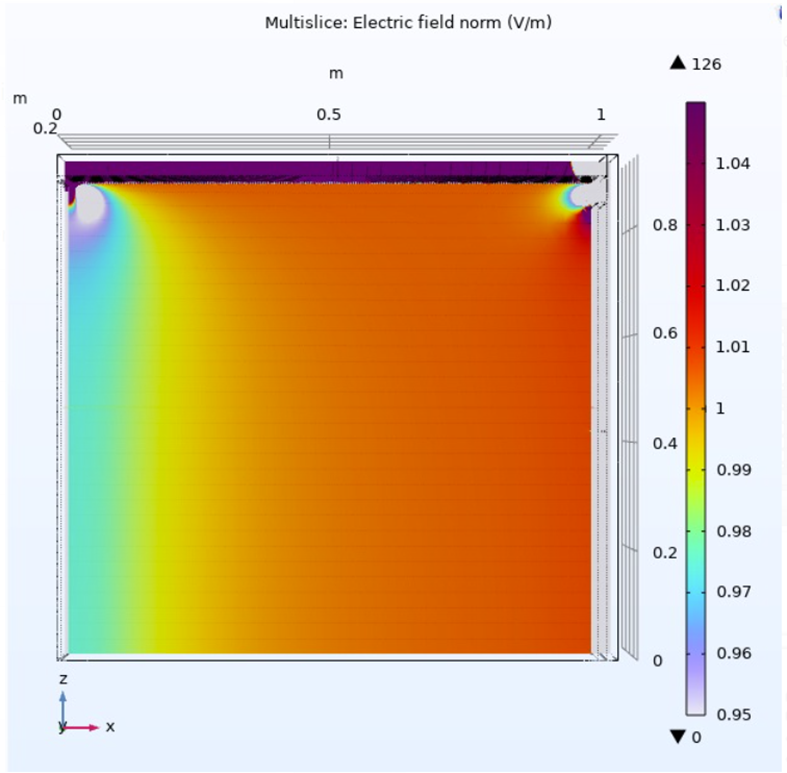}
  \caption{Electric field norm (\(\pm5\%\)) around the nominal 1~V/m value.}
  \label{fig:Efieldnorm}
\end{figure*}

However, shifting the first strip introduced localized field distortions. Due to tight commissioning schedules at CERN, these could not be fully characterized before installation. Consequently, adjustments to the voltage drop between the first strip and cathode were not implemented, and the distortions were only identified at J-PARC, where hardware intervention was no longer possible.  
Without laser data, cosmic rays provided the main tool for evaluating electric field distortions. Tracks recorded without a magnetic field revealed notable distortions, especially near the cathode.

A detailed 3D field model with accurate boundary conditions was developed in COMSOL Multiphysics~\cite{multiphysics1998introduction} (version 6.1), solving the static differential form of Maxwell’s equations in vacuum and dielectrics. The simulation geometry was constructed using CERN and Nexus Projectes S.L. metrology data, approximating each field cage as a parallelepiped and exploiting symmetry to simulate one-fourth of the volume.

After defining the geometry and potentials, a fine mesh was used to numerically solve the Laplace equation. From the computed potential, the electric field map was extracted. The solution for the plane \( y = \frac{y_{\text{max}}}{2} \), located midway between the upper wall and the central symmetry plane (respectively at  $y=y_{\text{max}}$ and $y=0$), is presented in Fig.~\ref{fig:Efieldnorm}. Significant deviations from homogeneity occur near the cathode, due to the potential difference between cathode, first strip, and external shielding, and near the anode, where grounded surfaces are unevenly distributed along \(x\).

The evaluation of the effects of the electric field non uniformities on the reconstruction are reported in Subsec.~\ref{subsec:Efielduniformity}.

%% file: ERAM.tex
\subsection{Readout production}
\label{subsec:eram_production}
\begin{figure}[ht]
    \begin{center}
        \includegraphics[width=0.45\textwidth]{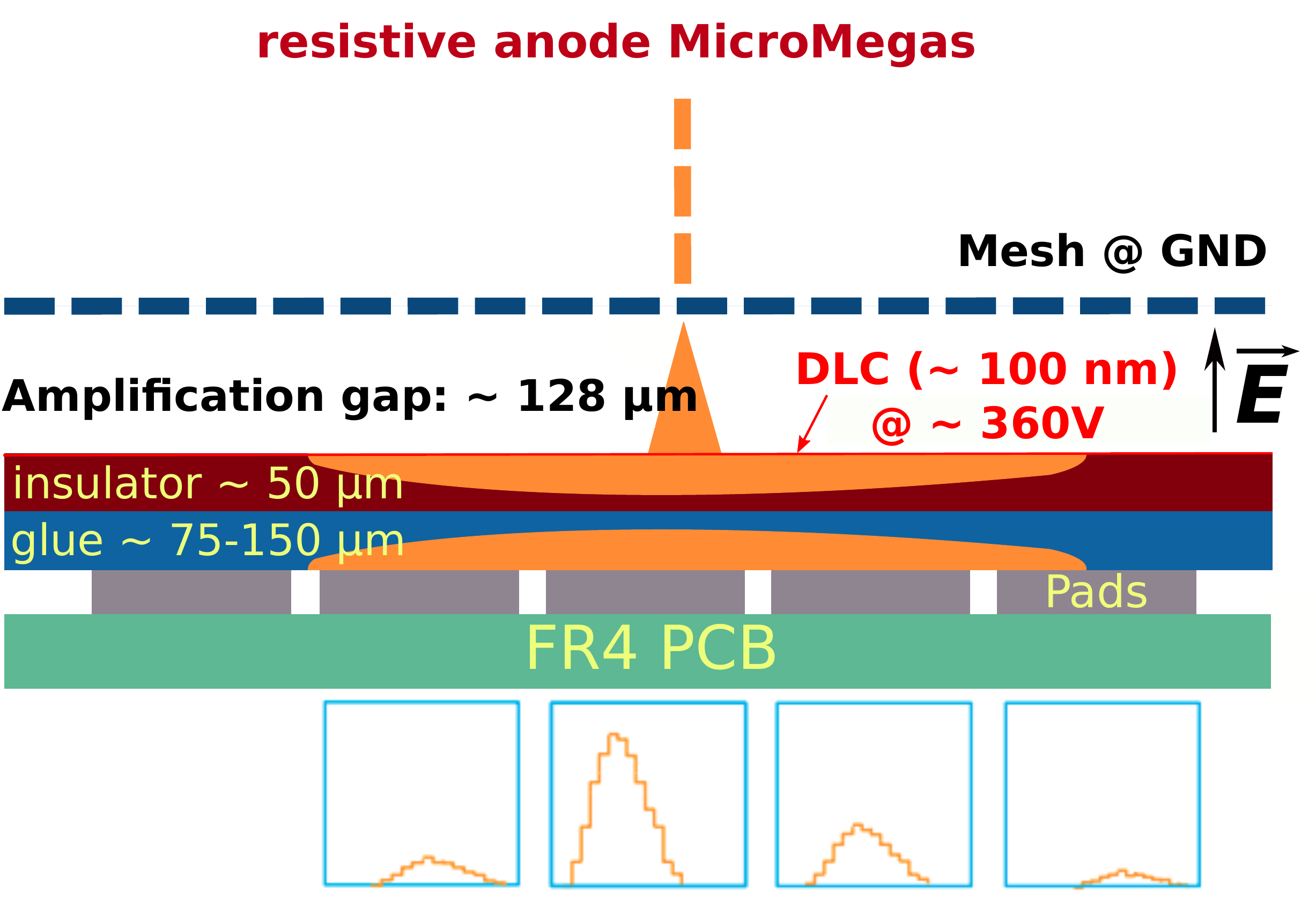} 
        \caption{Operating principle of Resistive Micromegas. The various detector elements are indicated. The glue thickness is optimized to achieve a consistent $RC$ value for each detector~\cite{Ambrosi:2023smx}. 
        \label{fig:ERAM_operating_principle}}
    \end{center}
\end{figure}

 The HA-TPC is equipped with a new readout system based on the innovative resistive Micromegas technology~\cite{Attie:2022smn}.  The working principle of the resistive Micromegas is illustrated in Fig.~\ref{fig:ERAM_operating_principle}. In this design, a resistive foil made of diamond-like carbon (DLC)~\cite{ROBERTSON1992185} is applied over the pad plane to achieve charge dispersion. As the induced charge spreads over multiple pads, this configuration improves position resolution without requiring smaller readout pads, offering a compact and cost-efficient solution. Furthermore, the resistive layer delays charge dissipation, effectively mitigating sparks and eliminating the need for additional spark protection in the front-end electronics.

A novel high-voltage powering scheme is employed in the operation of resistive anode Micromegas, where the mesh is grounded and the anode is held at a positive voltage. This configuration, referred to as Encapsulated Resistive Anode Micromegas (ERAM), helps equalize the gain across detectors by allowing individual high-voltage adjustments for each ERAM, addressing inherent variations in gain. Grounding all meshes---thereby maintaining a common potential---ensures a uniform electric field within the drift volume, even when neighboring detectors operate at different high-voltage values. Additionally, this approach improves operational safety by keeping the high voltage fully enclosed and electrically isolated, while keeping the mesh safely accessible.

Each ERAM is a bulk Micromegas detector featuring a $128~\mu$m amplification gap and utilizing the standard SD45/18 304L woven micromesh, mounted on a resistive anode PCB. The detectors were produced at the CERN EP-DT-EF PCB workshop and are installed in aluminum frames, with their associated readout electronics mounted on the rear side. Further construction details can be found in Refs.~\cite{Attie:2022smn, Ambrosi:2023smx, Attie:2021yeh, Attie:2019hua}.  

Each detector measures $42 \times 34~\text{cm}^2$, with a total of 1152 readout pads arranged in a matrix of 36 pads along the $x$-axis and 32 along the $y$-axis. Each pad measures $11.18 \times 10.09~\text{mm}^2$. The pad plane is covered with a $50~\mu$m thick Apical\textsuperscript{\textregistered}~\cite{ApicalWebpage} polyimide foil, laminated with $150~\mu$m of glue, onto which a DLC film is deposited. To ensure adequate charge spreading over at least two pads, a DLC surface resistivity of approximately $400~\text{k}\Omega/\square$ was chosen. A schematic cross-section and specifications of the ERAM are presented in Fig.~\ref{fig:ERAM_operating_principle}. The lateral charge dispersion is governed by the $RC$ constant of the resistive layer, where $R$ represents the surface resistivity of the DLC and $C$ is the capacitance per unit area of the stack, controlled by the glue thickness~\cite{Ambrosi:2023smx}.

Of the 32 ERAMs installed in the HA-TPC, 31 were built with a glue thickness of $150~\mu$m and a DLC surface resistivity of around $400~\text{k}\Omega/\square$. One exception, ERAM~29, was manufactured with half the glue thickness and half the surface resistivity. This design compensates for the lower resistivity by reducing the thickness of both the Apical\textsuperscript{\textregistered} and glue layers, ensuring a uniform and stable $RC$ constant across all 32 detectors.  Detailed performance studies and characterization of the detectors are provided in Refs.~\cite{Attie:2022smn, Ambrosi:2023smx, Attie:2021yeh, Attie:2019hua}. \\

\subsection{ERAMs Performance}
\label{subsec:eram_installation}
A dedicated X-ray test bench was employed to characterize the ERAMs, performing a pad-by-pad scan to assess gain uniformity and energy resolution. The setup consisted of an aluminium chamber with a 3~cm drift distance and a robotic $x$-$y$-$z$ arm, which positioned a 280~MBq $^{55}$Fe source inside a collimator. X-ray interactions produced electron avalanches in the Micromegas amplification gap above the selected pad, with the resulting charge collected on this pad and diffused to adjacent ones.

\begin{figure*}[ht]
    \begin{center}
        \includegraphics[width=0.95\textwidth]{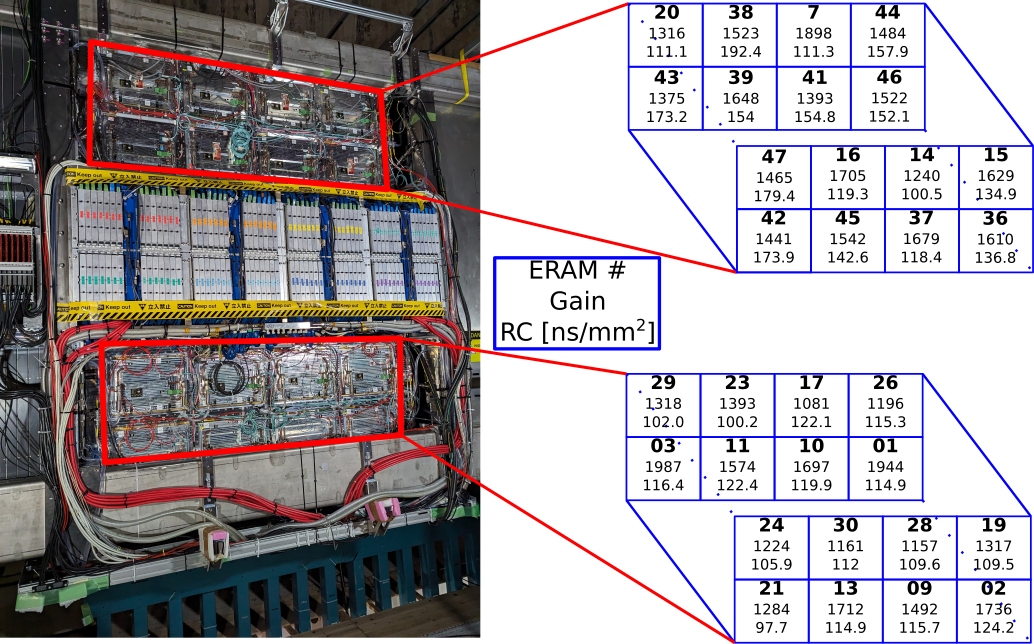}
        \caption{Left: a photo of the upgraded sub-detectors installed in ND280, Right: the various ERAM detectors with their location, average gain and RC values. 
        \label{fig:ERAM_production}}
    \end{center}
\end{figure*}

To analyze the measurements and extract the gain and charge dispersion parameters ($RC$) for each module, a detailed physical model~\cite{Ambrosi:2023smx}, accounting for ionization processes, diffusion effects, and readout electronics response, was applied. In Fig.~\ref{fig:ERAM_production}, a picture of the HA-TPCs surrounding the SuperFGD can be seen.
The right side of Fig.~\ref{fig:ERAM_production} illustrates the location of the various ERAM detectors that were installed in the HA-TPCs, along with their mean gain and $RC$ values. The statistical uncertainty on the values presented in Fig.~\ref{fig:ERAM_production} is negligible. A systematic uncertainty~\cite{joshi:tel-04836929} of 15\% for the gain and 13\% for the $RC$ is attributed to the choice of parameters in the electronics model, the noise model applied~\cite{Attie:2025wdt}, variations in temperature and pressure conditions, and the assumptions regarding a uniform versus discretized $RC$ distribution.

\begin{figure*}[h!]
    \begin{center}
        \includegraphics[width=0.9\textwidth]{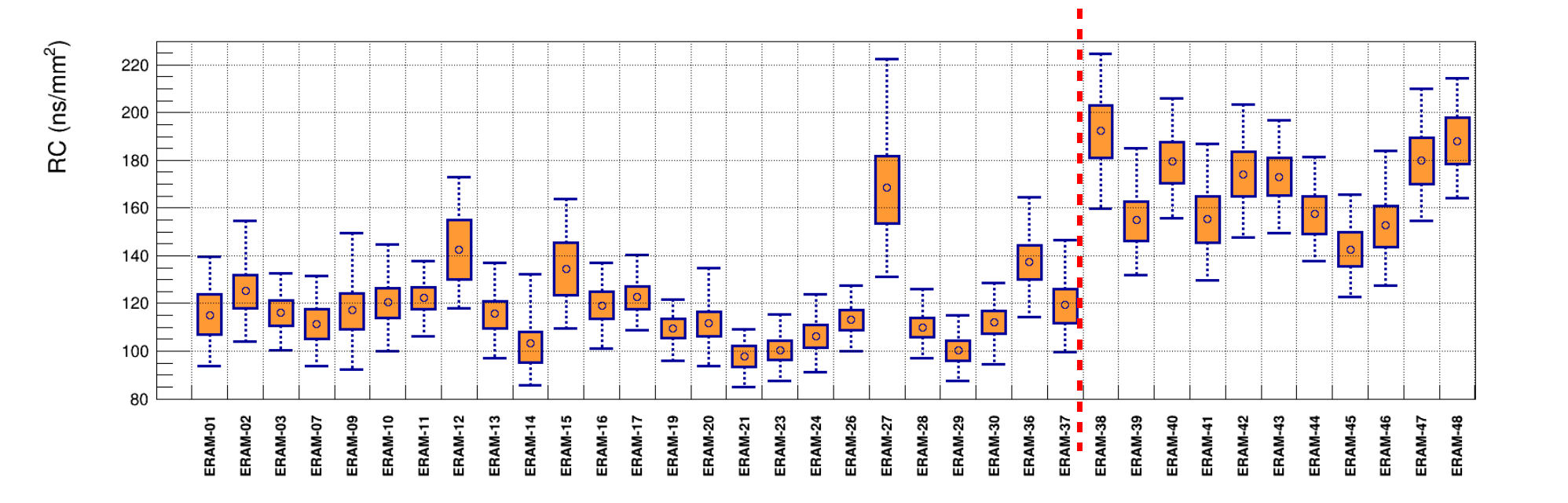}\\
        \includegraphics[width=0.9\textwidth]{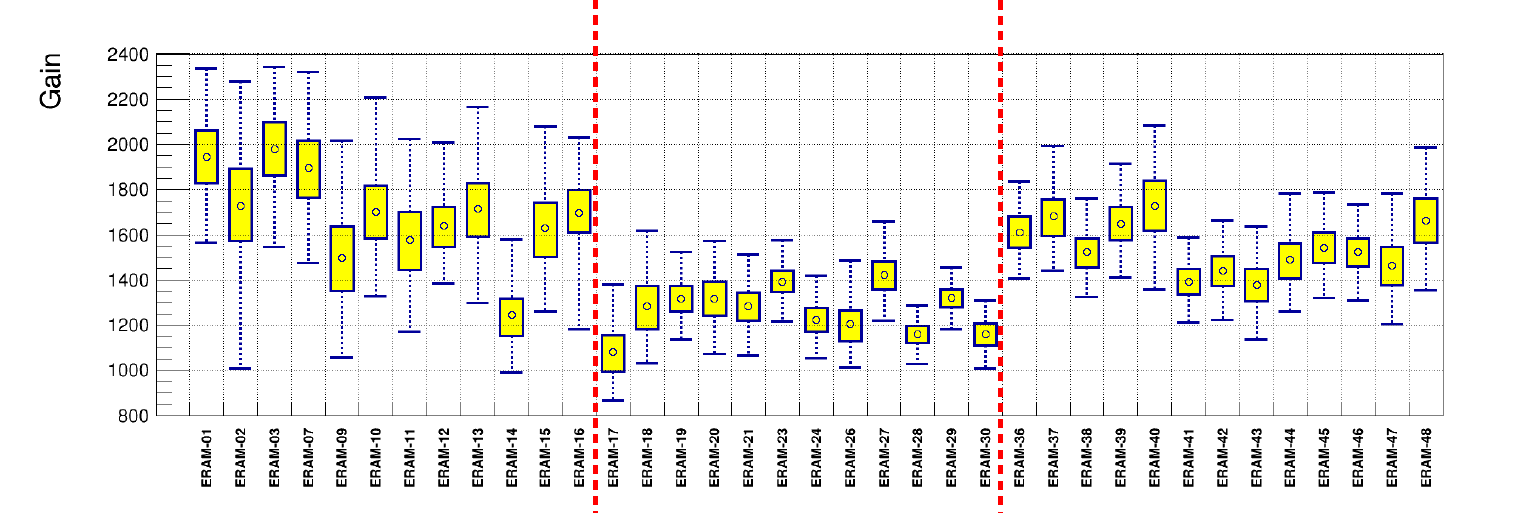}\\
        \includegraphics[width=0.85\textwidth]{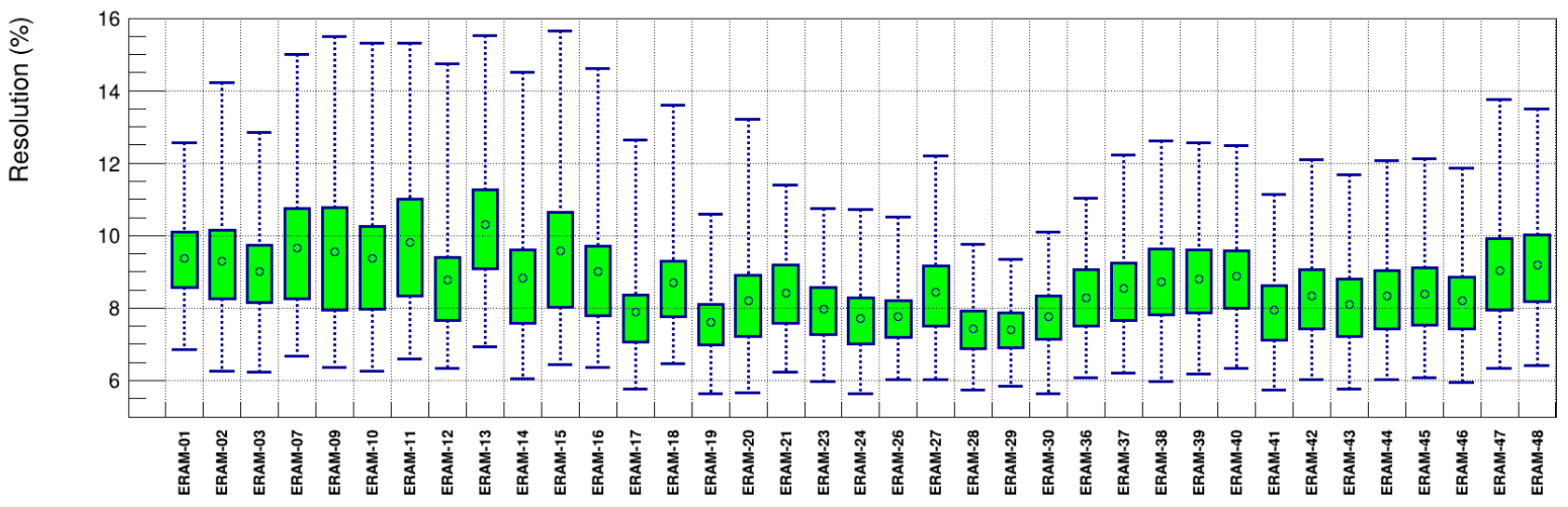}
        \caption{Summary of the performances of the different ERAM detectors produced for HA-TPC. 
        Top: $RC$ values, middle: gain, bottom: energy resolution.
       Different epochs in the production can be identified and are indicated by the red vertical dashed lines.
        \label{fig:ERAM_performances}
        }
    \end{center}
\end{figure*}

For each detector, the values of the $RC$, gain, and energy resolution for every pad were extracted. These results are summarized in Fig.~\ref{fig:ERAM_performances}, displayed as candle plots. In each plot, the circle within the box represents the mean value, with the box indicating a $\pm$~25\% deviation and the whiskers extending to $\pm$49\% around the mean. Such plots were essential for selecting and grouping ERAMs with similar mean values or dispersions of specific characteristics for installation in the HA-TPC, as well as for identifying defects or distinctive features in individual detectors. Additionally, the gain and $RC$ values serve as important inputs for both simulation and reconstruction.

\label{subsec:eram_performances}

From Fig.~\ref{fig:ERAM_performances}, several production trends can be identified. The $RC$ values are directly related to the batch of DLC foils used. While all detectors up to ERAM-37 exhibit similar $RC$ values, a noticeable increase appears from ERAM-38 onwards, corresponding to a final batch of DLC foils with significantly higher surface resistivity.

For the gain, three distinct production periods are observed. In the first phase, up to ERAM-16, a large pad-to-pad spread is visible within each detector. This was attributed to mechanical imprints from the stiffener structure, likely caused by local compression due to additional solder mask and copper layers during PCB pressing. Consequently, local gain inhomogeneities appeared, and the average gain was relatively high.
A second, improved period extends from ERAM-17 to ERAM-30, following the removal of the solder mask and the replacement of the plain copper grounding layer with a copper strip mesh. These changes eliminated mechanical patterns in the gain maps and reduced the spread. The mean gain also decreased, likely due to variations in the gap thickness, influenced by the Pyralux pillar height and the absence of the solder mask.
A final period begins at ERAM-36, marked by an increase in mean gain values. This is probably linked to a reduction in the Pyralux material thickness, leading to a smaller amplification gap and consequently higher gain.

Regarding energy resolution, the detectors consistently achieved values compatible with the target of 10\%. The spread in values was slightly larger for detectors produced before ERAM-17, likely due to the solder mask layer’s influence on the pad response. This spread reduced notably after the removal of the layer, resulting in more uniform performance across subsequent production batches.


%% file: ElectronicsNoise.tex
\begin{figure}[h]
    \begin{center}
        \includegraphics[width=0.75\linewidth]{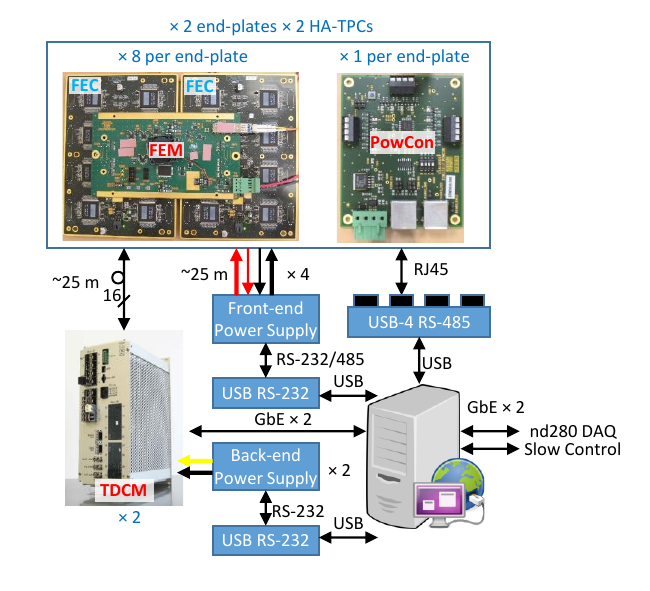}
        \caption{Schematic view of the readout system of the HA-TPCs. 
        \label{fig:hat_readout_diagram}}
    \end{center}
\end{figure}

\label{sec:Electronics}
The readout system of the HA-TPCs is composed of custom-made modular front-end electronics mounted at the back of each ERAM detector and back-end electronics placed at the floor located below the ND280 detector magnet. A schematic view of this readout system is shown in Fig.~\ref{fig:hat_readout_diagram}.
Each ERAM is read out by two Front-End Cards (FECs) hosting eight AFTER chips~\cite{Baron:2008zza}, controlled by a Front-End Mezzanine (FEM). The FEM communicates via optical links with Trigger and Data Concentrator Modules (TDCMs)~\cite{Calvet:2018lac}, each handling 16 detector modules (18,432 channels). TDCMs connect via Ethernet to a control PC interfacing with ND280’s DAQ and slow control systems. Custom PowCon cards manage remote power control of each ERAM, supplied by TDK-Lambda units through 25\,m cables. To operate within a 0.2\,T magnetic field, front-end boards use linear voltage regulators. Each HA-TPC end-plate consumes 35\,A at 4.5\,V, with total power dissipation of 630\,W (17\,mW/channel), managed by water-cooled aluminum plates on each FEC and FEM.
\subsection{Front-End Cards}
The FECs are the analog front-end of the HA-TPC readout, each housing eight AFTER chips (576 channels), an octal-channel 12 bit Analog to Digital Converter (ADC), a calibration circuit, linear regulators, and passive components. The AFTER chip includes 72 channels, each composed of a charge preamplifier, pole-zero cancellation, a Sallen-Key filter (with 16 selectable peaking times), and a 511-cell circular buffer (Switched Capacitor Array). It continuously samples signals and holds them until an external trigger freezes the buffer for readout.

Resistive Micromegas detectors offer higher spatial resolution via charge spreading, prevent sparks, and allow larger pads, reducing the required number of channels. As a result, FECs were made compact (26\,cm~$\times$~18\,cm) and mounted parallel to the ERAMs, unlike the older vertical TPCs needing more and larger boards. Low-force, reliable insertion was ensured using Hirose FX23 floating connectors with $\pm$0.6\,mm tolerance. A total of 84 FECs were produced, each tested on a dedicated test-bench and validated before integration on the detectors.
\subsection{Front-End Mezzanine cards}
The FEM hosts the digital electronics of an ERAM module, managing configuration, clock, triggers, and readout from 16 AFTER chips. It performs event buffering, pedestal equalization, zero-suppression, and data transfer to the back-end via an optical link. It also receives the global 100~MHz clock, triggers, and commands over the same link.

All functions are implemented in an AMD Xilinx Artix 7 FPGA, with a 36~Mbit SRAM for event buffering. Data from the 16 AFTER chips (digitized in parallel at 12.5~MHz, 12-bit) are stored and processed sequentially at 100~MHz. Event digitization takes about 3.2~ms, defining the minimum front-end dead-time. Total event processing lasts roughly 12.8~ms in zero-suppressed mode; for large events, transfer time over the 200~Mbps link becomes dominant.
\subsection{Back-end electronics}
The TDCM is a general-purpose module responsible for distributing clock and synchronization signals to up to 32 remote units via optical links, aggregating their data, and forwarding it to a control PC over a gigabit Ethernet connection. To maintain modularity and flexibility, one TDCM is dedicated to each HA-TPC. Its core is a commercially available Mercury ZX1 System-on-Module (SoM) from Enclustra, integrating an AMD Xilinx Zynq 7030 FPGA with a dual-core ARM processor, 1~GB SDRAM, and multiple interfaces. The SoM is mounted on a custom-designed carrier board providing power, connectivity, and mechanical integration. The TDCM can be equipped with up to two mezzanine cards, each hosting 16 optical transceivers.

The system’s functions are implemented through a combination of FPGA firmware and a command interpreter written in C running on the ARM processor without an operating system. This software decodes commands from the control PC and executes corresponding operations on the FEMs, FECs, and AFTER chips, while handling the transmission of event data via Ethernet using UDP/IP protocol. 

A comprehensive characterization of the electronic noise in the ERAM detectors is presented in \cite{Attie:2025wdt}, using data from 32 detectors operated without zero suppression. The analytical model reliably reproduces the observed noise, with Monte Carlo simulations in excellent agreement. All ERAM detectors exhibit quasi-identical and time-stable noise performance. These results are incorporated into the HA-TPC simulation, contributing to a more precise understanding and improved estimation of the detector resolution.

%% file: DAQ.tex
\begin{figure*}[t]
    \begin{center}
        \includegraphics[width=0.8\textwidth]{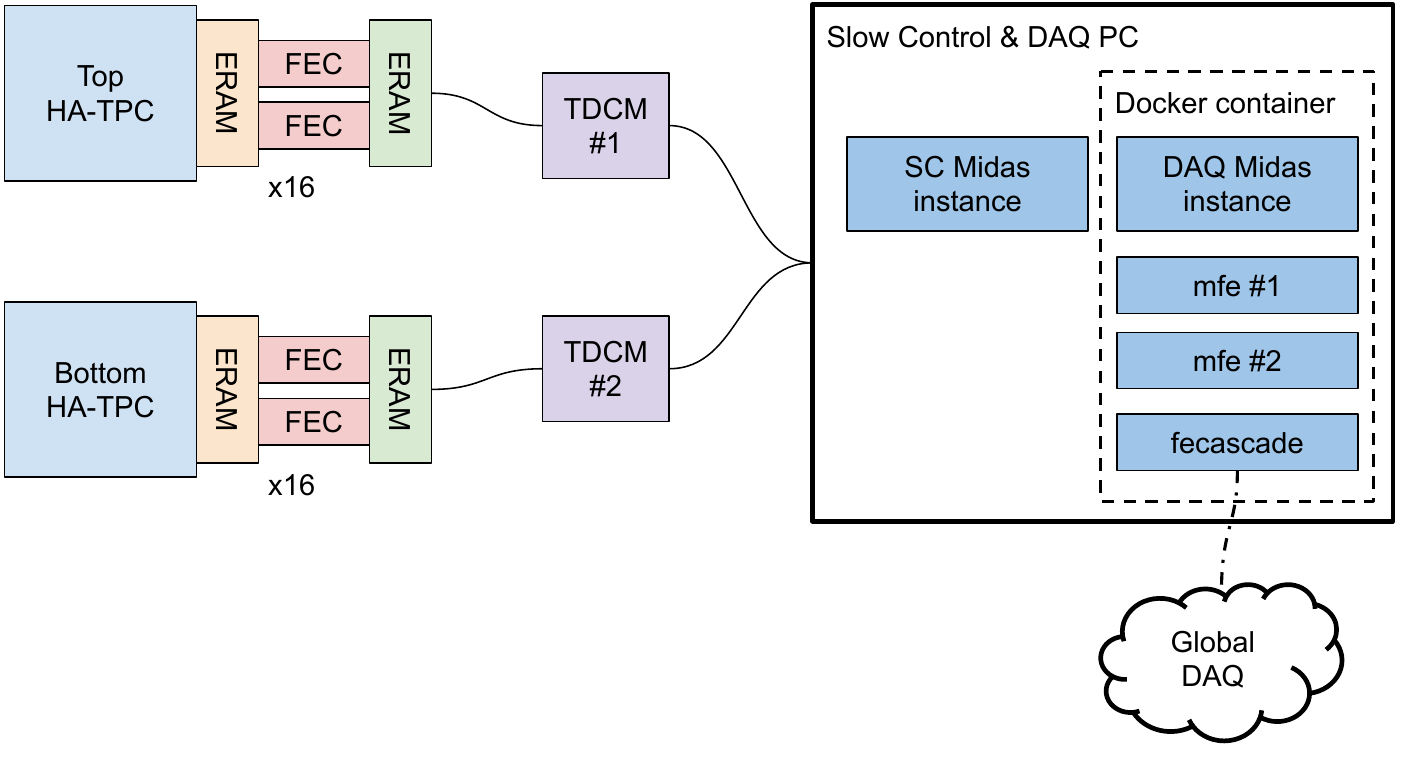}
        \caption{Scheme of the HA-TPC DAQ including the electronics and data concentration cards (TDCM) and the Midas instances for the slow-control and the DAQ. 
        \label{fig:hat_daq_sc}}
    \end{center}
\end{figure*}

The slow control and data acquisition system is based on the Midas framework \cite{Midas} and utilizes its core features (namely the Online Data Base (ODB) feature, "mhttpd" executable for web interface, "mlogger" executable for data recording and "mserver" executable for remote access to the Midas instance).
To avoid strong interference, one Midas instance is dedicated to the HA-TPC slow-control and another one is dedicated to the data acquisition as depicted in Figure~\ref{fig:hat_daq_sc}.
Both run on the same rack server, with the slow control instance running natively and the DAQ instance inside a Docker container, managed via docker-compose \cite{Merkel2014}.
This setup allows for two Midas instances to run independently and in fairly good isolation.
The Midas front-end program of the slow control part performs the configuration of the readout electronics with parameters stored in the global ODB of ND280 and periodically monitors the temperature, supply current and voltage of all boards.

For the DAQ part, a so-called "Midas front-end" interfaces between the TDCM data stream and the Midas data logging system.
This Midas front-end and its low-level interface with a TDCM are organized around three primary threads. 
The first thread manages communication with the TDCM and the connected FEM. 
It sends commands, receives replies, monitors their status, sets parameters, and controls the start and stop of data recording from the FEC to the TDCM.
The second one handles the reception of UDP packets containing data frames, such as monitoring information and physics data.
The third thread processes the received data frames, interprets them, and constructs physical events before sending them to the Midas ODB as Midas events.
Additionally, a so-called "cascade" front-end (fecascade) runs on the DAQ Midas instance and allows for the communication with the global DAQ Midas instance: in particular, this front-end forwards the run start or stop instructions and sends the data collected by the local front-ends to the global DAQ to be written to file.
Events consist of multiple Midas banks  containing event numbers, time stamps, channel IDs, waveform start and stop times, and ADC values. 

%% file: GasSystem.tex

The gas handling system (GHS) supplies gas to the HA-TPC and to the vertical TPC (v-TPCs) of the original ND280 tracking system. The drift regions are filled with the standard T2K gas mixture, composed of 95\% argon, 3\% carbon tetrafluoride (CF$_4$), and 2\% isobutane (iC$_4$H$_{10}$), whereas the insulating gaps of the v-TPCs are supplied with pure CO$_2$. Overall, the gas system must serve five T2K gas lines and three CO$_2$ lines.

To reduce operating costs and improve performance, a new GHS was designed and built by CERN in 2023. The system operates in closed-loop recirculation mode, continuously purifying and reusing the gas while maintaining stable pressure and composition. Flow rates are configured to achieve 3 to 5 complete volume exchanges per day in each TPC, corresponding to a total recirculation rate close to 3~m$^3$/h. Large buffer volumes are integrated into the system to compensate for rapid atmospheric pressure changes—common at the experimental site—which can reach up to 10~mbar/h. As a result, fresh gas consumption has been reduced by a factor of three compared to the previous system.

The gas mixture is produced in a dedicated mixing room on the surface of the experimental site, where fresh argon, CF$_4$, and iC$_4$H$_{10}$ are blended in precise proportions before entering the recirculation loop. Gas returning from the TPCs is first purified to remove contaminants and then mixed with fresh gas before re-entering the detector. The distribution system, located just beneath the detector, splits the gas to individual TPC lines and regulates their pressures through programmable valves. Two high-capacity, oil-free pumps—operating in active-standby configuration—ensure a stable differential pressure and continuous gas circulation.

The GHS incorporates multiple gas analysis stations at strategic points in the system, both in the mixing room and near the detector. These modules monitor gas composition and impurity levels, with the ability to automatically switch between different lines and locations for independent measurements. All operational data, including gas composition, pressure, and contaminant levels, are transmitted to the detector control system (DCS), ensuring proper monitoring, archiving, and integration with TPC calibrations. The GHS has demonstrated excellent reliability and performance, achieving oxygen contamination levels consistently below 1~ppm in the delivered gas. 

%% file: GMC.tex
\begin{figure}[bt!]
  \centering
  \begin{subfigure}[b]{0.49\textwidth}
    \centering
    \includegraphics[width=1.\linewidth]{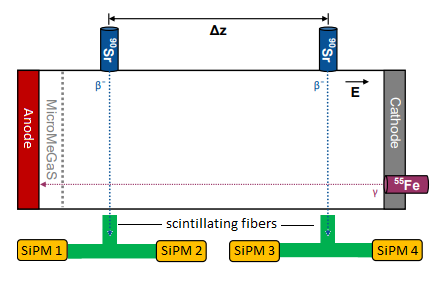}
  \end{subfigure}
  \hfill
  \begin{subfigure}[b]{0.49\textwidth}
    \centering
    \includegraphics[width=0.7\linewidth]{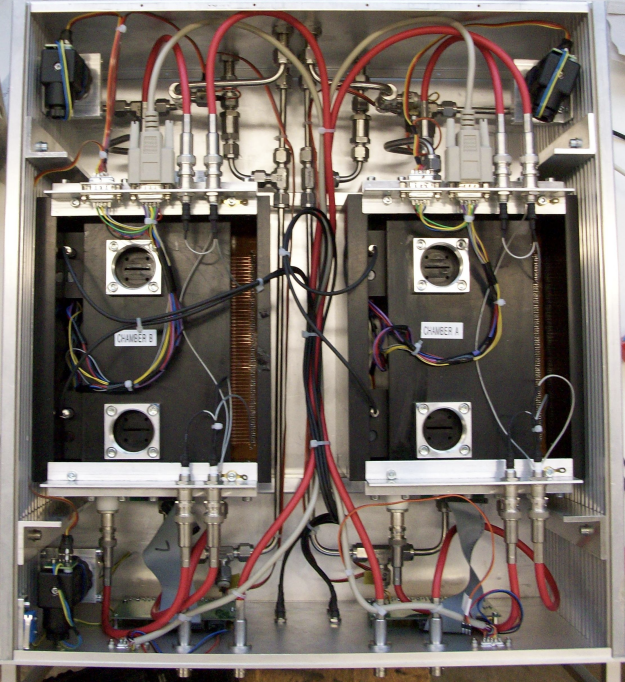}
  \end{subfigure}
  \caption{Schematic (left) and photograph (right) of the GMC system. The final setup includes two GMCs inside a crate equipped with temperature and pressure sensors. The ${}^{90}\text{Sr}$ source holders are shown empty, revealing the collimator.}
  \label{fig:GMCmodel}
\end{figure}

The gas used in the TPCs – i.e., both the HA-TPCs and the v-TPCs, is continuously monitored via two dedicated Gas Monitoring Chambers (GMCs). These compact TPCs measure two critical parameters: the electron drift velocity and relative changes in gas amplification (gain). Drift velocity is essential for accurate 3D track reconstruction, while gain monitoring verifies gas mixture stability and, with calibration, improves $\mathrm{d}E/\mathrm{d}x$ measurements.

\begin{figure}[hbt!]
  \centering
  \begin{subfigure}[b]{1.\linewidth}
    \centering
    \includegraphics[width=.9\linewidth]{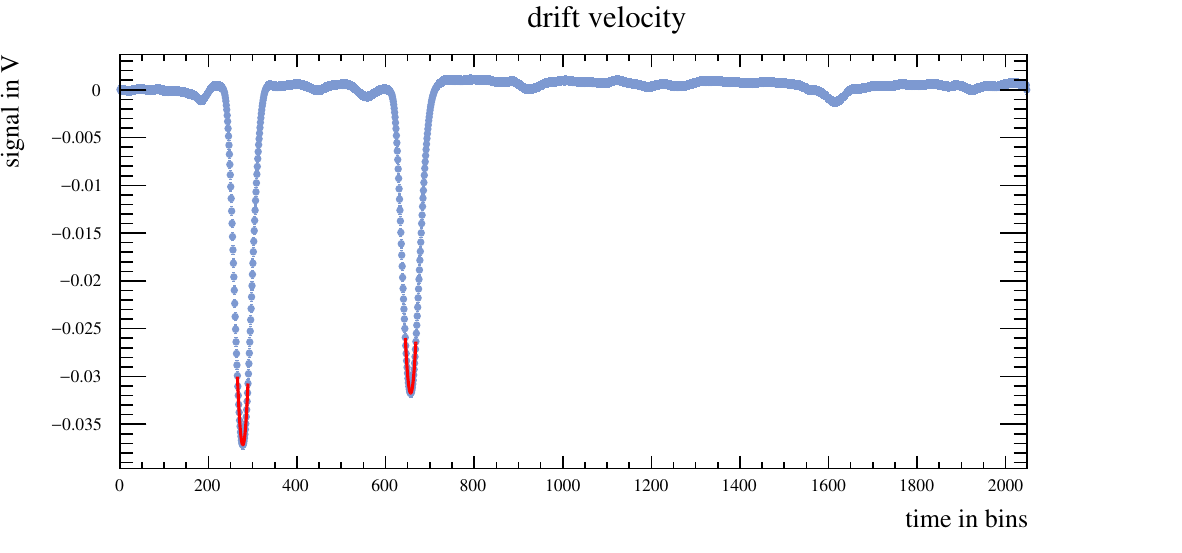}
  \end{subfigure}
  \hfill
  \begin{subfigure}[b]{1.\linewidth}
    \centering
    \includegraphics[width=.9\linewidth]{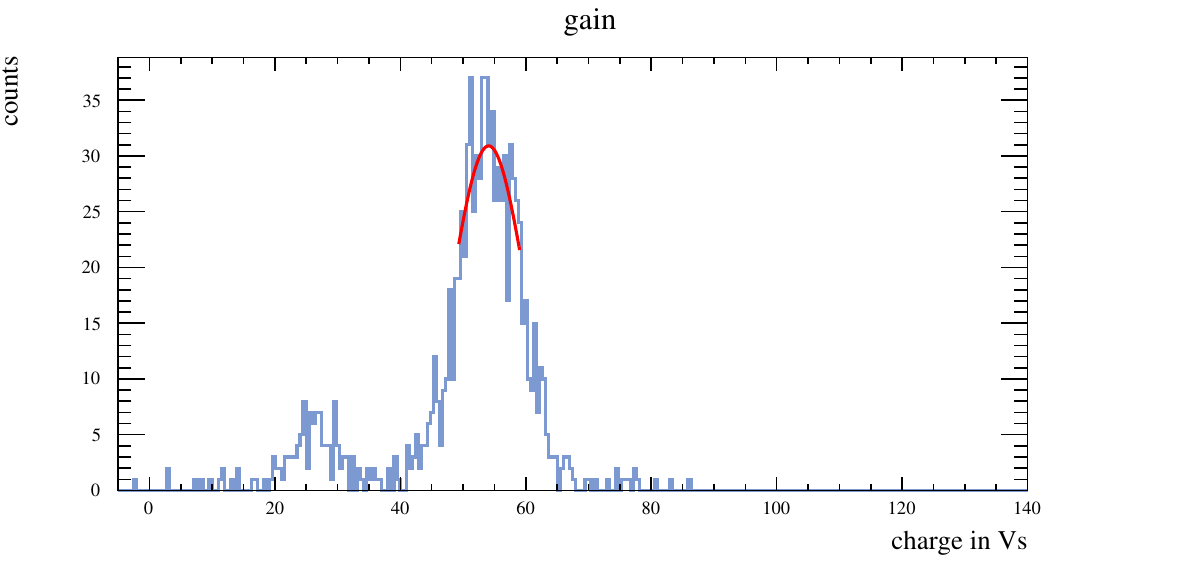}
  \end{subfigure}

\caption{Averaged waveforms from 1000 triggers for drift velocity (top) and gain (bottom) measurements.}
  
  \label{fig:exampleMeasurement}
\end{figure}

Each GMC, equipped with a Micromegas detector, contains two ${}^{90}\text{Sr}$ beta sources at known positions and a ${}^{55}\text{Fe}$ X-ray source providing a stable ionization signal for gain monitoring. Fig.~\ref{fig:GMCmodel} shows the operating principle and final setup.

Drift velocity is determined using two silicon photomultipliers (SiPMs) that monitor a scintillating fiber below each of the ${}^{90}\text{Sr}$ beta sources to trigger event recording. Upon trigger, the anode signal is recorded within a time window. Averaging 1000 such waveforms reveals two peaks corresponding to the ionization positions of the sources. The time difference between these peaks, combined with their known separation, yields the drift velocity (Figure~\ref{fig:exampleMeasurement}, top).

Gain is measured by integrating charge signals from ${}^{55}\text{Fe}$ X-rays on dedicated pads. A veto ring suppresses background from scattered beta electrons or incomplete signals. The position of the main peak in the charge distribution, averaged over 1000 events, indicates the gain, since we assume a constant primary charge. Relative changes of the peak position are tracked over time to monitor the gas gain (Figure~\ref{fig:exampleMeasurement}, bottom).

\begin{figure}[h!]
  \centering
  \includegraphics[width=0.9\linewidth]{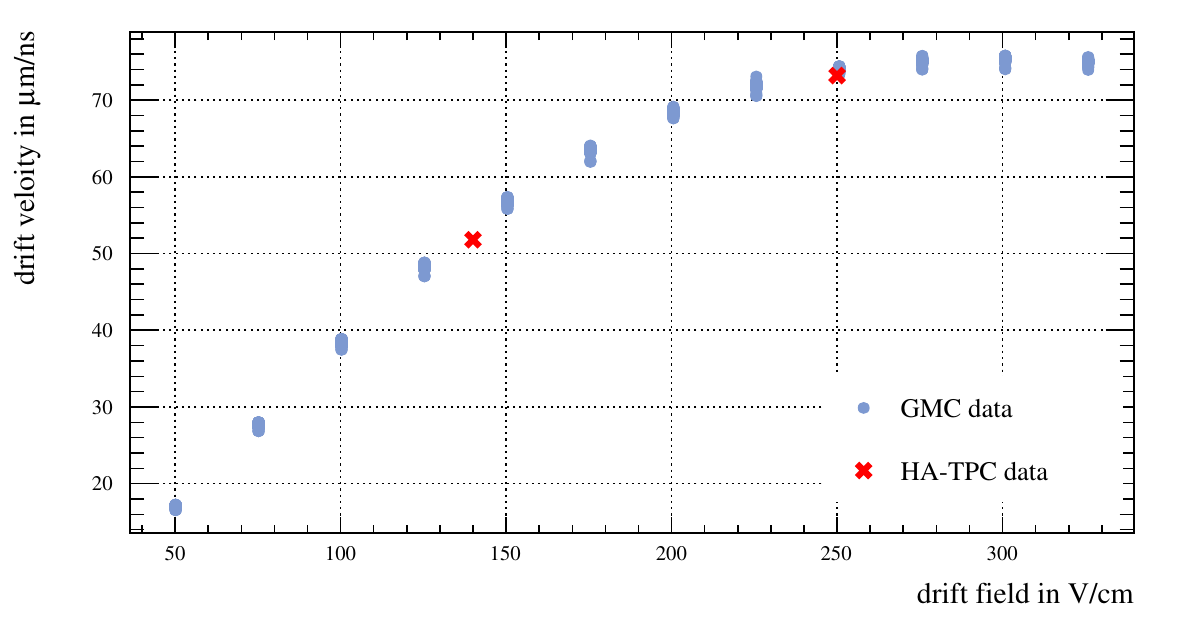}
  \caption{Comparison of drift velocity measurements from the HA-TPC and GMC over various drift fields. Their agreement confirms that GMC data reliably reflect the gas conditions inside the HA-TPCs.}
  \label{fig:GMCTPCcompare}
\end{figure}

To maintain consistent conditions, the same gas mixture flows through the TPCs and GMCs. Monitoring points are placed at the gas supply, after purification, and in the exhaust. The GMCs can also be repositioned to sample gas near specific TPCs. Temperature and pressure sensors inside the GMC crates enable corrections, as both drift velocity and gain depend on their ratio.

During commissioning, GMC drift velocity measurements were directly compared to those from the HA-TPCs across different electric fields. As shown in Fig.~\ref{fig:GMCTPCcompare}, the two systems are in good agreement, confirming the reliability of the GMCs for continuous gas parameter monitoring. The shown monitoring was done with an old-style GMC, which is described in~\cite{Abgrall:2010hi}.

%% file: HATRecoSim.tex

\begin{figure}[htbp]
	\centering
	\includegraphics[width=0.48\textwidth]{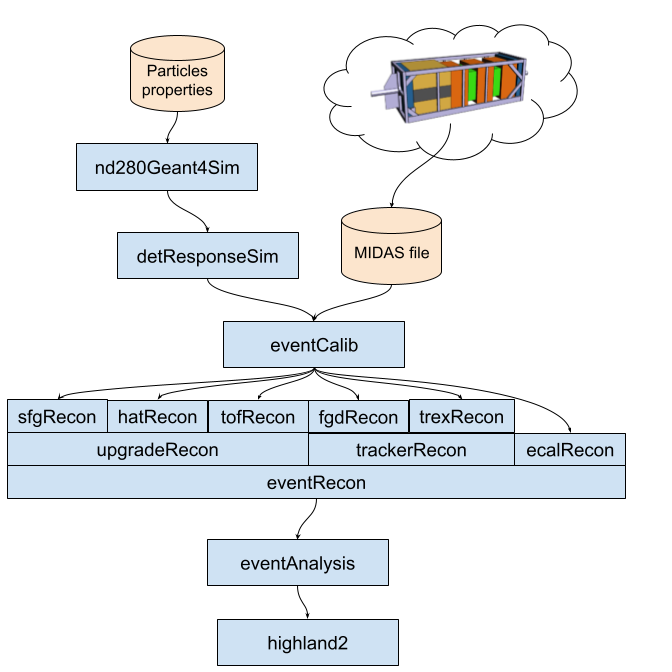}
	\caption{\label{fig:nd280dataflow}Simulated and real data flow in the ND280 software framework.}
\end{figure}

The data taken with the \hatpc are reconstructed with the ND280 offline software~\cite{T2K:2011qtm}, which processes both real data (MIDAS format) and simulated events through a common chain, as shown in Fig.~\ref{fig:nd280dataflow}.

\subsection{Simulation Framework}
\label{sec:hatsimu}

Simulations are based on \textsc{GEANT4}~\cite{GEANT4:2002zbu}, propagating charged particles through the full ND280 detector within a magnetic field. The field can be uniform ($0.2, 0, 0$~T), based on a detailed, position-dependent COMSOL map, or based on an in-situ campaign of B-field measurements done in 2009~\cite{T2K:2011qtm}. The \hatpc is modeled as a box containing a drift volume separated by a 12~mm cathode, with eight ERAM readouts (420 x 340 mm) on each endplate. The volume is filled with the T2K gas mixture (Ar:CF$_4$:iC$_4$H$_{10}$ = 95:3:2). Charged particles ionize the gas, with energy deposits (G4Segments) recorded every 1~mm step using the PAI model~\cite{Apostolakis2000}.


\subsubsection{Electron Drift and Diffusion}

Electrons from ionization drift towards the anode plane under combined electric and magnetic fields, with velocity described by the Langevin equation:

\begin{equation}
\vec{V_d} = \frac{\mu}{1 + (\omega\tau)^2}\left( \vec{E} + (\omega\tau)\frac{\vec{E}\times\vec{B}}{|\vec{B}|} + (\omega\tau)^2\frac{(\vec{E}\cdot\vec{B})\vec{B}}{|\vec{B}|^2}\right)
\end{equation}
where $\mu = \frac{e}{m}\tau$ is the electron mobility in the gas, $\omega = \frac{eB}{m}$, and $\tau$ is the time between two collisions. The value of $\omega\tau$ used in the simulation is 0.42.
A fourth-order Runge-Kutta method propagates electrons, fully accounting for non-uniform $\vec{E}$ and $\vec{B}$ fields.

Electron clouds are generated based on deposited energy and a 26~eV ionization threshold. Transverse and longitudinal diffusions are applied with default coefficients $\sigma_{\mathrm{trans}} = 286\, \mu\mathrm{m}/\sqrt{\mathrm{cm}}$
  and $\sigma_{\mathrm{long}} = 210\, \mu\mathrm{m}/\sqrt{\mathrm{cm}}$.

\subsubsection{ERAM Amplification}

Upon reaching the ERAM plane, electrons undergo avalanche multiplication, modeled using a Polya distribution:
\begin{equation}
P_m(g) = \frac{m^m}{\Gamma(m)} \frac{g^{m-1}}{G^m} \exp\left(-m \frac{g}{G}\right)
\end{equation}
The average gain $G$ for each ERAM is determined from X-ray test bench measurements using a $^{55}$Fe source~\cite{Ambrosi:2023smx}  while a value of $m=0.5$ is used to match the  $\sim9\%$ energy resolution (Fig.~\ref{fig:ERAM_performances}). 

\subsubsection{Resistive Layer Charge Spread}
The next step in the simulation is to introduce the time-dependent spread of the charge induced by the ERAM resistive layer. 
The charge induced on the resistive layer propagates according to a 2D diffusion equation:
\begin{equation}
\rho(\vec{r},t) = \frac{\mathrm{RC}}{4\pi t} \exp\left(-\frac{r^2 \mathrm{RC}}{4t}\right)
\end{equation}

The total charge on a pad is obtained by integrating this density over the pad’s area.
\begin{eqnarray}
    Q_{unit}(t) &= & \int _{x_{min}}^{x_{max}}  \int _{y_{min}}^{y_{max}}  \rho(\vec{r},t)~\mathrm{d}x~\mathrm{d}y  \nonumber\\
     &= & \frac{1}{2}\pi\left(\mathrm{Erf}\left[\frac{\sqrt{\rm RC}(x_{max}-x_0)}{2\sqrt{t}}\right]
    -\mathrm{Erf}\left[\frac{\sqrt{\rm RC}(x_{min}-x_0)}{2\sqrt{t}}\right]\right)\nonumber \\
    & \times& \left(\mathrm{Erf}\left[\frac{\sqrt{\rm RC}(y_{max}-y_0)}{2\sqrt{t}}\right]-\mathrm{Erf}\left[\frac{\sqrt{\rm RC}(y_{min}-y_0)}{2\sqrt{t}}\right]\right)
\label{eq:qresp}
\end{eqnarray}
where $\rm Erf$ is the error function and $x_{min},~x_{max},~y_{min},~y_{max}$ are pad borders coordinates.
To optimize computation, pre-computed charge density and signal shapes are stored for discrete sub-pad positions (e.g., a 25×25 grid) and for two representative $RC$ values (158~ns/mm$^2$ and 112~ns/mm$^2$) corresponding to two ERAM groups identified via X-ray test bench data, following the method introduced in~\cite{Attie:2022smn}.

\subsubsection{Electronics Response and Signal Formation}

The time evolution of the charge signal on each pad is shaped by the AFTER ASIC electronics, with the response function~\cite{Attie:2025wdt}:
\begin{equation}
    \begin{aligned}
        E(t, \omega_{s}, Q) = e^{-\omega_{s} t} + e^{-\omega_{s} t/2Q} \bigg[ \sqrt{\frac{2Q-1}{2Q+1}} \sin\left(\frac{\omega_{s} t}{2} \sqrt{4-\frac{1}{Q^2}}\right) \\
        - \cos \left(\frac{\omega_{s} t}{2} \sqrt{4-\frac{1}{Q^2}}\right) \bigg]
    \end{aligned}
    \label{eq:electronicsResponse}
\end{equation}
The parameters $\omega_{s}$ and $Q$ are the natural frequency of the electronics and the quality factor. 
Their values are determined by the circuit's resistance and capacitance values. Both parameters have been measured through a fit to electronics calibration data~\cite{Ambrosi:2023smx} yielding the values $Q=0.64$ and $\omega_{s}= \frac{2}{409.88 \;{\rm ns}}$ ~\cite{joshi:tel-04836929} for a 412~ns peaking time. The final pad waveform is the convolution of the charge arrival distribution with the electronics response derivative, digitized at 40~ns intervals. The electronic noise is added to the waveform following the model described in~\cite{Attie:2025wdt}.

This simulation framework models ionization, electron drift, diffusion, amplification, resistive charge dispersion, and electronics shaping within the \hatpc. 
\subsection{Reconstruction}
\label{sec:hatreco}
The reconstruction process involves several steps, transforming raw waveforms into reconstructed tracks.
First, waveforms are converted into hits by identifying the maximum of each waveform, defining the hit's time and charge from this peak.

All hits are then passed to a pattern recognition algorithm, adapted from the one developed for the ND280's vertical TPCs. The algorithm groups hits into patterns, identifies edges and junctions (points where tracks intersect), and connects them using an A* pathfinding algorithm applied independently to each ERAM.

Tracks crossing several ERAMs generate multiple segments, which are subsequently merged. The merging is based on the quality of the fitted trajectories: clusters belonging to two different segments are first fitted independently, yielding fit chi-squared values $\chi^2_1$ and $\chi^2_2$. They are then fitted together, producing a combined fit chi-squared $\chi^2_J$.  The two segments are merged if the combined fit satisfies the following criterium
\begin{equation}
    \chi^2_J < 1.3 \times \sqrt{\chi^2_1 \times \chi^2_2}
\end{equation}

The $x$-position (drift direction) of each pattern is typically reconstructed by multiplying the hit time by the drift velocity plus a constant offset. While sufficient for cosmic tracks, this method is unsuitable for neutrino-induced events, where neutrinos arrive in eight bunches separated by $\sim$600~ns. In those cases, a $\mathrm{T}0$ finder associates patterns with timing information from the surrounding Super-FGD and TOF detectors, identifying the correct bunch number and time offset for precise reconstruction of $x$-coordinate.

Next, hits within a pattern are grouped into clusters using a clustering algorithm adapted from~\cite{Attie:2022smn}. The cluster orientation depends on the local track angle: horizontal clustering is used for vertical tracks, vertical clustering is applied for horizontal tracks, and diagonal clustering is used for inclined tracks. Unlike earlier work~\cite{Attie:2022smn}, which was optimized for test-beam data with clustering orientation pre-defined by the beam geometry, the local track angle is here estimated from a preliminary track fit and used to select the clustering strategy.

Within each cluster, pads are ordered by charge, and the leading pad is the one with the highest charge. The track position is then determined using the logarithm of charge ratios between the leading pad and its neighbours, profiting of the properties of the resistive layers. 

\begin{figure}[tp]
    \centering
    \includegraphics[width=0.49\textwidth]{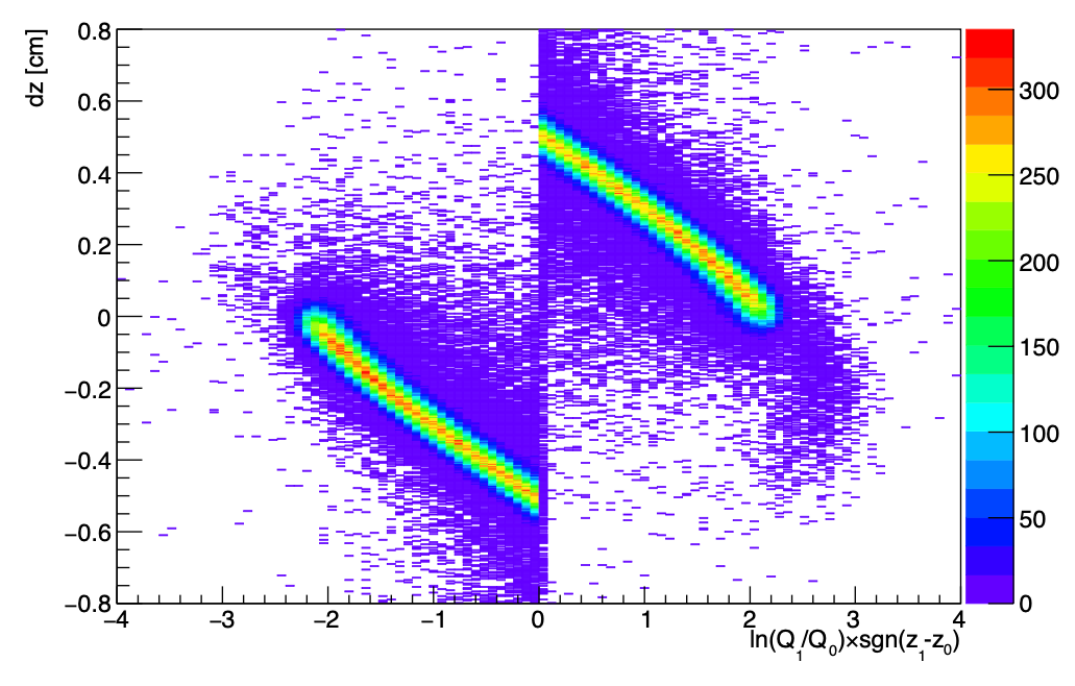}
	\includegraphics[width=0.49\textwidth]{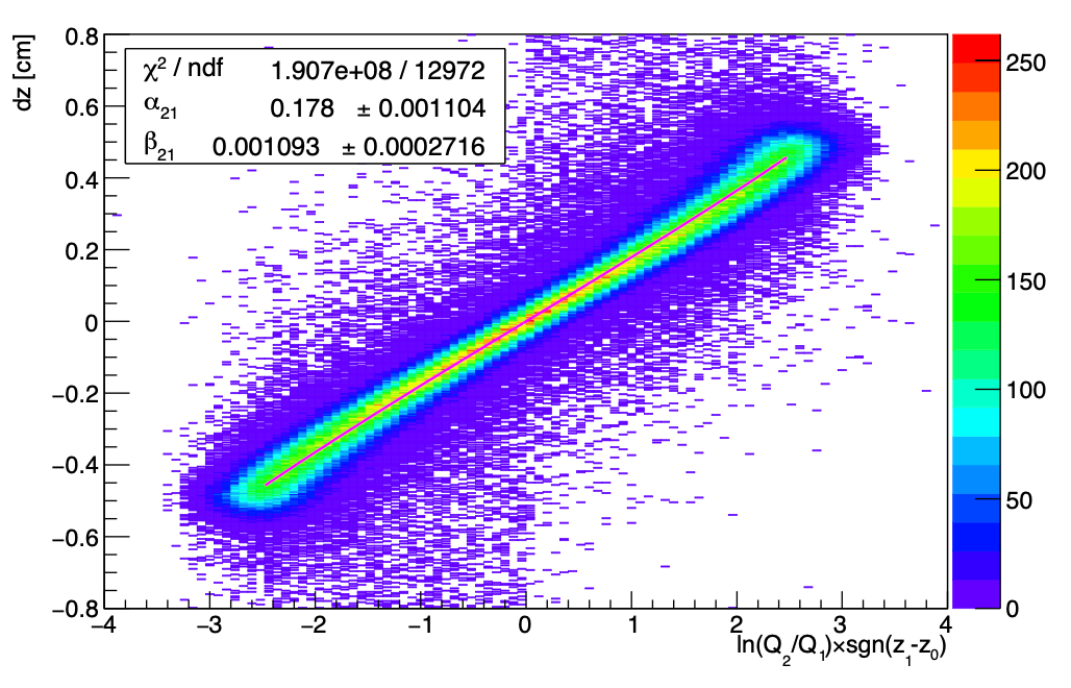}
    \caption{Distributions of the correlation between track position offset ($dz$) and the logarithmic charge ratios $\ln(Q_1/Q_0)$ and $\ln(Q_2/Q_1)$ for vertical tracks.}
    \label{fig:LogQFitDESY}
\end{figure}

Defining $Q_0$ as the charge in the leading pad and $Q_1$, $Q_2$ in neighbouring pads, the relation between track position offset ($dz$) and the logarithmic charge ratios $\ln(Q_1/Q_0)$ and $\ln(Q_2/Q_1)$ is illustrated in Fig.~\ref{fig:LogQFitDESY} for vertical tracks.

For both ratios, the logarithmic value is multiplied by the sign of the relative pad position $ \mathrm{sign}(z_1 - z_0)$ to indicate whether the neighbouring pad lies to the left or right of the leading pad.  
\begin{itemize}
    \item When the track is near a pad center ($dz\sim0$): $Q_0 \gg Q_1 \sim Q_2$; thus, $\ln(Q_1/Q_0)$ and $\ln(Q_2/Q_1)$ are large and near zero, respectively.
    \item When near a pad boundary: $Q_0 \sim Q_1 \gg Q_2$; so $\ln(Q_1/Q_0)\sim0$, and $\ln(Q_2/Q_1)$ is large.
\end{itemize}

Depending on track topology:
\begin{itemize}
    \item If near a leading pad edge, $\ln(Q_1/Q_0)$ gives the best position estimate.
    \item If closer to the pad center, $\ln(Q_2/Q_1)$ is more reliable.
\end{itemize}

To convert these ratios into continuous position estimates, $dz$ is scaled by $ \mathrm{sign}(z_1 - z_0) \times \mathrm{pad}_{\mathrm{width}}/2$.

The relationship between $dz$ and the charge ratios is parametrized iteratively:
\begin{enumerate}
    \item Initially, cluster positions are estimated via charge barycenters, and a track fit is performed.
    \item The track positions within clusters provide a first estimate of $dz$ for parametrization.
    \item A third-degree polynomial is fitted:
    \begin{equation}
        f(z) = \alpha \ln(Q) + \beta (\ln(Q))^3
    \end{equation}
    \item This process is repeated ten times to converge.
\end{enumerate}

Separate parameterizations are produced for each ERAM (to account for $RC$ variations) and for 10~cm bins of drift distance. The same procedure is applied to vertical and inclined tracks.

Before final track fitting, reconstructed cluster positions are corrected for distortions from non-uniform electric and magnetic fields by back-propagating the positions to their $x$-position inside the chamber using the fourth-order Runge-Kutta method. 

Finally, the track is fitted using a helix model, extracting the curvature radius (and its sign), starting position, and initial direction.

More details on the HA-TPC reconstruction algorithms and their performance can be found in Ref.~\cite{UlyssePhD}.

\subsubsection{Charge Spreading Model for Energy Loss Reconstruction}
\label{sec:hatdEdxmethod}
As introduced in Section~\ref{sec:hatsimu}, the ERAM response can be modeled using point-like charge deposits for simulation. However, for energy loss ($dE/dx$) reconstruction, it is preferable to smooth out local effects such as delta-rays. In this context, the deposited charge is treated as a continuous linear density evolving according to a diffusion equation:
\begin{equation} \label{eq:diffusion}
    \frac{\partial \rho_{1D}}{\partial t} = \frac{1}{RC}\left(\frac{\partial^2 \rho_{1D}}{\partial x^2} + \frac{\partial^2 \rho_{1D}}{\partial y^2}\right),
\end{equation}
where $RC$ characterizes the resistive-capacitive behavior of the ERAM layer. For a linear track projected in the $yz$ plane with $y = mz + q$, the solution is:
\begin{equation}\label{eq:rholineic}
  \rho_{1D}(z, y, t; m, q, \lambda) = \frac{\lambda}{\sqrt{2\pi}\sigma} \exp\left(-\frac{(-y + mz + q)^2}{2(1+m^2)\sigma^2}\right),
\end{equation}
with $\lambda$ the linear charge density and $\sigma$ representing the combined contribution of resistive charge spreading and transverse diffusion:
\begin{equation}\label{eq:sigma}
  \sigma(t; x, D_T, RC) = \sqrt{\frac{2t}{RC} + x D_T^2},
\end{equation}
where $D_T$ is the transverse diffusion coefficient.

The total charge collected by a pad with boundaries $z_L$, $z_R$, $y_B$, and $y_T$ is then obtained by integrating this density over the pad’s area: 
\begin{equation}
\label{eq:Qpadinclined}
Q_{\text{pad}}(t) = \iint_{\text{pad}} \rho_{1D}(x, y, t) \, dx \, dy
\end{equation}
This model naturally accounts for track inclination ($m$), the impact parameter ($q$), diffusion through the gas, charge spreading on the resistive layer, and contributions from neighboring pads, assuming an infinitely long track. However, several effects are neglected. First, longitudinal diffusion is omitted, as it reaches at most \SI{2.1}{mm} over a 1~m drift (with $D_L = \SI{210}{\micro m/\sqrt{cm}}$), corresponding to \SI{27}{ns}, which is smaller than the TPC time bin size of \SI{40}{ns}. Second, local track curvature is considered negligible within a \SI{1}{cm} pad, given a \SI{0.2}{T} magnetic field and a particle momentum of \SI{600}{MeV/c}, for which the curvature radius is about \SI{10}{m}. Finally, local fluctuations of charge deposition are not modeled, as the diffusion equation assumes a continuous charge distribution.

In previous work~\cite{Attie:2022smn}, energy loss in the ERAMs was estimated by taking the maximum of the sum of the waveforms associated with the same cluster. While effective for vertical and horizontal tracks, this method was less reliable for diagonal tracks. A new reconstruction algorithm, based on the above model, addresses this limitation without requiring clustering. For each pad crossed by a track, the expected signal is computed using Eq.~(\ref{eq:Qpadinclined}) convoluted with the derivative of the electronics response (Eq.~\ref{eq:electronicsResponse}). The ratio between the model’s peak signal $S^{max}_{model}$ and total charge $Q_{model}$ serves as a scaling factor to estimate the deposited charge $Q_{real}$ from the observed peak $S^{max}_{real}$:

\begin{equation}\label{eq:XP}
    Q_{real} = S^{max}_{real} \times \frac{Q_{model}}{S^{max}_{model}}.
\end{equation}

This technique, referred to as the XP (Crossed Pads) method~\cite{TristanPhD}, reconstructs the deposited charge using only the peak signals from pads crossed by the track, ignoring uncrossed pads and the remaining waveform samples. To optimize computational performance, the scaling factors are precomputed for various track geometries and stored in Look-Up Tables (LUTs).

 The XP method applies a truncated mean at the pad level, discarding the highest 30\% of $dE/dx$ values~\cite{RieglerChap10}. The event-level energy loss is then estimated by summing the reconstructed charges per pad and normalizing by the track segment length within each pad:

\begin{equation}\label{eq:dedxestimator}
    \left\langle\frac{dE}{dx}\right\rangle_{event} = \frac{\sum_{pad} dE_{pad}}{\sum_{pad} dx_{pad}}.
\end{equation}

Pads with track lengths below \SI{2}{mm} are excluded, as their signals are predominantly influenced by charge spreading from neighboring pads. Although small track segments may still exhibit fluctuations, the summation in Eq.~(\ref{eq:dedxestimator}) mitigates their overall impact.

\subsection{Simulation and Reconstruction performances}
\label{sec:simrecperf}

\begin{figure*}[p]
       \centering
       \begin{subfigure}{0.47\linewidth}
           \centering
           \includegraphics[width=\linewidth]{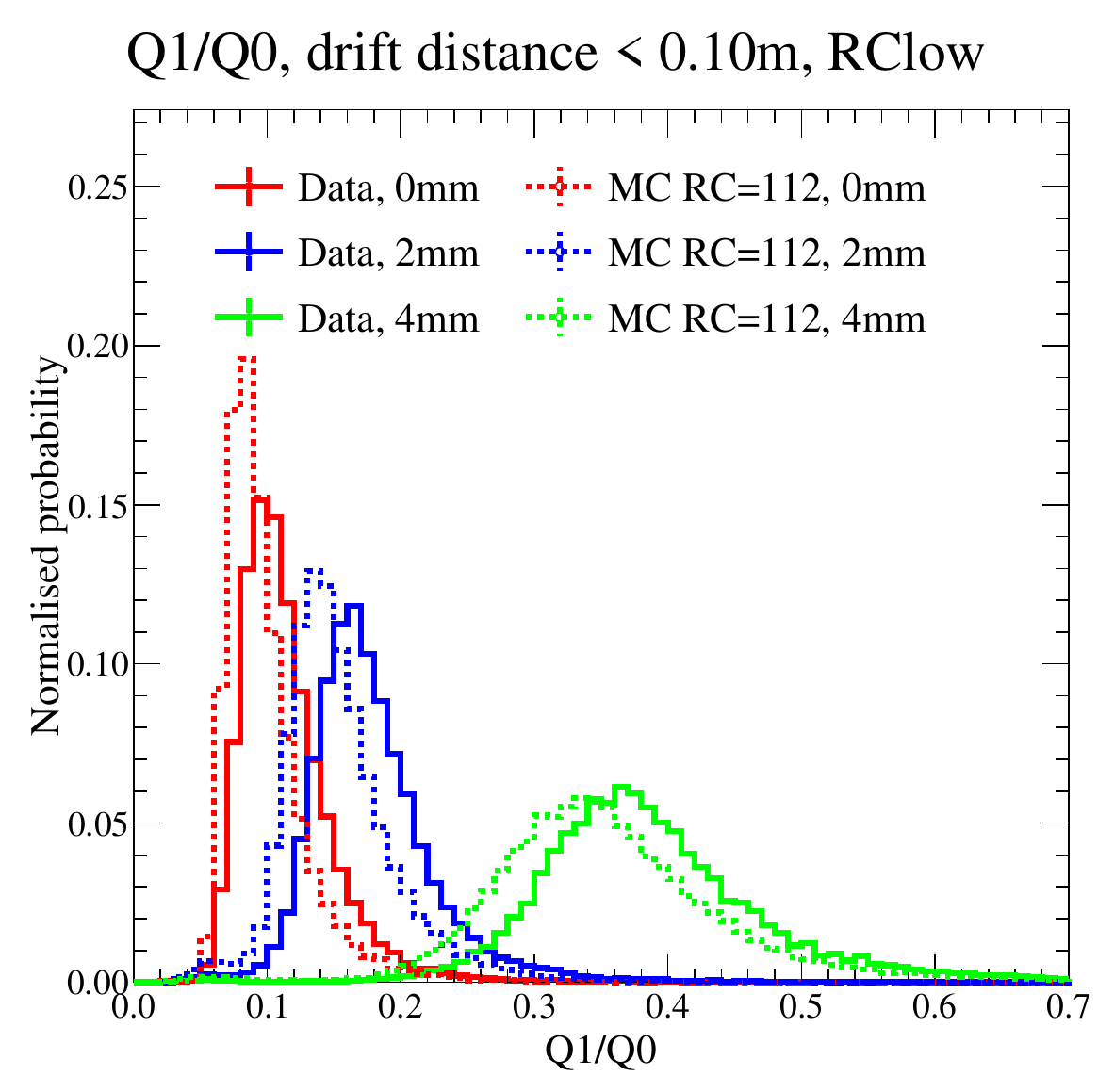}
           \caption{}
           \label{fig:chargeratio_a}
       \end{subfigure}
       \hspace{0.03\linewidth}
       \begin{subfigure}{0.47\linewidth}
           \centering
           \includegraphics[width=\linewidth]{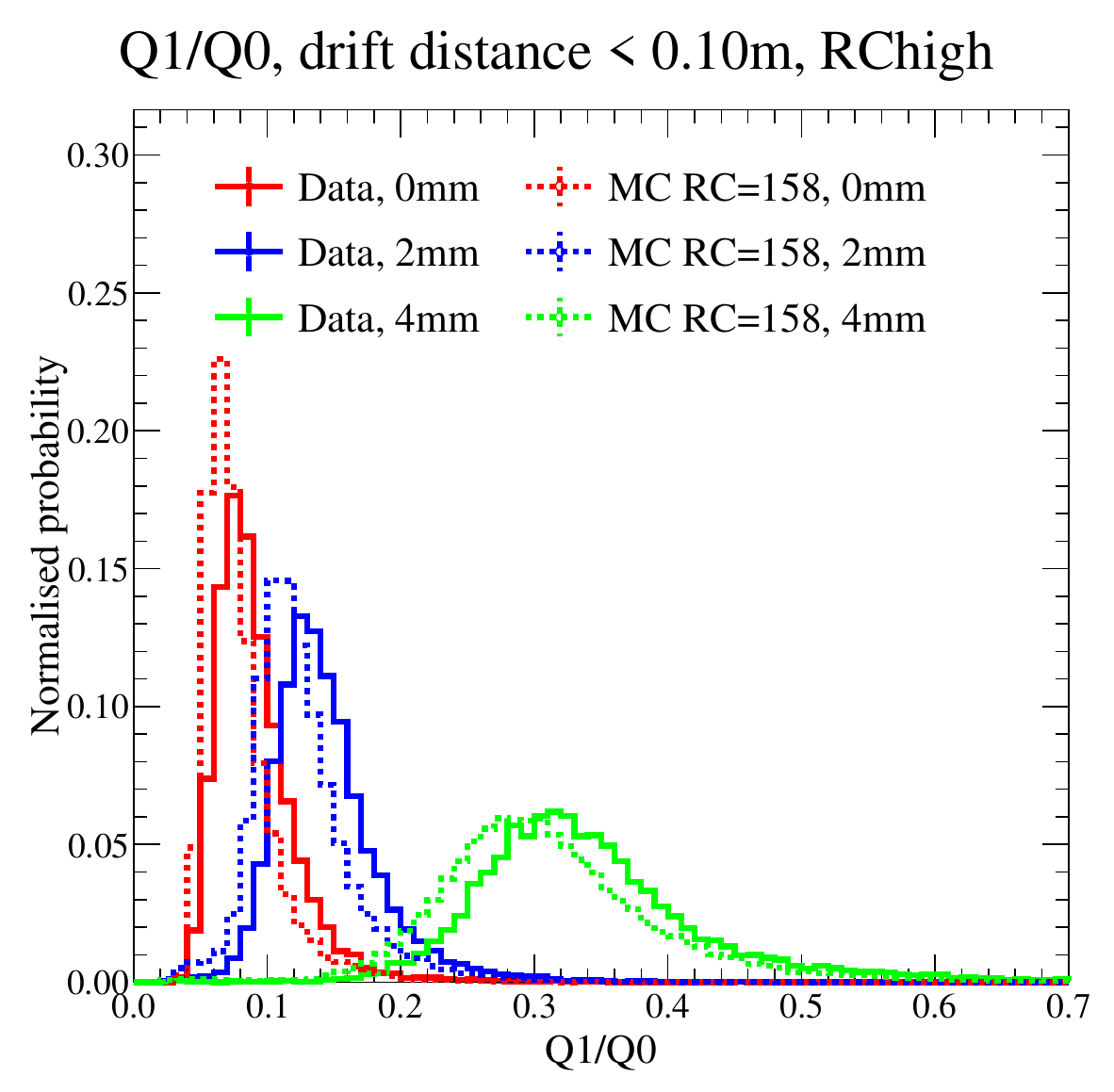}
           \caption{}
           \label{fig:chargeratio_b}
       \end{subfigure}
       \begin{subfigure}{0.47\linewidth}
           \centering
           \includegraphics[width=\linewidth]{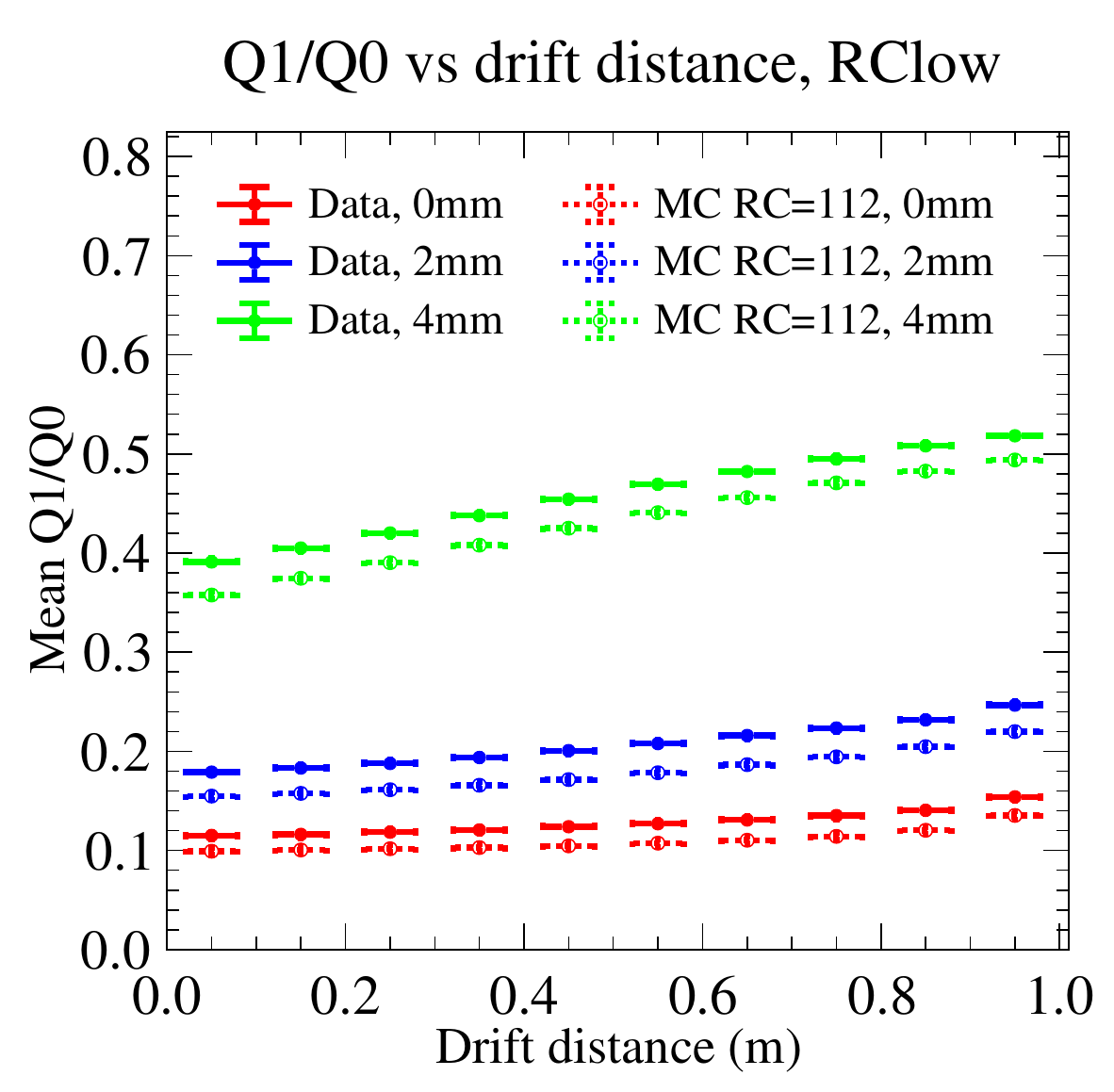}
           \caption{}
           \label{fig:chargeratio_c}
       \end{subfigure}
       \hspace{0.03\linewidth}
       \begin{subfigure}{0.47\linewidth}
           \centering
           \includegraphics[width=\linewidth]{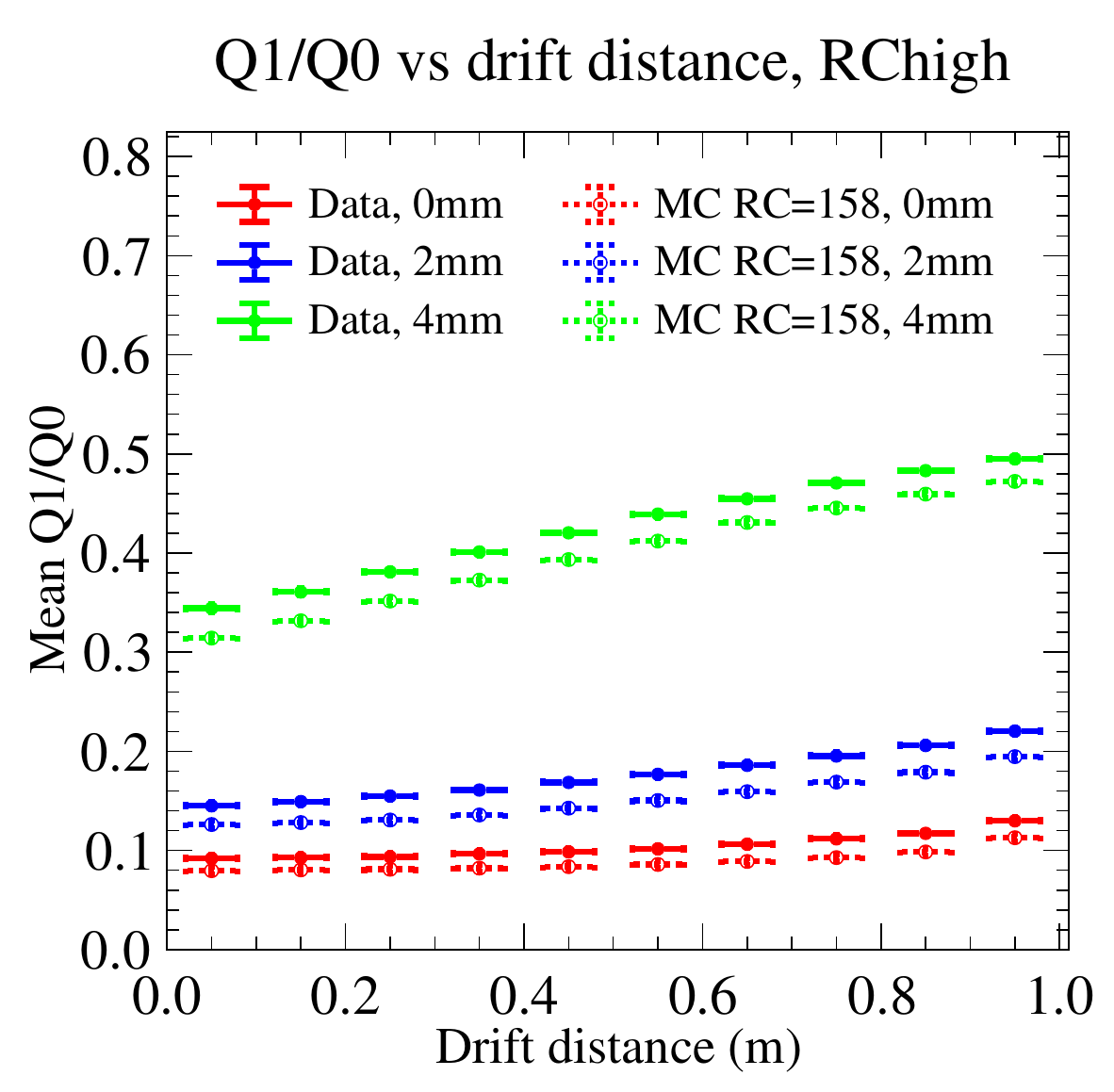}
           \caption{}
           \label{fig:chargeratio_d}
       \end{subfigure}
       \caption{Top: $Q_1/Q_0$ normal distribution considering small drift distances for cosmics data (continuous) and simulated $\mu^+$ and $\mu^-$ (dashed) for vertical tracks in HA-TPCs, with reconstructed $z$ distance 0 mm (red), 2 mm (blue) and 4 mm (green) from the center of the leading pad.  Bottom: mean $Q_1/Q_0$ obtained from the normal distributions considering different ranges of drift distance. The ERAMs results are averaged considering the group they are classified in: \textit{RClow} or \textit{RChigh} (Section \ref{sec:hatsimu}). The number of events considered is of the order of $10^7$ to minimise the statistical fluctuation.}
       \label{fig:all_charge_ratio}
   \end{figure*}

   \begin{figure*}[p]
       \centering
       \begin{subfigure}{0.47\textwidth}
           \centering
           \includegraphics[width=\linewidth]{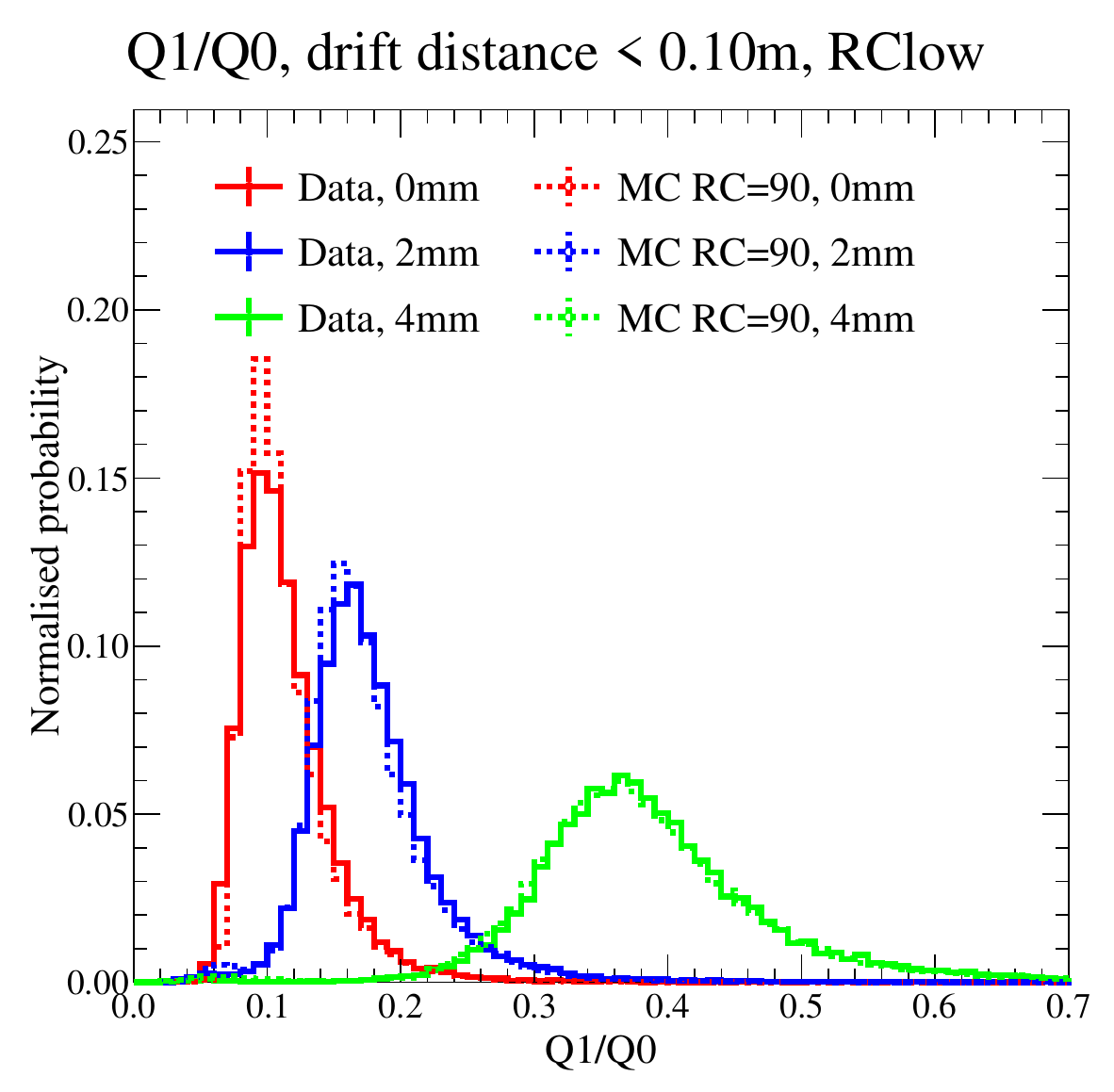}
           \caption{}
           \label{fig:chargeratio_a_lowRC}
       \end{subfigure}
       \hspace{0.03\textwidth}
       \begin{subfigure}{0.47\textwidth}
           \centering
           \includegraphics[width=\linewidth]{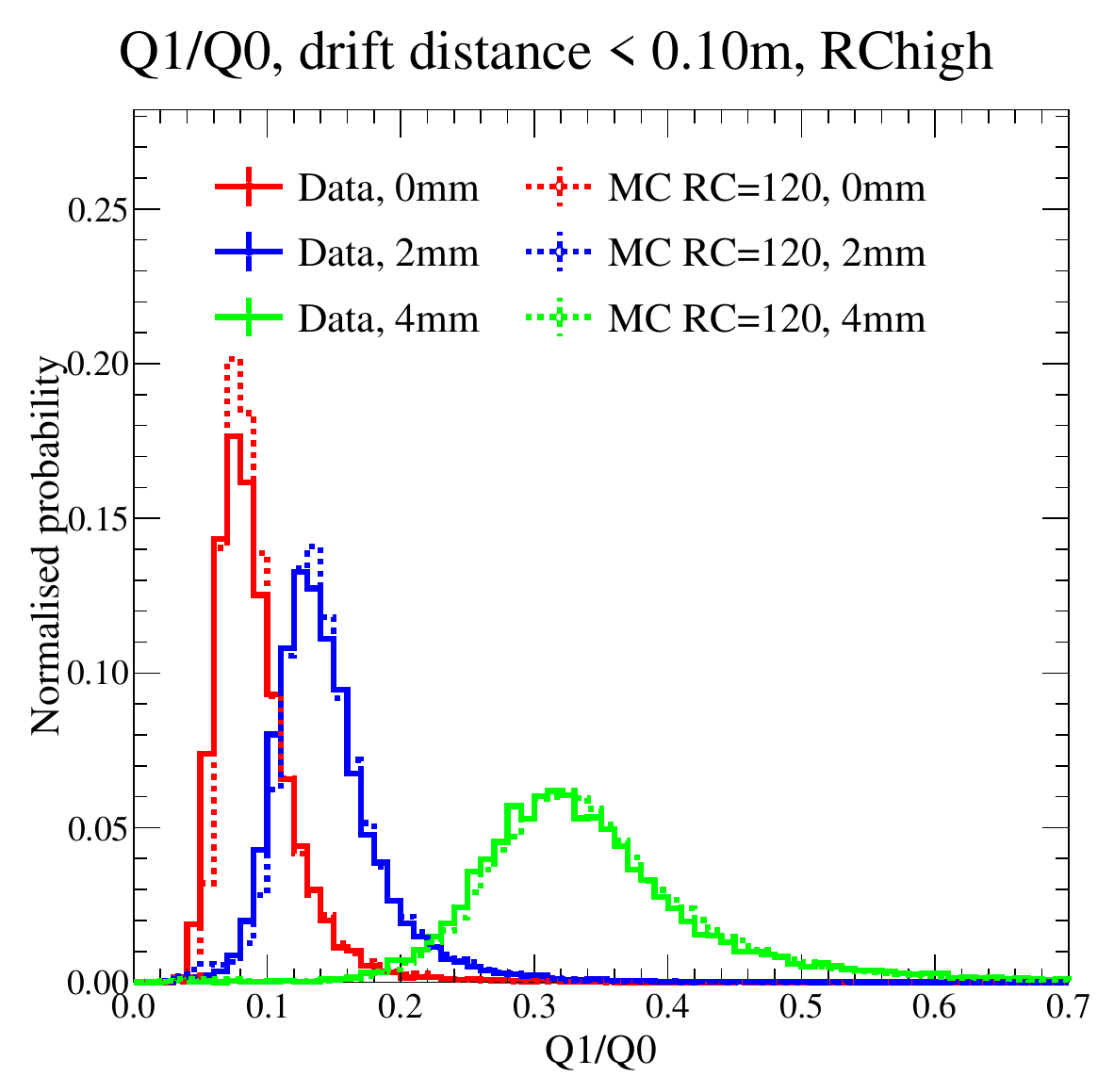}
           \caption{}
           \label{fig:chargeratio_b_lowRC}
       \end{subfigure}

       \begin{subfigure}{0.47\textwidth}
           \centering
           \includegraphics[width=\linewidth]{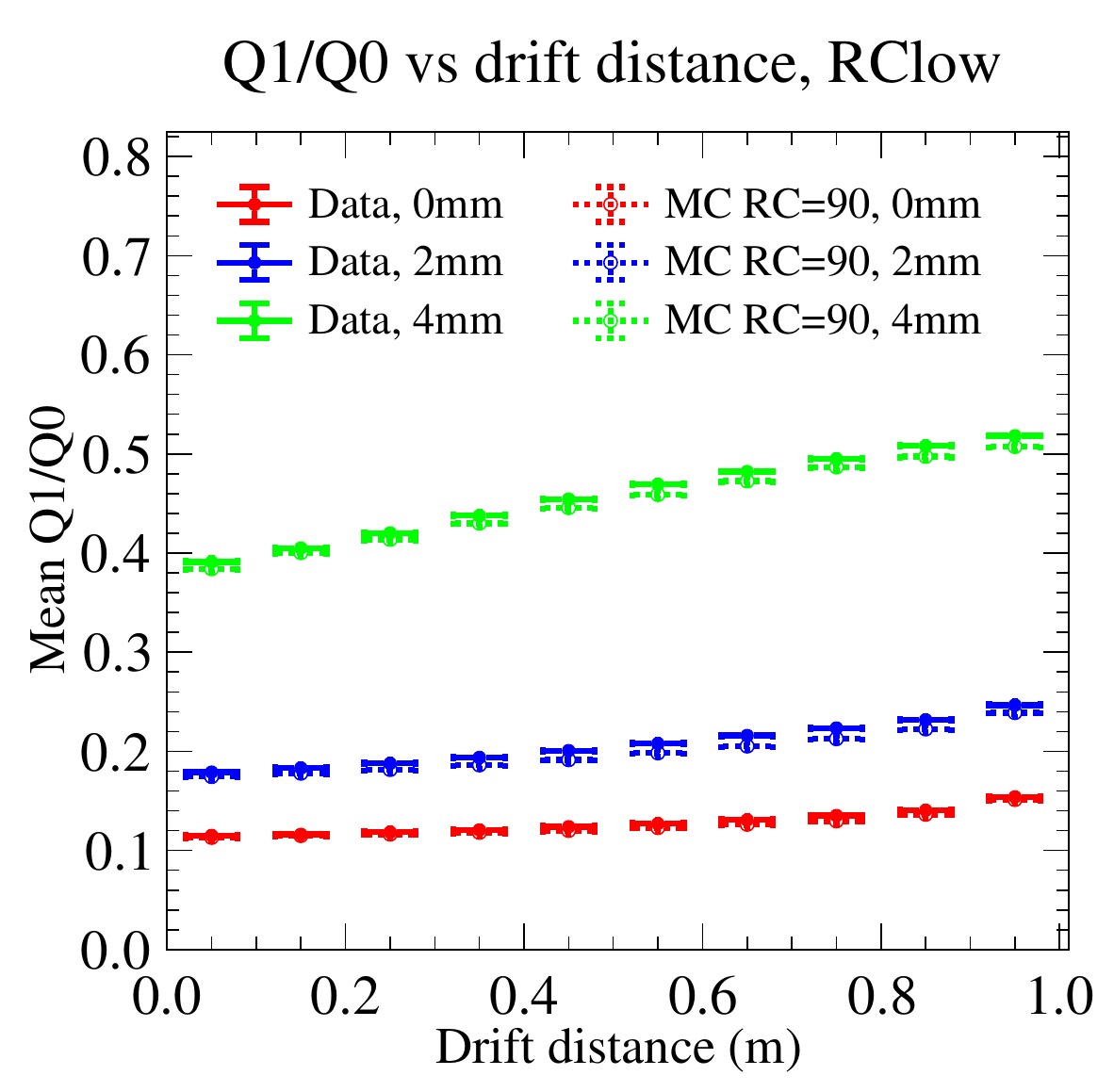}
           \caption{}
           \label{fig:chargeratio_c_lowRC}
       \end{subfigure}
       \hspace{0.03\textwidth}
       \begin{subfigure}{0.47\textwidth}
           \centering
           \includegraphics[width=\linewidth]{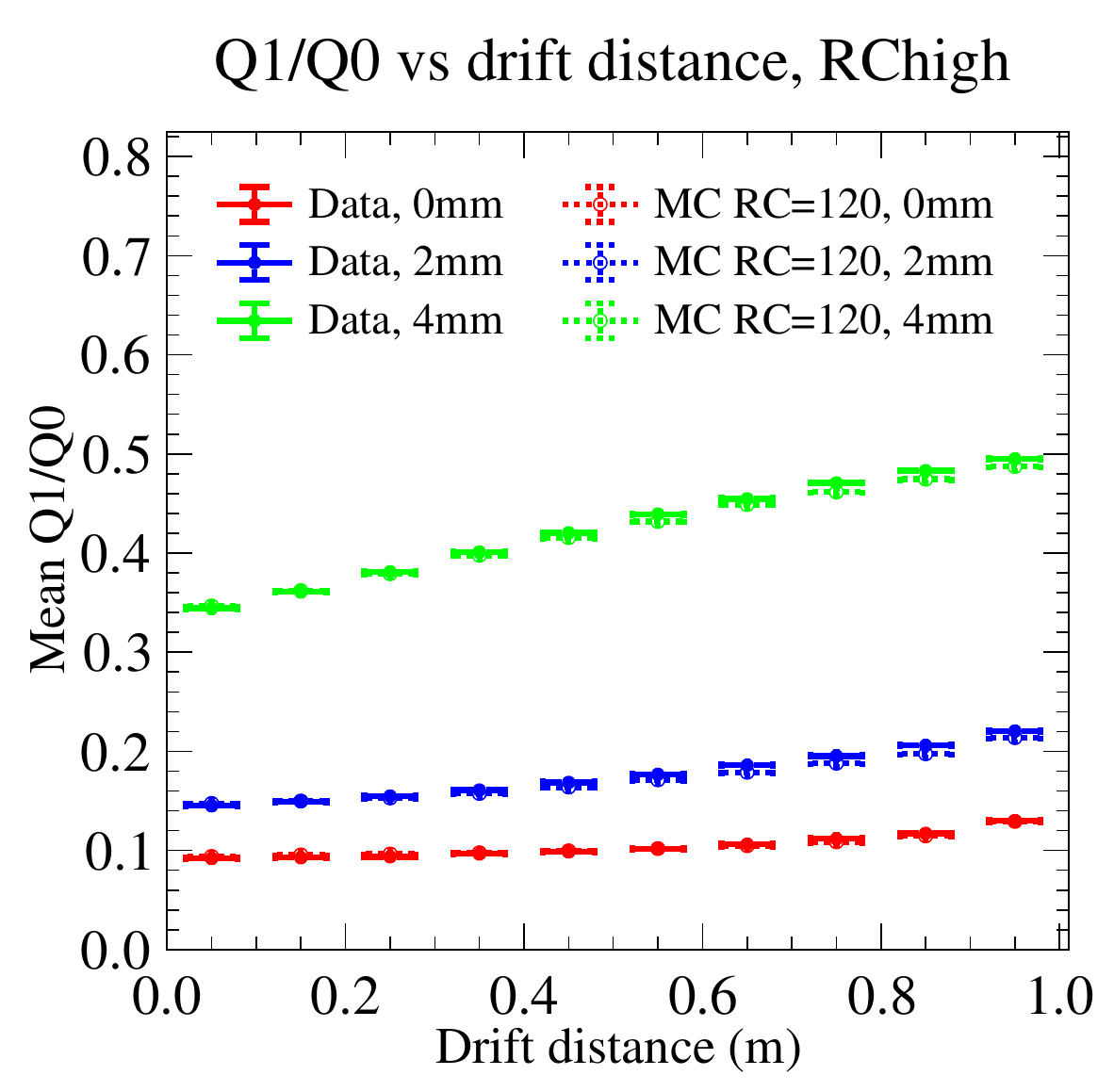}
           \caption{}
           \label{fig:chargeratio_d_lowRC}
       \end{subfigure}
       \caption{Same layout as presented in Fig.~\ref{fig:all_charge_ratio}, with a configuration assuming RC$_{low}$=\SI{90}{\nano\second\per\square\milli\meter}  and RC$_{high}$=\SI{120}{\nano\second\per\square\milli\meter} for MC. In this approach simulation achieves a better agreement with cosmics data.}
       \label{fig:all_charge_ratio_lowRC}
   \end{figure*}
   
In order to validate the simulations presented in the previous sections, a series of sanity checks were performed. 
Since the reconstruction algorithms rely on charge ratios between neighbouring pads, it is essential to evaluate their performance both on data and on simulations. 
To this end, Figs.~\ref{fig:chargeratio_a} and \ref{fig:chargeratio_b} show the $Q_{1}/Q_{0}$ distributions for HA-TPC vertical tracks, where the reconstructed $z$ position is displaced by 0, 2, and 4~mm with respect to the center of the leading pad, averaged over groups of ERAMs (as discussed in Sec.~\ref{sec:hatsimu}). 
Small drift distances are selected in order to suppress the impact of diffusion. 
For all the considered positions the simulations consistently underestimate the charge ratio. 
This trend persists when extending the comparison to the full drift distance: as illustrated in Figs.~\ref{fig:chargeratio_c} and \ref{fig:chargeratio_d}, the discrepancy remains across the whole drift region.
A possible improvement to the modelling of the charge ratio was investigated by reducing $RC$ by approximately 20\%. 
As shown in Fig.~\ref{fig:all_charge_ratio_lowRC}, this modification leads to a noticeably better agreement between data and Monte Carlo. 
It should be noted, however, that the corresponding $RC$ values are smaller than those obtained from X-ray measurements (see Section~\ref{subsec:eram_installation}), and the origin of this discrepancy is currently under investigation.

%% file: HATPerformance.tex
This section presents the initial performance of the \hatpc, compared with corresponding Monte Carlo (MC) simulations. Two data samples are considered: beam data, producing predominantly horizontal tracks, and cosmic data, yielding mostly vertical tracks. Dedicated MC simulations are generated for each. Both data and simulations are reconstructed using the ND280 software framework (presented in Fig.~\ref{fig:nd280dataflow}).

We report the achieved spatial resolution (Sec.~\ref{sec:spatresPerformances}), a study on the momentum resolution (Sec.~\ref{sec:hatpc_mom_res}) and $\mathrm{d}E/\mathrm{d}x$ resolution (Sec.~\ref{sec:depodEdxresolution}). The spatial resolution directly affects momentum measurement precision, while the $\mathrm{d}E/\mathrm{d}x$ resolution, combined with momentum, enables particle identification. The HA-TPC design requirements are:

\begin{itemize}
    \item Momentum resolution better than \SI{10}{\percent} at \SI{1}{\giga\electronvolt\per\clight}, corresponding to a spatial resolution of \SI{600}{-}\SI{1000}{\micro\meter}, as achieved in the vertical TPCs, to allow precise neutrino energy reconstruction.
    \item $\mathrm{d}E/\mathrm{d}x$ resolution better than \SI{10}{\percent}, to ensure reliable separation between electrons and muons. This is essential for measuring the $\nu_e$ contamination in the beam, which represents the primary background to the $\nu_e$ appearance signal at Super-Kamiokande. For comparison, the vertical TPCs achieve an \SI{8}{\percent} resolution, providing a $4\sigma$ separation between electrons and muons in the \SIrange{200}{2000}{\mega\electronvolt\per\clight} range.
\end{itemize}
\subsection{Spatial resolution}
\label{sec:spatresPerformances}
The spatial resolution is computed from residuals between reconstructed cluster positions and the fitted track, as described in Sec.~\ref{sec:hatreco}. After fitting tracks with a helix, the residual for each cluster is defined as:
\begin{equation}
    \text{residual} = \sqrt{(y^{\text{rec}} - y^{\text{fit}})^2 + (z^{\text{rec}} - z^{\text{fit}})^2} - R
    \label{residuals}
\end{equation}
where $y^{\text{rec}}$ and $z^{\text{rec}}$ denote the reconstructed cluster coordinates, $y^{\text{fit}}$ and $z^{\text{fit}}$ the fitted track coordinates, all on the ERAM plane, and $R$ is the helix radius.

\begin{figure}[h!]
    \centering
    \includegraphics[width=0.5\linewidth]{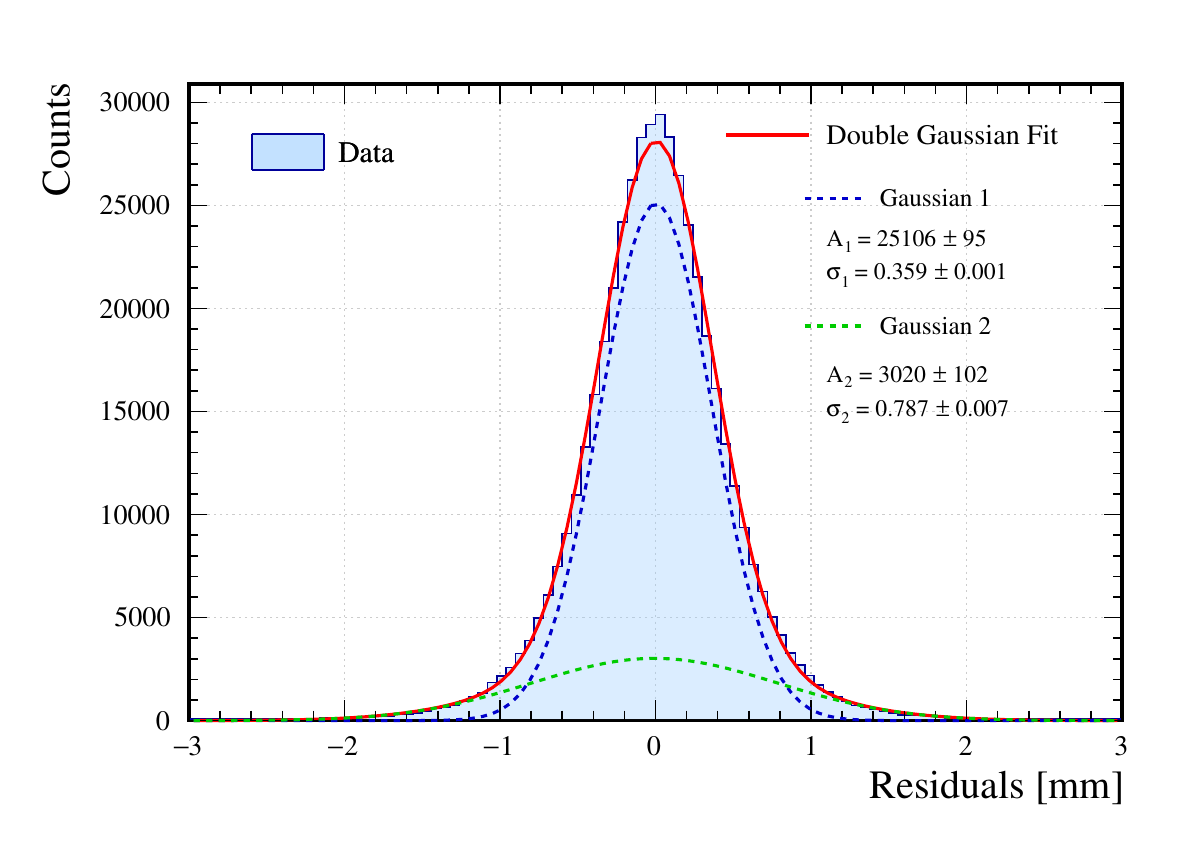}
    \caption{Residual distribution for beam data at drift distance \SI{150}{-}\SI{200}{\milli\meter}. The red curve is the total fit, while the blue and green dotted curves show the two Gaussian components. The amplitude and the standard deviation values of each Gaussian are reported in the top right panel.}
\label{fig:residualsDoubleGaus}
\end{figure}

Residuals from all tracks fill a histogram, fitted with a sum of two Gaussian functions. The resulting fit for a drift distance range of \SI{150}{-}\SI{200}{\milli\meter} is shown in Fig.~\ref{fig:residualsDoubleGaus}.

The spatial resolution is taken from the standard deviation of the narrower Gaussian component, which in Fig.~\ref{fig:residualsDoubleGaus} corresponds to $\sigma_1$.

\begin{figure}
    \centering
    \includegraphics[width=0.49\linewidth]{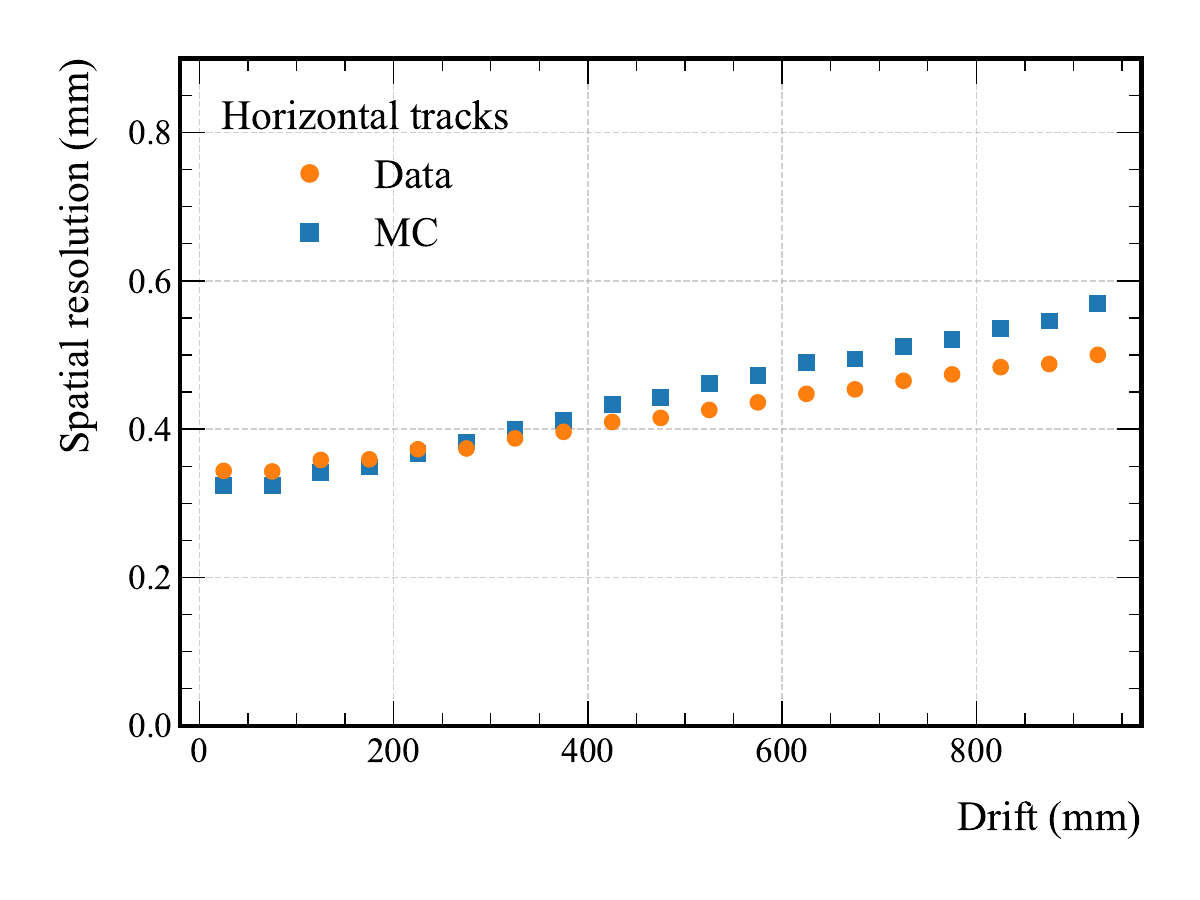}
    \includegraphics[width=0.49\linewidth]{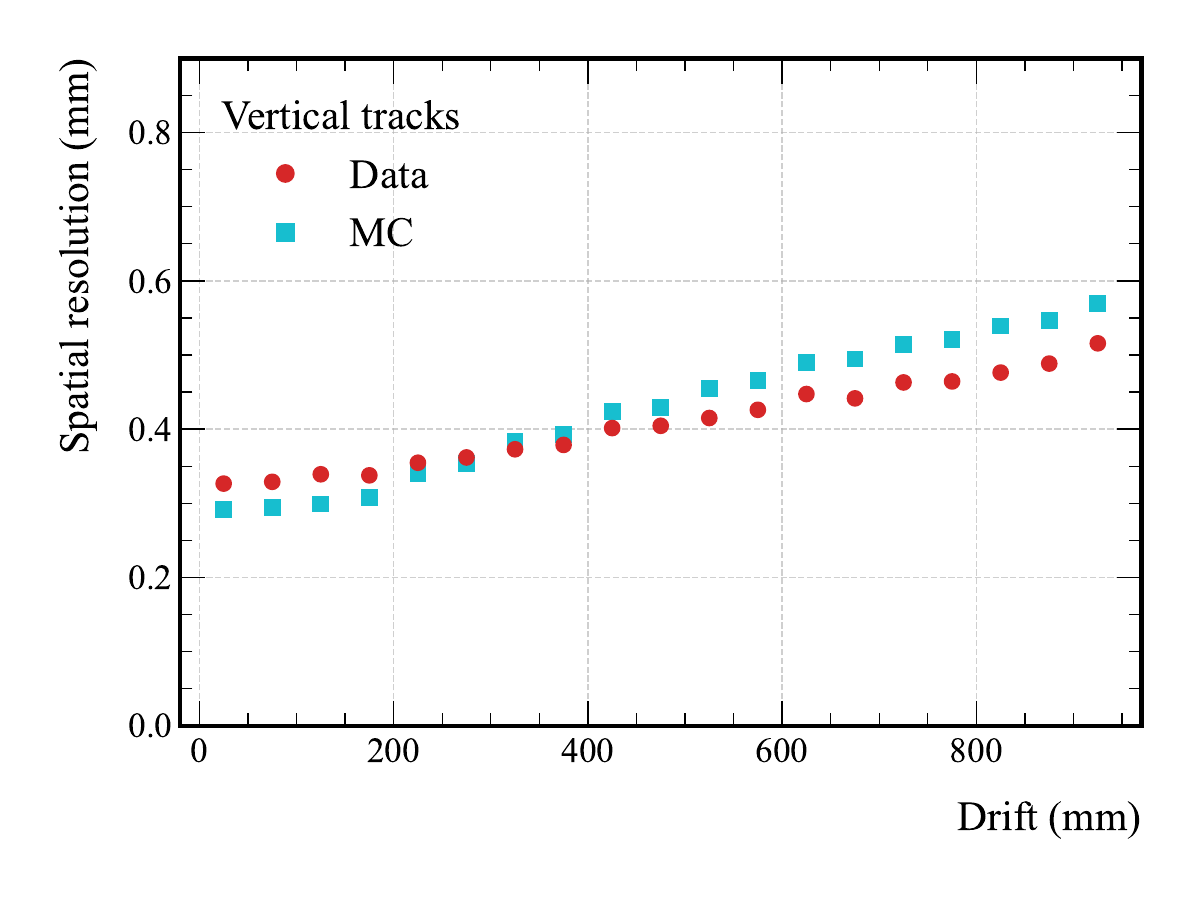}
    \caption{Spatial resolution versus drift distance for beam data (left) and cosmic data (right), compared with the corresponding MC simulations.}
    \label{fig:spatresVSdrift_beam_cosmic_sigma1}
\end{figure}

This procedure is repeated in bins of drift distance. Fig.~\ref{fig:spatresVSdrift_beam_cosmic_sigma1} shows the resulting spatial resolution as a function of drift distance for beam (left) and cosmic (right) data, compared to MC simulations.
As a better data/MC agreement on the charge ratios had been observed with lower values of $\mathrm{RC_{low}}$ and $\mathrm{RC_{high}}$ (see Sec. \ref{sec:simrecperf}), this lower RC configuration has hence been chosen for performing the study.

As expected from transverse diffusion, the spatial resolution worsens with increasing drift distance in both data and simulation. The MC/data agreement is particularly good at short drift distances -- especially for the beam case -- and gradually degrades with increasing drift. The spatial resolution remains better than \SI{0.6}{\milli\meter} throughout the full drift range in both data and simulations.

\begin{figure}[htbp]
    \centering
   
        \includegraphics[width=0.5\linewidth]{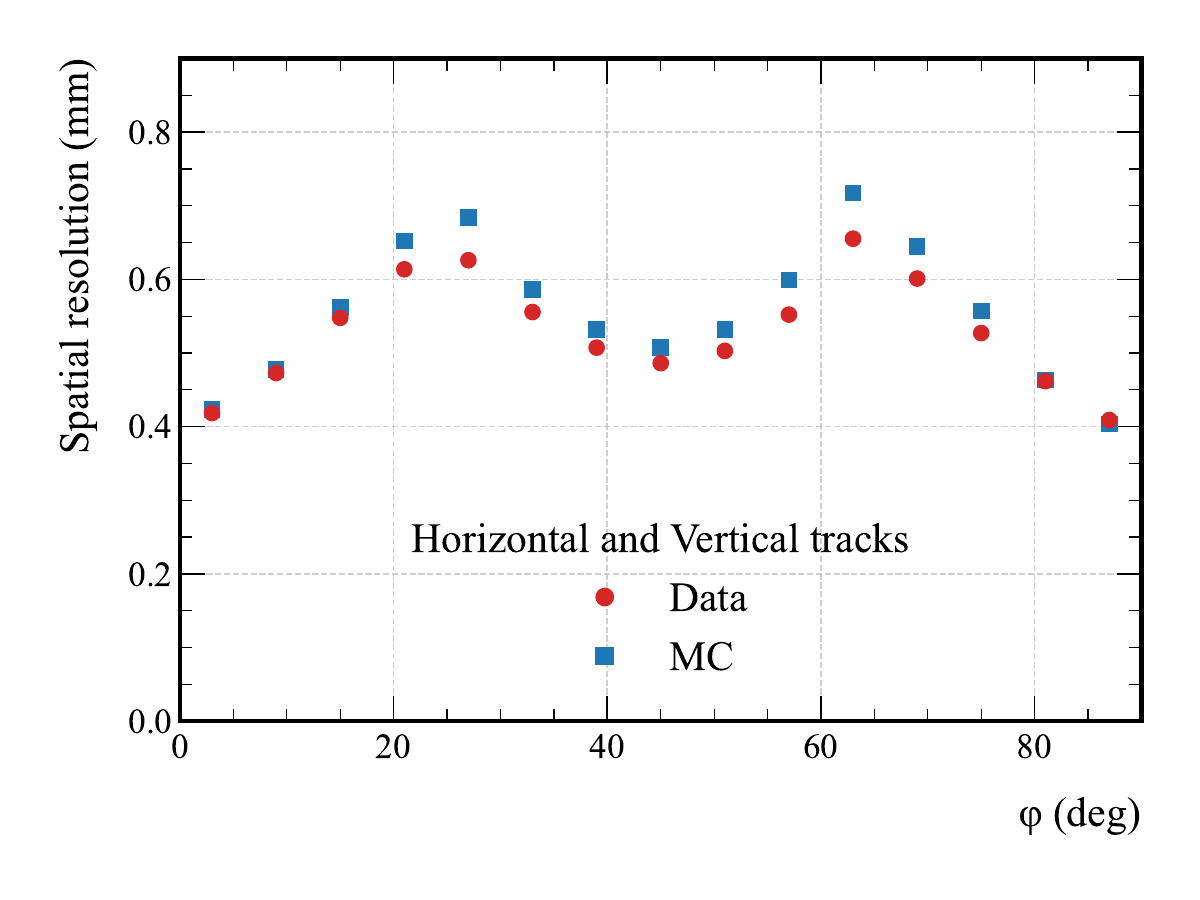}
    \caption{Spatial resolution versus track angle combining beam and cosmic data with the corresponding MC simulations.}
    \label{fig:SRvsPhiAngle}
\end{figure}

The spatial resolution has then been characterized as a function of the angle of the tracks defined as $\varphi=\tan^{-1}(|dir_y|/|dir_z|)$, combining both beam and cosmic data. The results are presented in Fig.~\ref{fig:SRvsPhiAngle} integrated over the full drift distance.

Three local minima can be seen at $0^\circ$, $45^\circ$ and $90^\circ$, where the orientation of the cluster is optimized for the given track direction (see Sec.~\ref{sec:hatreco}), in other words, when they are perfectly perpendicular to each other. Two local maxima are observed at $30^\circ$ and $60^\circ$, which are the pivot values between the different clusters' orientations: vertical, diagonal, and horizontal.

\begin{figure}
\includegraphics[width=0.49\linewidth]{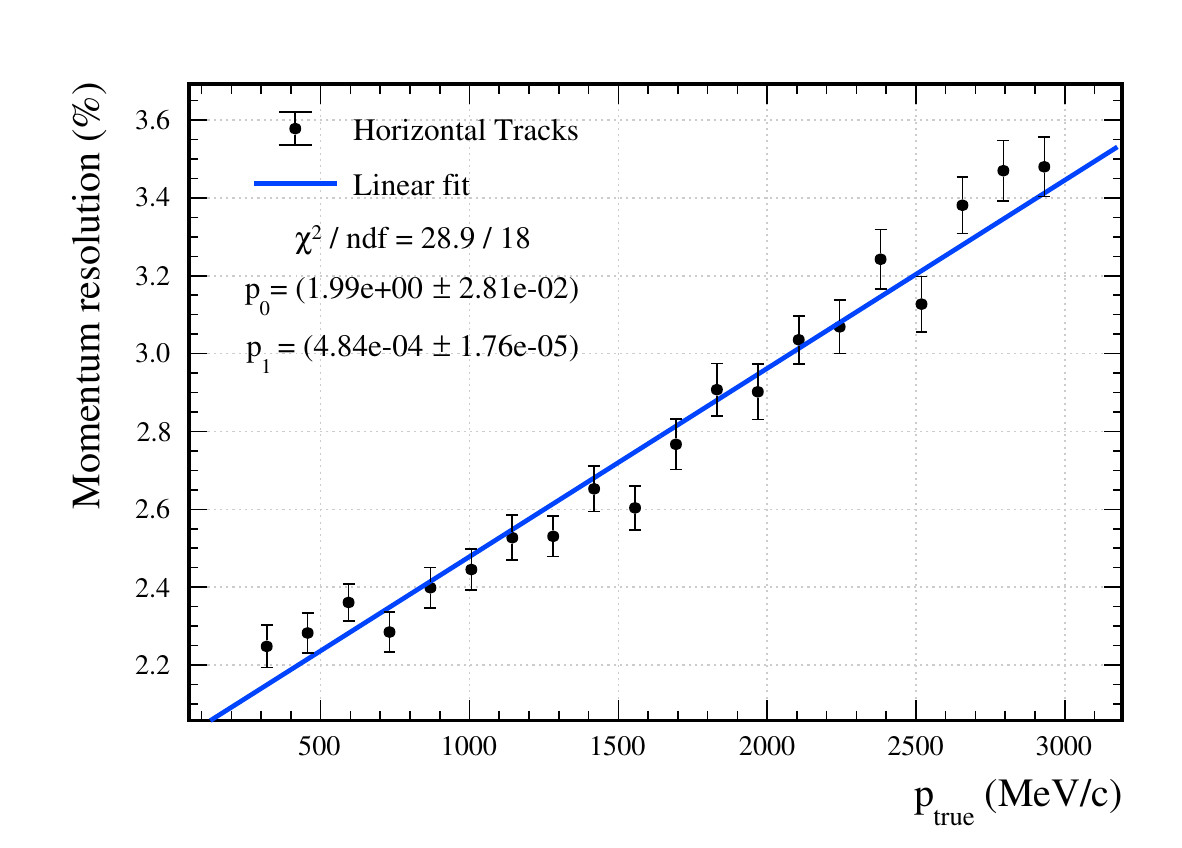}
\includegraphics[width=0.49\linewidth]{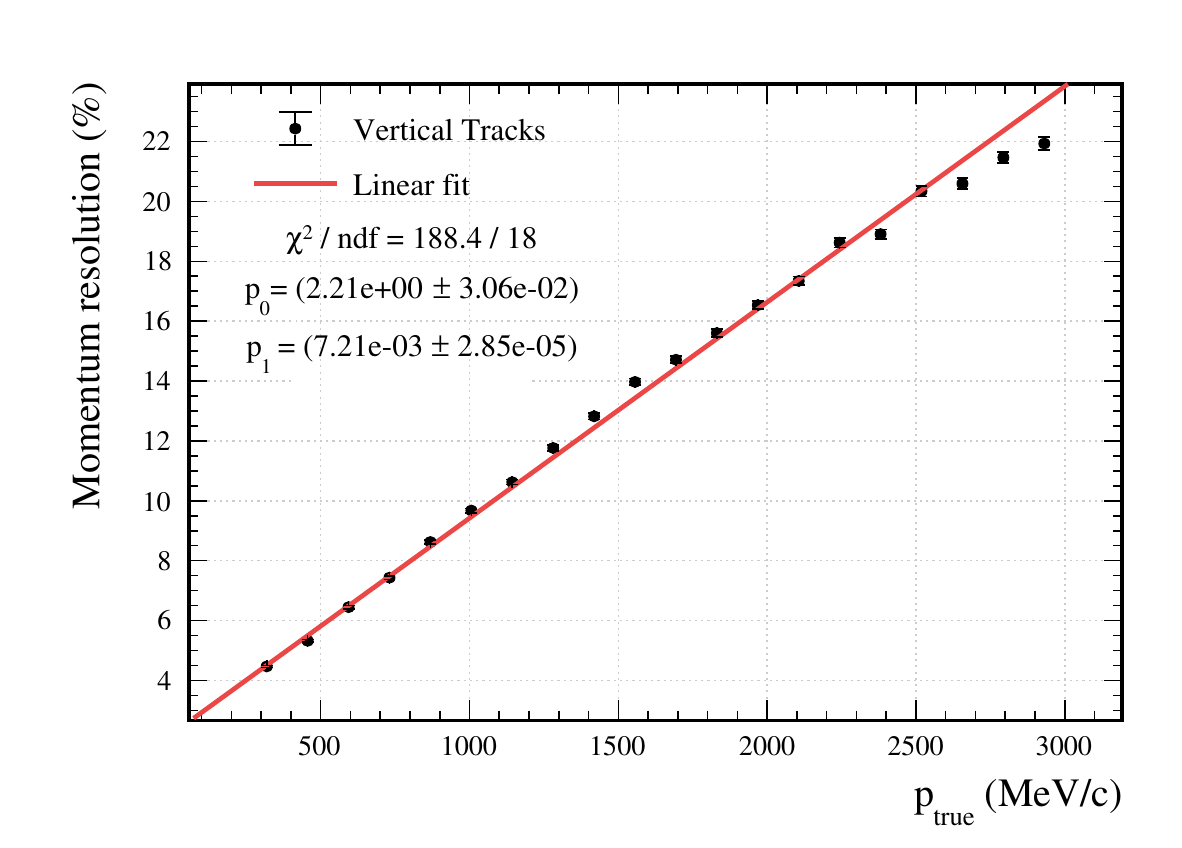}
\caption{Momentum resolution as a function of the true momentum for horizontal (left) and vertical (right) tracks. A linear fit to the points is shown, with the fit parameters reported in the legend.}
    \label{fig:momres_VSmom}
\end{figure}

Momentum resolution is estimated using Monte Carlo simulations of horizontal and vertical tracks crossing the two HA-TPCs. The simulated events are grouped in momentum bins, and for each bin, the distribution of the relative difference $\frac{p^{\mathrm{rec}} - p^{\mathrm{sim}}}{p^{\mathrm{sim}}}$ -- where $p^{\mathrm{rec}}$ and $p^{\mathrm{sim}}$ denote the reconstructed and simulated momentum, respectively -- is computed and fitted with a Gaussian. The standard deviation of the fit is taken as the momentum resolution in that bin.
Figure~\ref{fig:momres_VSmom} shows the resulting momentum resolution as a function of true momentum for horizontal (left) and vertical (right) tracks. A linear fit to the points is overlaid, with the fit parameters reported in the legend.

These results are consistent with expectations from the Gluckstern formula~\cite{gluckstern}, which describes the transverse momentum resolution for a charged particle in a magnetic field as:

\begin{equation}
\frac{\sigma_{p_t}}{p_t} = \sigma_{xy}[\SI{}{\meter}] \frac{p_t~[\SI{}{\giga\electronvolt\per\clight}]}{e ~ B~[\SI{}{\tesla}] ~ l^2~[\SI{}{\meter^2}]} \sqrt{\frac{720}{N_p + 4}}
\label{eq:momres}
\end{equation}
where 
\(\sigma_{p_t}\) is the absolute uncertainty on the transverse momentum,  
\(p_t\) is the transverse momentum,  
\(\sigma_{xy}\) is the single-hit spatial resolution in the bending plane,  
\(e\) is the elementary electric charge,  
\(B\) is the magnetic-field strength,  
\(l\) is the track lever arm (i.e. the distance between the first and last measured point in the bending plane), and  
\(N_p\) is the number of measured points along the track used in the fit.  

The linear dependence of the relative momentum resolution on the particle momentum, observed in both horizontal and vertical tracks, is expected from the formula due to the proportionality to $p_t$ in the numerator. The inverse quadratic dependence on the track length $l$ explains the better resolution observed for horizontal tracks, which are typically longer due to the detector geometry. For vertical tracks, the momentum resolution reaches slightly less than 10\% at \SI{1}{\giga\electronvolt\per\clight}.

\subsection{HA-TPC momentum reconstruction}
\label{sec:hatpc_mom_res}

The momentum resolutions shown in Fig.~\ref{fig:momres_VSmom} are based on a comparison between the reconstructed and the true momentum that is only accessible in the simulations. In order to validate the \hatpc performance on the data and investigate possible biases due to $E\times B$ effects produced by  mis-modeling of the electric and magnetic field maps, samples of vertical tracks from cosmics and of horizontal muons produced by neutrino interactions occurring outside the ND280 magnet have been selected.

With the cosmics sample, we selected vertical tracks crossing both \hatpc, parallel to the ERAM planes ($|dir X|<0.1$) and we computed the momentum difference between top and bottom TPC as:

\begin{equation}
    \Delta p_V = \frac{p_{top} - \mathrm{d}E/\mathrm{d}x_{sFGD} - p_{bot}}{p_{bot}}
    \label{eq:deltapV}
\end{equation}

Here the term $\mathrm{d}E/\mathrm{d}x_{sFGD}$ is included in order to account for the energy loss in the Super-FGD. A fixed value of \SI{120}{\mega\electronvolt}, corresponding to the typical energy loss of a MIP in a scintillator (\SI{2}{\mega\electronvolt\per\centi\meter}) multiplied by the height of the Super-FGD, is used. 

In absence of biases, we expect both distributions to be centered at zero with a width corresponding to $\sqrt 2$ times the resolution of a single track.

In order to reconstruct the momentum with the method described in Sect.~\ref{sec:hatreco}, electric field distortions have been taken into account using the maps described in Sect.~\ref{FieldCageElectricField} while for the magnetic field we considered three options:

\begin{itemize}
    \item COMSOL: B-field map produced via a COMSOL simulation of the ND280 geometry;
    \item Perfect: B-field components of (0.2~T, 0, 0)
    \item Mirror: B-field map obtained using the measurements of the magnetic field in ND280 that were done in 2009 in the region of the vertical TPCs~\cite{T2K:2011qtm}. Since the \hatpc region was not probed in that campaign we assumed a symmetry of the B-field components with respect to the center of the magnet. 
\end{itemize}

\begin{figure}
\includegraphics[width=0.49\linewidth]{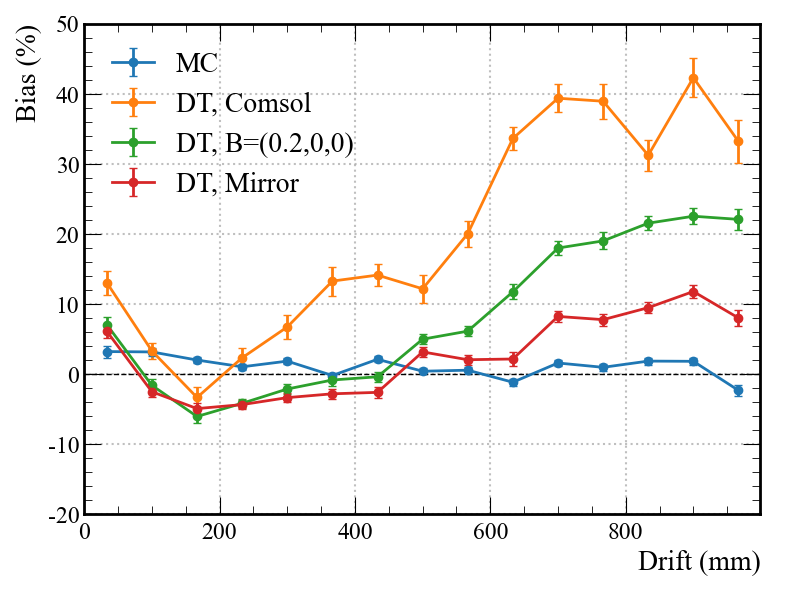}
\includegraphics[width=0.49\linewidth]{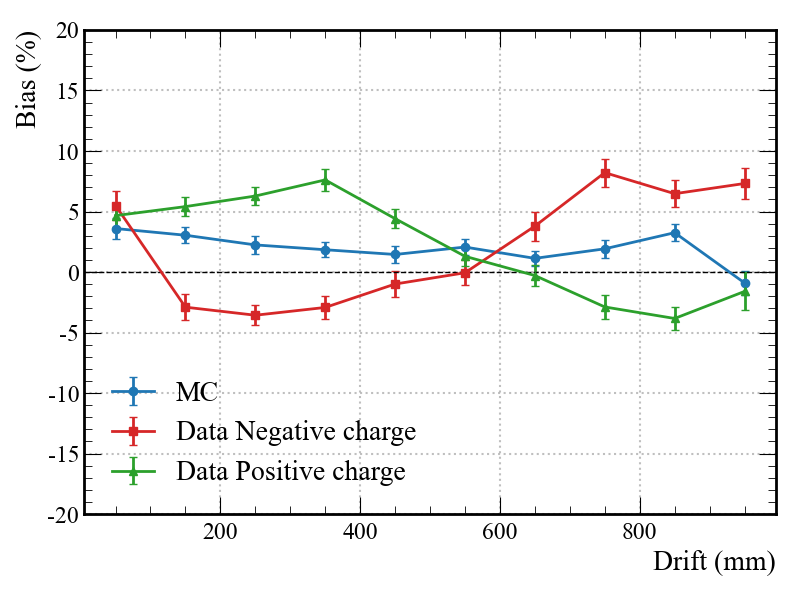}
\caption{Bias on the reconstructed momentum as a function of the drift distance for different B-field maps (left) and for positively and negatively charged tracks with the Mirror B-field map (right).}
    \label{fig:cosmics_bias}
\end{figure}

The bias as a function of the drift distance for these three options in the data for a sample of negatively charged tracks, compared with a MC simulation, is shown in Fig.~\ref{fig:cosmics_bias}. From the figure we notice that the COMSOL simulation introduces large biases. This was understood to be due to predicted $B_y$ components of the magnetic field that are much larger ($\sim60$~G) than the ones observed during the B-field measurements campaign in the mapped region ($\le 5$~G). Better results are obtained in the "Mirror" case, with biases not larger than 10\% for the whole drift region and with opposite sign for positively and negatively charged tracks.

\begin{figure}
\includegraphics[width=0.49\linewidth]{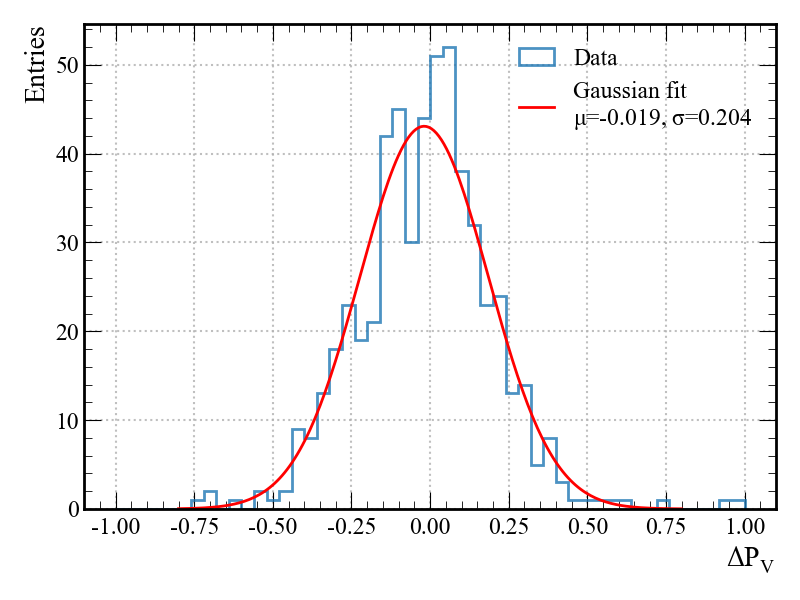}
\includegraphics[width=0.49\linewidth]{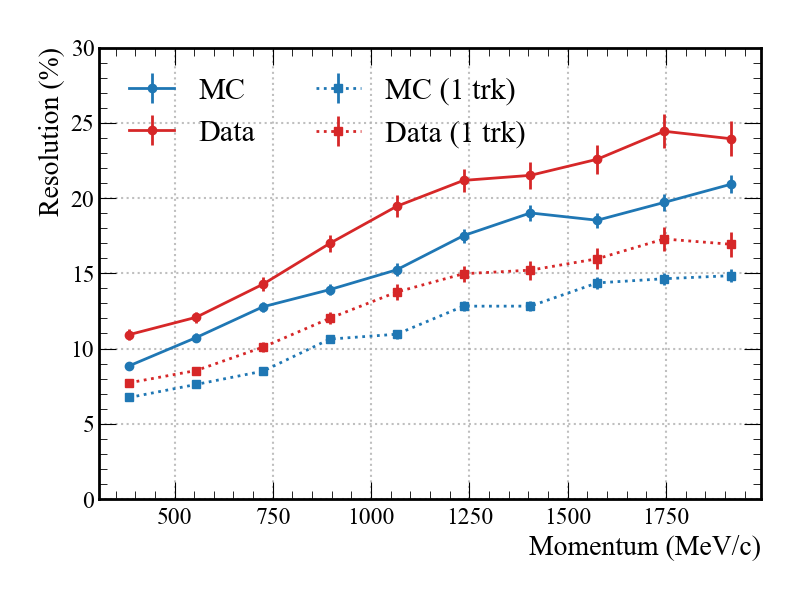}
\caption{$\Delta p_V$, Eq.~(\ref{eq:deltapV}), distribution for 1~GeV/c tracks (left) and momentum resolution as a function of the recostructed momentum for data and MC.}

    \label{fig:deltaPV_momentum}
\end{figure}

Eq.~(\ref{eq:deltapV}) allows to compute the momentum resolution for data and simulations as a function of the momentum, integrating over the whole drift distance. 
An example for vertical tracks with reconstructed momentum between 900~MeV/c and 1100~MeV/c and the resolution as a function of momentum are shown in Fig.~\ref{fig:deltaPV_momentum}. The remaining bias observed in Fig.~\ref{fig:cosmics_bias} degrades the resolution in the data with respect to MC. A resolution of 13\% for vertical tracks of 1~GeV/c is obtained in the data while, for the MC, this method return resolutions similar to the one shown in Fig.~\ref{fig:momres_VSmom}.

\begin{figure}
\includegraphics[width=0.49\linewidth]{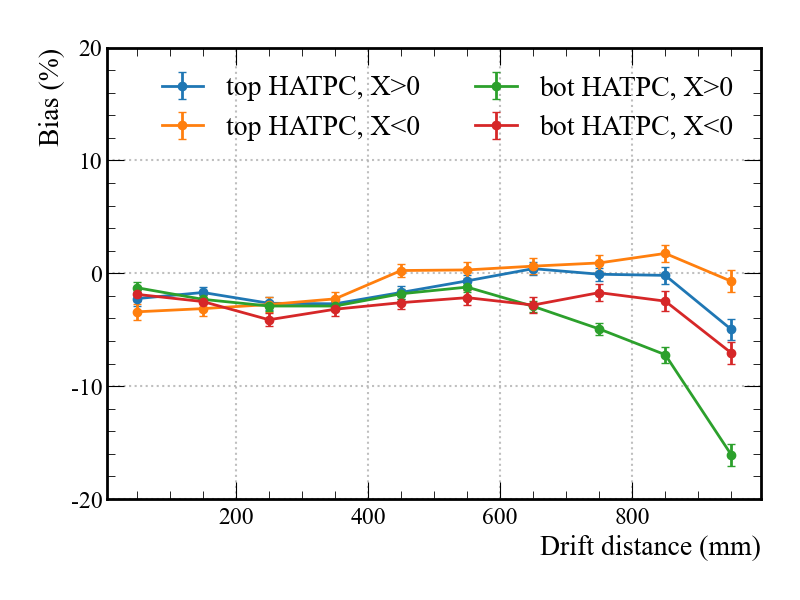}
\includegraphics[width=0.49\linewidth]{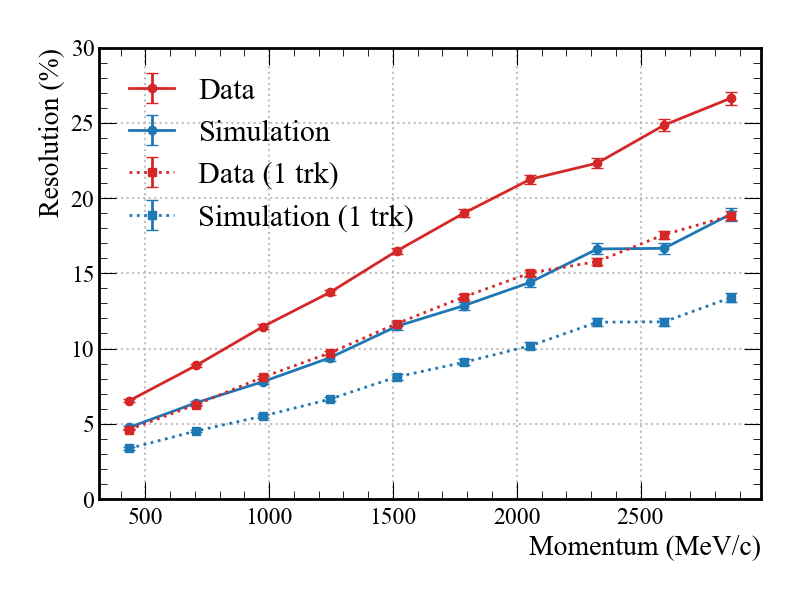}
\caption{Bias as a function of drift distance for the horizontal tracks (left) and resolution as a function of reconstructed momentum in data and simulations (right)}

    \label{fig:bias_sandmu}
\end{figure}

A similar study is done with horizontal tracks. In this case we selected long tracks crossing the entire \hatpc with $|dirZ|>0.9$. Before reconstruction, we separated the pattern into two halves and we reconstructed the momentum separately for each segment. For each pair of segments we computed:

\begin{equation}
    \Delta p_H = \frac{p_{left}  - p_{right}}{p_{right}}
    \label{eq:deltapH}
\end{equation}
where left (right) corresponds to the upstream (downstream) track segment. Similarly to Eq.~(\ref{eq:deltapV}), the mean of this distribution represents the bias and the width provides an estimate of the resolution. The bias as a function of the drift distance in the data, reconstructed with the Mirror B-field is shown in Fig.~\ref{fig:bias_sandmu} separately for the 2 \hatpc and the two endplates. The momentum resolution as a function of the reconstructed momentum for data and simulation is also shown. In the case of horizontal tracks, each track segment has a length corresponding to two ERAMs in the horizontal direction (72 clusters) and a resolution of $8\%$ is obtained in the data for 1~GeV/c tracks.

\subsection{Energy Loss Resolution and Particle Identification}
\label{sec:depodEdxresolution}
\begin{figure}[ht]
    \centering
    \includegraphics[width=0.49\linewidth]{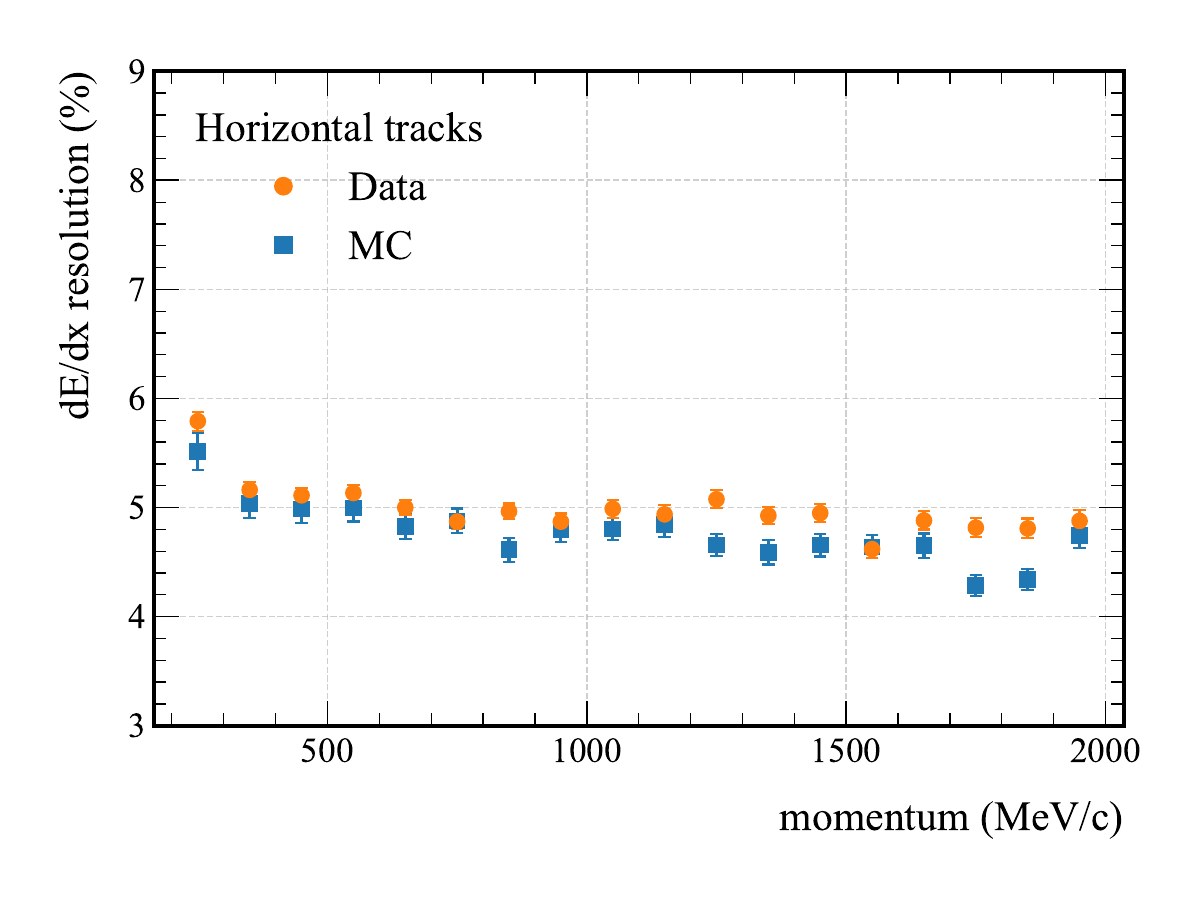}
    \includegraphics[width=0.49\linewidth]{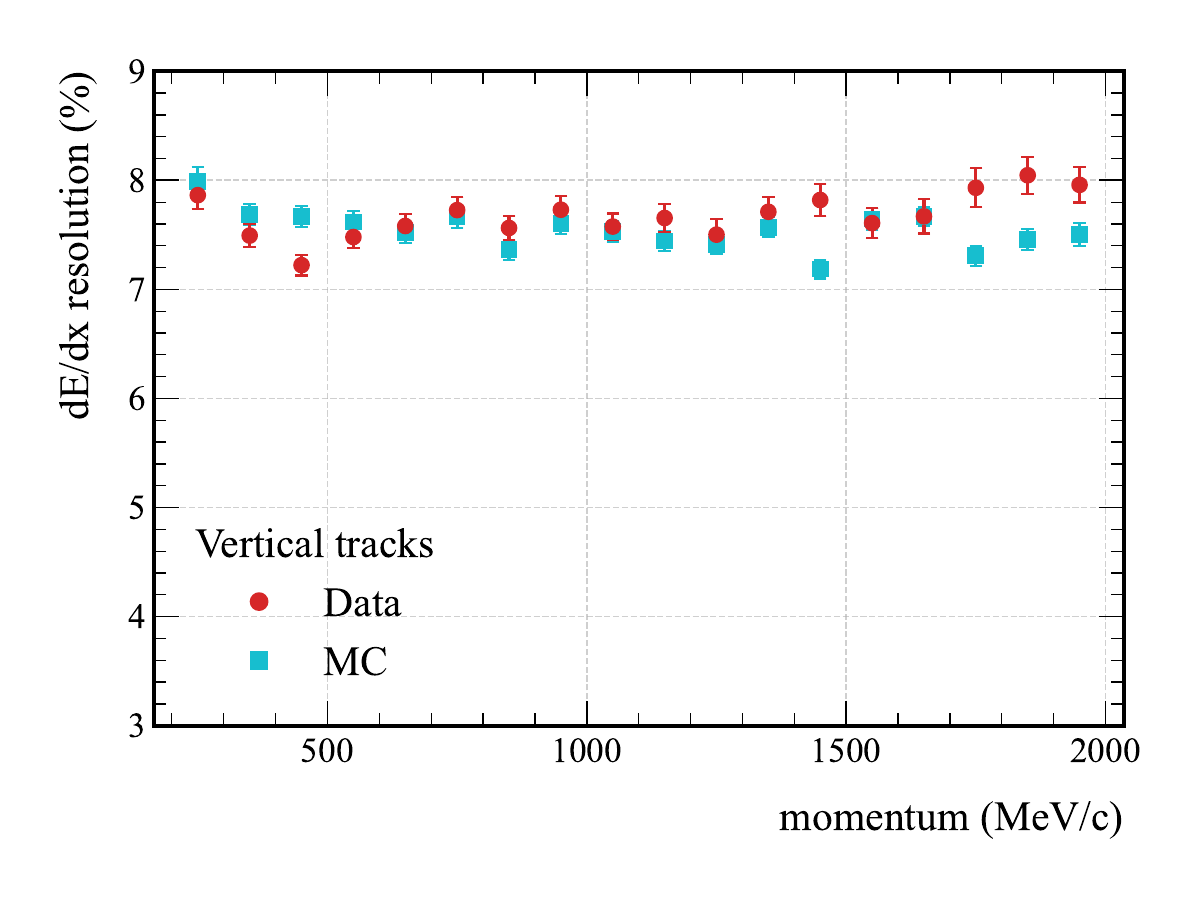}
    \caption{$\mathrm{d}E/\mathrm{d}x$ resolution as a function of the momentum for data and MC for beam (left) and cosmic (right) samples.}
    \label{fig:dEdxResolutionvsMomentum}
\end{figure}

Particle identification in the \hatpc relies on combining momentum information with the energy loss per unit length ($\mathrm{d}E/\mathrm{d}x$), calculated using the algorithm in Sec.~\ref{sec:hatdEdxmethod}. 
The analysis presented here uses tracks from neutrino interactions upstream of the \hatpc. Only long, nearly horizontal tracks with at least 130 clusters were selected. The only exception is the right panel of Fig.~\ref{fig:dEdxResolutionvsMomentum}, where a cosmic dataset is analyzed and nearly vertical tracks with at least 60 clusters are selected.

\begin{figure}[ht]
    \centering
    \includegraphics[width=0.49\linewidth]{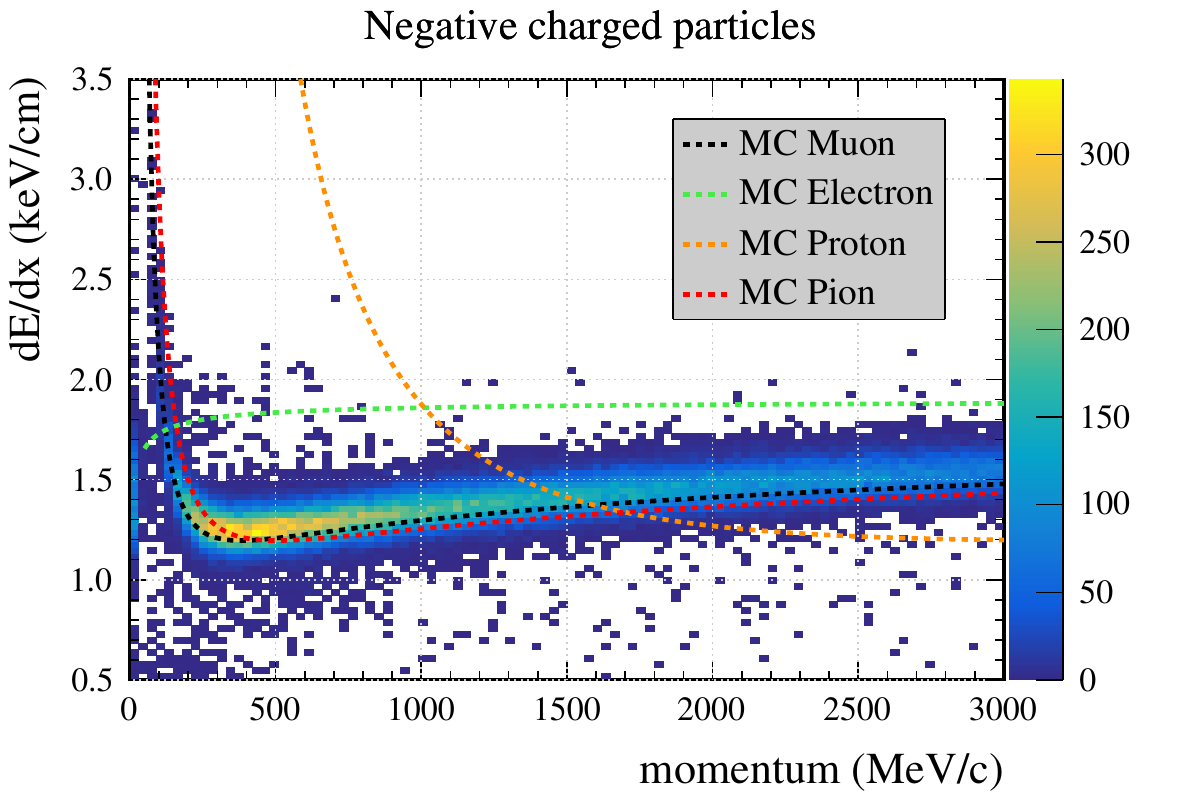}
    \includegraphics[width=0.49\linewidth]{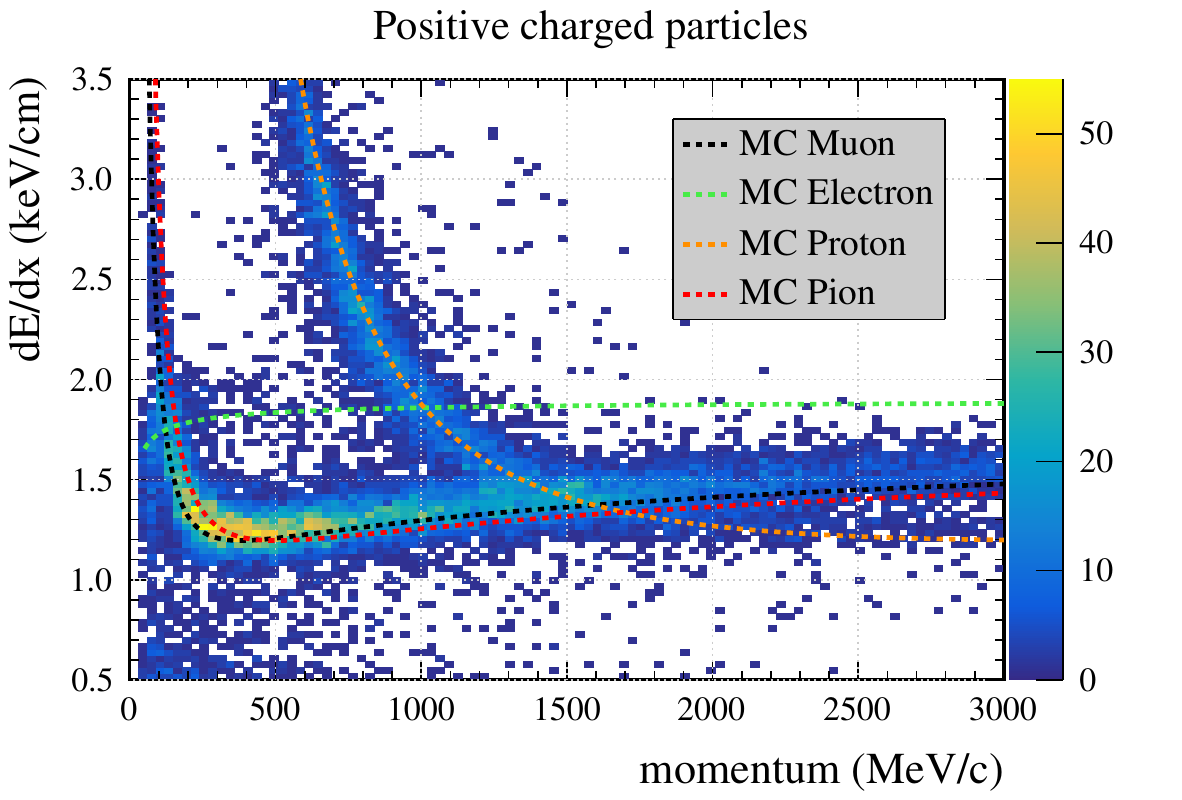}
    \caption{Reconstructed $\mathrm{d}E/\mathrm{d}x$ vs reconstructed momentum for horizontal beam tracks: negatively charged (left) and positively charged (right). MC predictions for muons, electrons, and protons are overlaid.}
    \label{fig:dEdxMomentum}
\end{figure}

Figure ~\ref{fig:dEdxMomentum} shows the reconstructed $\mathrm{d}E/\mathrm{d}x$ as a function of momentum for negatively (left) and positively (right) charged horizontal tracks, with MC predictions for muons, electrons, and protons based on the Bethe-Bloch equation for gaseous detectors~\cite{RieglerChap10}. The muon band is clearly visible in both plots. Protons appear only in the positively charged sample, while electrons and positrons populate the low-momentum region in both cases. These originate from neutral pion decays into photons, which subsequently convert into low-momentum electron-positron pairs. Overall, the data agree well with the simulation.

\begin{figure}[ht]
    \centering
    \includegraphics[width=0.55\linewidth]{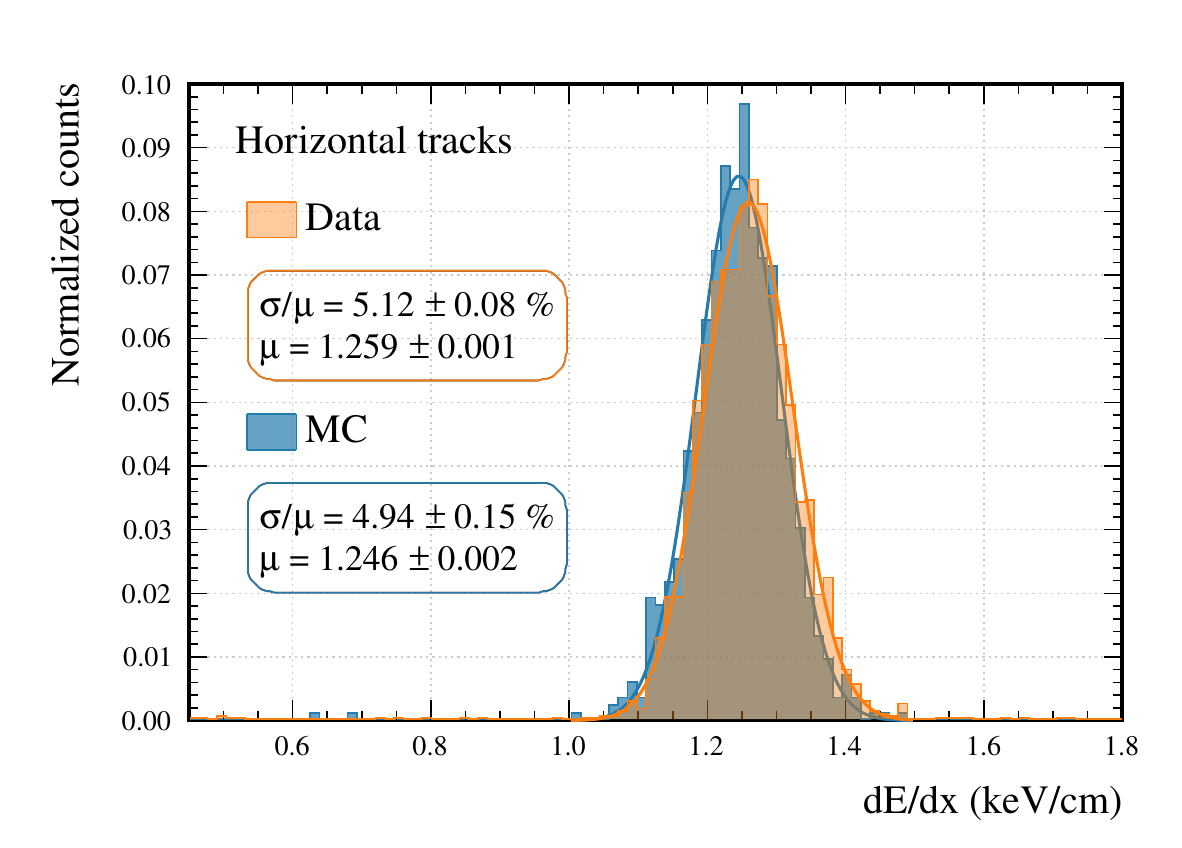}
    \caption{Deposited energy distribution for long horizontal beam tracks and reconstructed momenta between \SI{300}{-}\SI{500}{\mega\electronvolt\per\clight}, corresponding to the peak of the T2K beam muon spectrum. The corresponding Monte Carlo distribution is also shown. A Gaussian fit is performed for both data and simulation, with the fit results displayed in the left panels.}
    \label{fig:dEdxDistrib}
\end{figure}

For particle identification, both the $\mathrm{d}E/\mathrm{d}x$ value and its resolution are critical. To evaluate the resolution, data are binned in momentum slices. In each slice, the $\mathrm{d}E/\mathrm{d}x$ distribution is fitted with a Gaussian, and the resolution is defined as $\sigma/\mu$, with $\sigma$ the standard deviation and $\mu$ the mean. Fig.~\ref{fig:dEdxDistrib} shows the $\mathrm{d}E/\mathrm{d}x$ distribution for beam tracks between 300 and \SI{500}{\mega\electronvolt\per\clight}. The beam data yield a $\mathrm{d}E/\mathrm{d}x$ resolution of $(5.12 \pm 0.08)\%$, well within the 10\% requirement and in good agreement with the MC result of $(4.94 \pm 0.15)\%$.

The $\mathrm{d}E/\mathrm{d}x$ resolution is also evaluated over the full momentum range using \SI{100}{\mega\electronvolt\per\clight}-wide momentum bins. Figure~\ref{fig:dEdxResolutionvsMomentum} shows the results for the beam (left) and cosmic (right) datasets, along with their corresponding MC simulations. For the beam data, the $\mathrm{d}E/\mathrm{d}x$ resolution remains below 6\% and improves at higher momenta, as expected from decreasing relative fluctuations. In the cosmic case, the $\mathrm{d}E/\mathrm{d}x$ resolution follows the same trend but is slightly worse, between 7\% and 8\%. This can be attributed to the shorter vertical tracks, which produce fewer energy deposits. In both cases, the corresponding MC simulations are in good agreement with the data. Overall, not only in the \SI{300}{-}\SI{500}{\mega\electronvolt\per\clight} range shown previously, but across the entire momentum spectrum, the measured $\mathrm{d}E/\mathrm{d}x$ resolutions remain well within the 10\% requirement.

\begin{figure}[h!]
    \centering
    \includegraphics[width=0.55\linewidth]{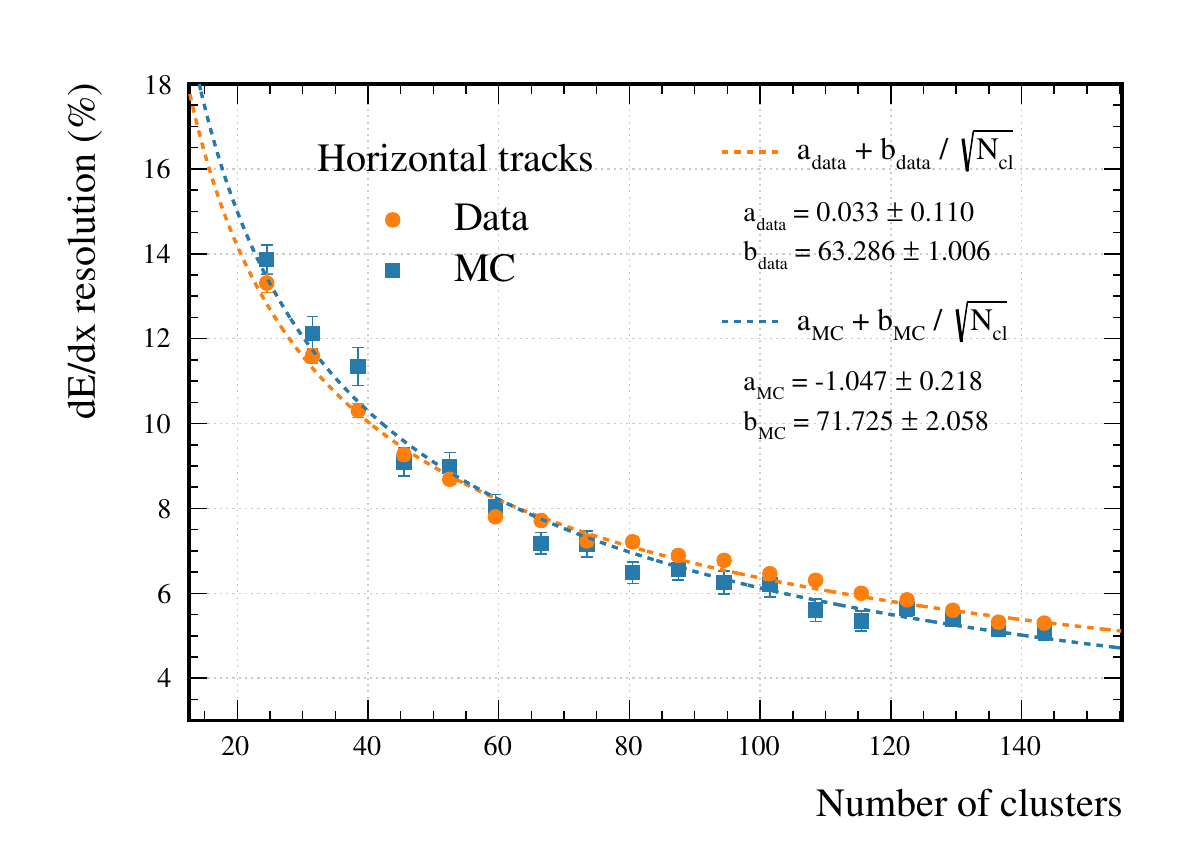}
    \caption{$\mathrm{d}E/\mathrm{d}x$ resolution as a function of the number of clusters ($N_{cl}$) for beam data and MC within the momentum range \SI{300}{-}\SI{500}{\mega\electronvolt\per\clight}. Results from the fit using the function $a_{fit} + b_{\rm fit}/\sqrt{N_{cl}}$ are shown in the right panel both for data and MC.}
    \label{fig:dEdxResolutionvsNcl}
\end{figure}

Finally, the dependence of the $\mathrm{d}E/\mathrm{d}x$ resolution on the number of clusters ($N_{cl}$) was studied for beam tracks with momenta between \SI{300}{-}\SI{500}{\mega\electronvolt\per\clight}. Figure~\ref{fig:dEdxResolutionvsNcl} presents the results for both data and MC samples. A dependence of the form $a + b/\sqrt{N_{cl}}$ is expected, as the $\mathrm{d}E/\mathrm{d}x$ resolution scales with the statistical uncertainty on the mean of the individual energy deposits, which decreases as $1/\sqrt{N_{cl}}$. Both data and MC are fitted with this function and are found to follow the expected trend. Only very short tracks with fewer than 40 clusters fail to meet the 10\% requirement, as expected, since these tracks provide a limited number of energy deposits and are therefore more affected by statistical fluctuations.
\subsection{Drift Direction Resolution and Bias}
To study the electric field behavior in the \hatpc, the resolution and bias along the drift ($x$) direction were measured using cosmic data and simulation. Vertical tracks with at least 64 clusters were selected to minimize geometric distortions. Since the magnetic field is aligned along $x$, these tracks appear approximately straight in the $yx$ plane. A linear fit was performed in this projection, and residuals $x_{\text{fit}} - x_{\text{data}}$ were calculated. Outliers with residuals above \SI{10}{\milli\meter} were removed, followed by a second fit and recalculation of residuals.

Residual distributions were analyzed separately for the bottom and top \hatpc. The resolution was defined as the fitted Gaussian width $\sigma_{\text{fit}}$, and the bias as its mean $\mu_{\text{fit}}$. The study was performed in bins of cluster positions along $x$, $y$, and $z$, and extended to subdivisions such as individual ERAM modules and endplates.

\begin{figure*}
    \centering
    \begin{subfigure}[b]{0.48\linewidth}
        \centering
        \includegraphics[width=\linewidth]{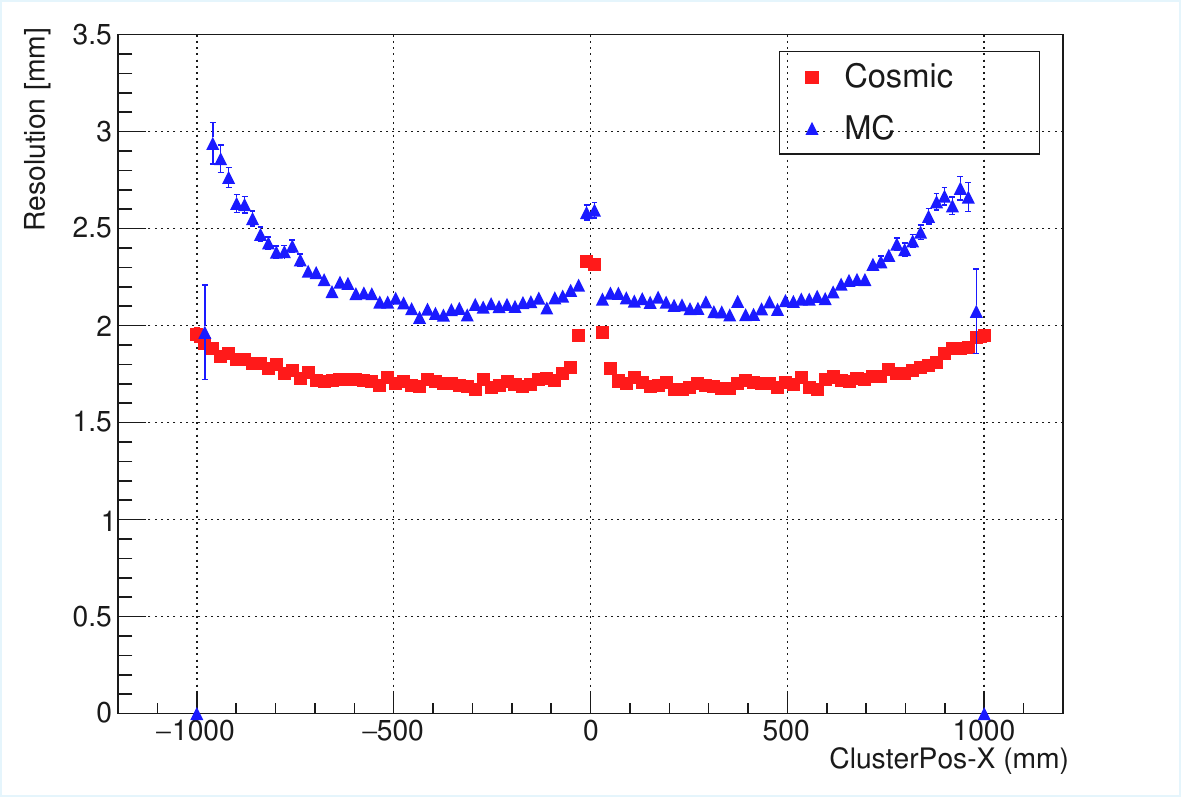}
        \caption{Resolution vs drift position $x$. Data: red squares, MC: blue triangles.}
        \label{fig:drift_res}
    \end{subfigure}
    \hfill
    \begin{subfigure}[b]{0.48\linewidth}
        \centering
        \includegraphics[width=\linewidth]{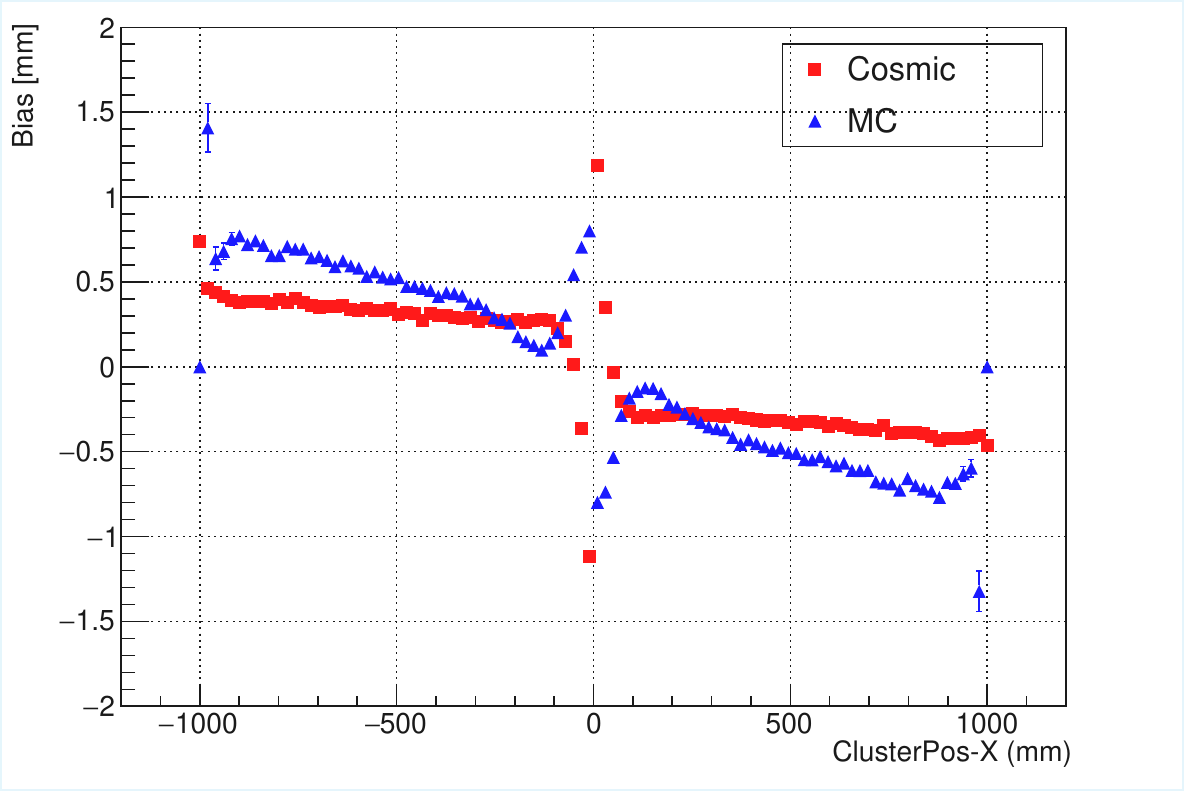}
        \caption{Bias vs drift position $x$. Data: red squares, MC: blue triangles.}
        \label{fig:drift_bias}
    \end{subfigure}
    \caption{Comparison of drift direction resolution and bias for the bottom HA-TPC (bHAT) using cosmic data and simulation.}
    \label{fig:drift_res_bias}
\end{figure*}

Fig.~\ref{fig:drift_res_bias}(a) shows the drift resolution as a function of $x$. In the detector center, the resolution is stable in data, while simulation exhibits a pronounced U-shape and overall higher values, likely reflecting differences in noise treatment. Both degrade at the edges, more sharply in MC, consistent with expected electric field non-uniformities.

Fig.~\ref{fig:drift_res_bias}(b) presents the bias distribution. Across most of the volume, both data and MC remain within \SI{\pm0.5}{\milli\meter}, but deviations appear near the cathode and edges. Real data shows more pronounced asymmetry, indicating possible field distortions or drift velocity mis-calibration near boundaries.

The qualitative agreement between data and simulation in overall trends validates the MC model, though discrepancies in edge behavior and resolution magnitude suggest areas for further refinement.

\subsection{Electric field uniformity}
\label{subsec:Efielduniformity}

\begin{figure*}
    \centering
    \subfloat[]{%
        \includegraphics[width=0.475\textwidth]{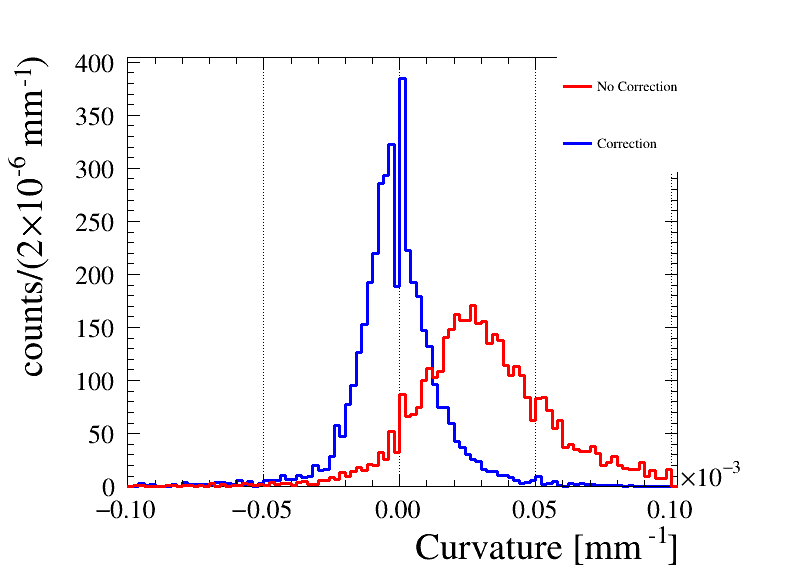}
        \label{subfig:cathoderegioncurv}
    }
    \hfill
    \subfloat[]{%
        \includegraphics[width=0.475\textwidth]{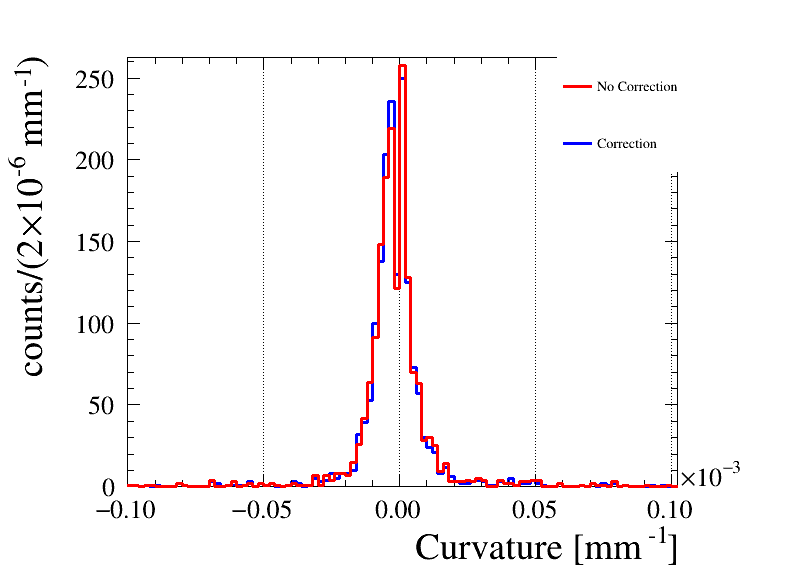}
        \label{subfig:anoderegioncurv}
    }
    \caption{Curvature distribution in bottom HA-TPC endplate 1, at \(z=0\), without (red) and with (blue) electric field correction. (a) cathode region; (b) anode region.
    }
    \label{fig:curvaturedistrib}
\end{figure*}

Electric field distortions were assessed via the apparent curvature \( \mathrm{curv}\propto 1/R_\perp \) of cosmic rays tracks recorded without a magnetic field, where only multiple scattering should affect curvature. Events were selected with (i) at least 50 clusters and (ii) a vertical track component \( |u_y| > 0.9 \). The TPC volume was divided into a 10×4 grid along \(x\) and \(z\), with the most downstream \(z\)-bin labeled 0. Curvature was assigned positive for tracks consistent with negatively charged forward-going muons.

\begin{figure*}
    \centering
    \subfloat[]{%
        \includegraphics[width=0.475\textwidth]{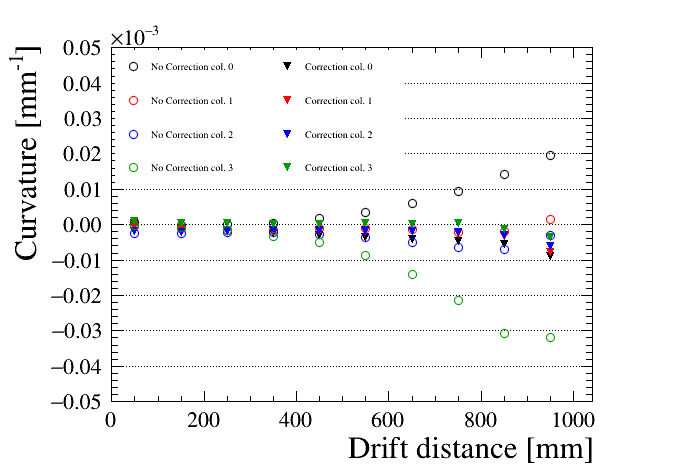}
        \label{subfig:topEP0}
    }
    \hfill
    \subfloat[]{%
        \includegraphics[width=0.475\textwidth]{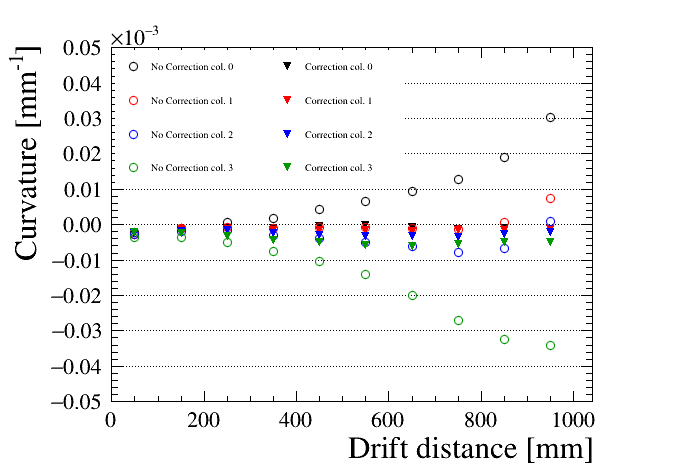}
        \label{subfig:topEP1}
    }

    \vspace{0.5cm}

    \subfloat[]{%
        \includegraphics[width=0.475\textwidth]{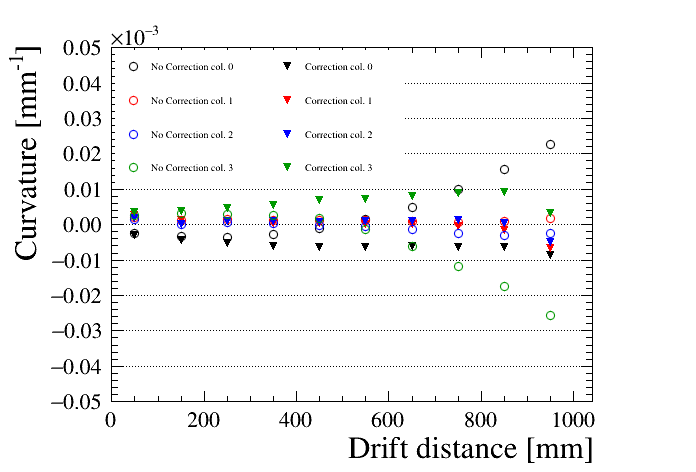}
        \label{subfig:botEP0}
    }
    \hfill
    \subfloat[]{%
        \includegraphics[width=0.475\textwidth]{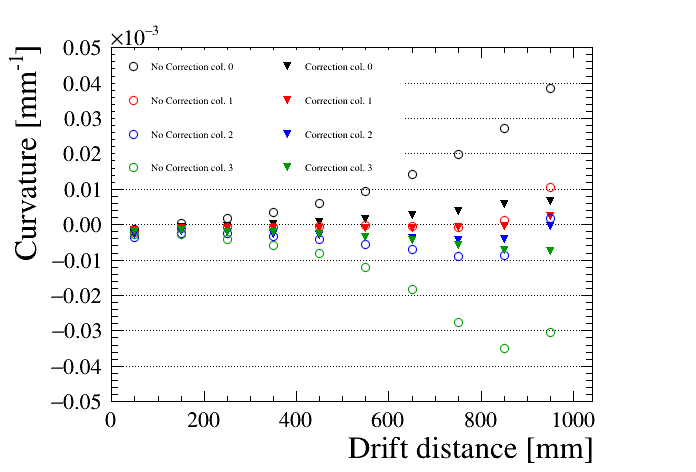}
        \label{subfig:botEP1}
    }
    \caption{Centroid of Gaussian fits to curvature distributions in top HA-TPC endplates 0 (\ref{subfig:topEP0}) and 1 (\ref{subfig:topEP1}), and bottom HA-TPC endplates 0 (\ref{subfig:botEP0}) and 1 (\ref{subfig:botEP1}) before (circles) and after (triangles) the application of Electric field correction. Column 0 refers to the upstream column of two ERAMs, while column 3 refers to the most downstream column.
    }
    \label{fig:curvaturecorrection}
\end{figure*}

As an example, Fig.~\ref{fig:curvaturedistrib} shows curvature distributions for the bottom HA-TPC endplate 1 at \(z=0\), comparing tracks near the cathode and anode with and without electric field correction from the COMSOL map. Distributions were fitted with Gaussians to extract centroids.

The centroids for each endplate, before and after correction, are summarized in Fig.~\ref{fig:curvaturecorrection}. While differences between endplates persist, the corrections significantly reduce cathode-region biases. For comparison, a 1~GeV/\(c\) muon in a 0.2~T field would have a curvature of \(6 \times 10^{-5}\,\mathrm{mm}^{-1}\). The observed apparent curvatures therefore represent a non-negligible effect in momentum reconstruction.

%% file: conclusions.tex
This paper presented the construction details and the initial performances of the new HA-TPCs installed in the upgraded T2K off-axis near detector. The field cages successfully met all mechanical and electrical specifications, with precise geometrical tolerances and reliable structural properties achieved through rigorous production and validation procedures.

Field studies showed that shifting the first field strip by \SI{8}{\milli\meter} toward the anode introduced localized distortions, most pronounced near the cathode. A detailed 3D electric field model with accurate boundary conditions, validated using cosmic ray data without a magnetic field, allowed the characterization and correction of these distortions, ensuring reliable momentum reconstruction.

The ERAM production campaign was summarized, including gain and $RC$ measurements correlated with production parameters. The description of the front-end electronics, data acquisition system, and new gas system was also provided. Drift velocity measurements from the HA-TPCs were found to be consistent with those from the dedicated gas monitoring chamber, validating system integration.

The reconstruction and simulation framework developed for the HA-TPCs was detailed, alongside a comprehensive performance evaluation using cosmic ray and neutrino beam data at J-PARC. Comparisons with Monte Carlo simulations demonstrated reasonable agreement, achieving a spatial resolution better than \SI{800}{\micro\meter} and a $\mathrm{d}E/\mathrm{d}x$ resolution better than 10\% over the full range of track angles and drift distances.

These results confirm that the HA-TPCs meet their design performance goals, providing precise tracking and energy loss measurements essential for the T2K near detector upgrade program.